\documentclass[12pt, a4paper]{article}
\usepackage[text={17cm,25cm},top=3cm]{geometry}  
\usepackage{amsmath,amsfonts,latexsym,mathtools}
\usepackage{mathrsfs} 
\usepackage{graphicx, subfigure}
\usepackage[hidelinks]{hyperref} 
\numberwithin{equation}{section}

\usepackage{xcolor}

\graphicspath{{Figures/}}

\usepackage[title, titletoc, page]{appendix}

\usepackage[super]{nth}

\usepackage{comment}

\usepackage{tabularx}
\usepackage{hhline}
\newcolumntype{Y}{>{\centering\arraybackslash}X}
\usepackage{array, multirow}

\DeclarePairedDelimiter{\abs}{\lvert}{\rvert}

\allowdisplaybreaks[1]

\newcommand{\lb}{\left (}
\newcommand{\rb}{\right )}
\newcommand{\lset}{\left \{}
\newcommand{\rset}{\right \}}
\newcommand{\lsq}{\left [}
\newcommand{\rsq}{\right ]}

\newcommand{\eqtext}[1]{\quad \text{#1} \quad}
\newcommand{\RA}{\quad \Rightarrow \quad}

\newcommand{\sechn}[2]{\mathrm{sech}^{#1} \lb #2 \rb}
\newcommand{\tanhn}[2]{\mathrm{tanh}^{#1} \lb #2 \rb}

\newcommand{\dd}[1]{\; \mathrm{d} #1}

\newcommand{\diffn}[3]{\dfrac{\mathrm{d}^{#1} #2}{\mathrm{d} #3^{#1}}}
\newcommand{\pinv}[1]{\; \partial_{#1}^{-1}}

\renewcommand{\O}[1]{O \lb #1 \rb}

\begin{document}

\baselineskip=15pt
\vspace{-2.5cm}
\title {D'Alembert-type solution of the Cauchy problem for \\a Boussinesq-Klein-Gordon equation}
\date{}
\maketitle
\vspace{-22mm}
\begin{center}
{\bf K.R. Khusnutdinova$^{*}$
\footnote{Corresponding author: K.Khusnutdinova@lboro.ac.uk}, M.R. Tranter$^{*}$} \\[2ex]
$^{*}$ Department of Mathematical Sciences, Loughborough University, \\
Loughborough LE11 3TU, UK.\\[3ex]
{\it Dedicated to Roger Grimshaw on the occasion of his 80th Birthday.}
\vspace{4mm}

\end{center}

\abstract{In this paper we construct a weakly-nonlinear d'Alembert-type solution of the Cauchy problem for a Boussinesq-Klein-Gordon equation. Similarly to our earlier work based on the use of spatial Fourier series, we consider the problem  in the class of periodic functions on an interval of finite length (including the limiting case of an ``infinite" interval with zero boundary conditions), and work with the equation describing a deviation from the mean value.

Unlike our earlier paper, 
here we develop a novel multiple-scales procedure involving fast characteristic variables and two slow time scales, which allows us to construct an explicit and compact d'Alembert-type solution of the nonlinear problem in terms of solutions of two Ostrovsky equations emerging at the leading order and describing the right- and left-propagating waves. 
Validity of the constructed solution follows from our earlier results, and is illustrated numerically for a number of instructive examples, both for periodic solutions on a finite interval, and the well-studied scenario for localised solutions on a large (``infinite") interval. 

Importantly, in all cases the initial conditions for the leading-order Ostrovsky equations by construction have zero mass. Thus, the so-called  ``zero-mass contradiction" has been completely by-passed.
\bigskip
\bigskip
\bigskip


\newpage

\section{Introduction}
\label{sec:Intro}
In recent decades there has been a lot of research associated with the Ostrovsky equation
\begin{equation}
\lb \eta_t + \nu \eta \eta_x + \mu \eta_{xxx} \rb_x = \lambda \eta,
\label{Oe}
\end{equation}
which is a modification of the Korteweg - de Vries (KdV) equation. Here, $\eta_t = \frac{\partial \eta}{\partial t}, \eta_x = \frac{\partial \eta}{\partial x}$, etc.
The equation has initially emerged as a model for long weakly-nonlinear internal and surface waves in the rotating ocean \cite{O} and was extensively studied in this context, where $\eta$ describes the amplitude of a dominant linear long-wave mode in the reference frame moving with the linear wave-speed of this mode, and $\nu, \mu, \lambda$ are the nonlinearity, dispersion and rotation coefficients, respectively  (see \cite{GOSS, H, GH2012} and references therein). In the absence of currents we have $\lambda \mu > 0$, and there are no solitary wave solutions \cite{L, GS, GHO}. Instead, a localised wave packet associated with the extremum of the group velocity emerges as a dominant feature in the long-time asymptotics of a localised initial condition on the infinite line \cite{GH, GHJ}. The emergence of the wavepacket associated with the extremum of the group speed was also reported in a separate study \cite{YK}, devoted to waves in a Toda chain on an elastic substrate, which can be related to the Ostrovsky equation.  More precisely, the speed of the emerging wavepacket is only initially close to the extremum of the group velocity, but deviates from this value in the course of its subsequent evolution, dominated by modulational instability \cite{WJ, WJ15, NSS15}. 
 In the presence of a parallel depth-dependent shear flow, the modified formula for the rotation coefficient $\lambda$ has been derived in \cite{AGK2013, G2013}. In \cite{AGK2014} examples were found when  $\lambda \mu < 0$ for oceanic waves propagating over a shear flow, leading to the emergence of steady wave packets associated with the extremum of the phase velocity. Earlier, such solutions were studied in the context of plasma, but there were no examples for ocean waves \cite{OS}. Long-time evolution of the solutions was investigated numerically and analytically in \cite{AGK2014} and \cite{GSA}, respectively. We also note recent extensions accounting for higher-order nonlinearities and weak dependence on the transverse coordinate  \cite{GPTK, GSM2015, ORS, G2015}. 

The Ostrovsky equation and coupled Ostrovsky equations have also emerged in the studies of long nonlinear longitudinal bulk strain waves in layered elastic waveguides with soft (imperfect) interfaces \cite{KM, GKM, KT2017}, described by coupled Boussinesq-type equations \cite{KSZ}.  Averaging with respect to the fast time was used in \cite{KM} to obtain either uncoupled or coupled Ostrovsky equations, depending on the difference between the linear characteristic speeds in the layers, and the weakly-nonlinear solution of the initial-value problem for localised initial conditions on the infinite line has been constructed in both cases. The behaviour of solutions was shown to be very different in these two cases, resulting in the emergence of radiating solitary waves in the first case, and several wave packets in the second case. Analytical estimates for the amplitude of the tail of the radiating solitary wave solution of coupled Boussinesq-type equations were obtained in \cite{GKM}. Scattering of a radiating solitary wave in bi-layers with delamination has been studied in \cite{KT2017}, using both direct and semi-analytical (weakly-nonlinear) approaches. Coupled Ostrovsky equations have been extensively studied in the oceanic context in \cite{AGK2014}. In particular, the study has shown that in most cases the dominant features of the long-time asymptotics of solutions with localised initial data on the infinite line can be inferred from the linear dispersion curves of the system (extrema of group and phase speed curves and various resonances). 

The validity of long-wave approximations of the KdV type has been studied in many works in the context of water waves, for example see \cite{Benjamin72, Bona02, Bona04, Bona05} and references therein. The emergence of right- and left-propagating KdV and coupled KdV equations as leading order approximations to Boussinesq-type equations and systems was discussed in \cite{Whitham, Johnson97, Gear84}, and justified, including some higher-order corrections, in  \cite{Craig85, Kano86, Kalyakin89,  Schneider98, BenYoussef00, Schneider00, Wayne02}. A weakly-nonlinear extension of d'Alembert's formula for the solution of the Cauchy problem for the linear wave equation has been extended  to the Cauchy problem for the Boussinesq equation and illustrated by several examples with integrable initial conditions  in \cite{KM12} (see also \cite{KMP}).
 
In this paper we reconsider the initial-value problem for a Boussinesq-type equation 
previously studied in our paper \cite{KMP}. This equation  arises as a limit in the case of a layered lattice model \cite{KSZ}, when the particles in one layer of the chain are significantly heavier than in the other layer, which is similar to the case of a chain on an elastic substrate \cite{YK}. The equation governing displacements in the chain takes the form of the regularised Boussinesq-type equation with an additional term ($\sim u$). It is natural to refer to this equation as a Boussinesq-Klein-Gordon (BKG) equation, since it is a combination of a regularised Boussinesq and the linear Klein-Gordon equations.
The equation is given by
\begin{equation}
u_{tt} - c^2 u_{xx} = \epsilon \lsq \frac{\alpha}{2} \lb u^2 \rb_{xx} + \beta u_{ttxx} - \gamma u \rsq,
\label{BousOst}
\end{equation}
where $\gamma > 0$, $c$, $\alpha$ and $\beta$ are constants and $\epsilon$ is a small parameter. 
In this context, $u$ describes the longitudinal strain, $c$ is the linear longitudinal wave speed, $\alpha$ and $\beta$ are the nonlinearity and dispersion coefficients, respectively, and $\gamma$ is defined by elastic properties of the bonding layer or an elastic substrate (see \cite{KSZ} and \cite{YK}). Note that the Benjamin-Bona-Mahony (BBM)-type regularisation \cite{Benjamin72} can be applied to various versions of Boussinesq-type equations in order to bring them to the form (\ref{BousOst}) (see \cite{KSZ, GKS} and references therein). 

We note that up to a scaling of variables one can assume that $c = \alpha = \beta = 1$. Keeping the constants in the model is preferable from the viewpoint of applications. Equation (\ref{BousOst}) has also arisen in the context of oceanic waves in a rotating ocean \cite{Gerkema}. While the accuracy of the Boussinesq-type equation does not exceed the accuracy of uni-directional models in the water-wave context, it is a valid two-directional model in the context of waves in various solid waveguides  (see \cite{Maugin, Samsonov, Porubov, KS, Engelbrecht, Peets} and references therein). We also note that, in the oceanic context, the valid two-directional rotation-modified strongly-nonlinear and Boussinesq-type systems have been derived and discussed in \cite{O, Shrira81, Shrira86, Helfrich2007} (see also \cite{GOSS}).

The Ostrovsky equation (\ref{Oe}) implies that for any regular localised solution on the infinite interval (or periodic solution on a finite interval) the mass is zero for any $t>0$:
\begin{equation*}
	\int_{-\infty}^{\infty} \eta \dd{x} = 0 \eqtext{or} \int_{-L}^L \eta \dd{x} = 0.
\end{equation*}
However, the original physical equations (e.g., Euler equations in fluids) or the equation (\ref{BousOst}) in the context of solids do not impose similar restrictions on the solutions for the respective physical variables. 
If one tries to use the Ostrovsky equation as a uni-directional model directly, say, for a uni-directional initial condition of the equation (\ref{BousOst}) with non-zero mass, there appears a contradiction. Of course, this contradiction emerges from our will to use the equation in order to solve a given initial-value problem. Thus, it is a mathematical contradiction rather than a physical contradiction, because from the physical point of view the equation should only be used to model solutions with zero mass.
 This contradiction has been resolved on the infinite line in \cite{Grimshaw99} by considering a regularised Ostrovsky equation. The regularisation was similar to the regularisation used for the Kadomtsev-Petviashvili equation in \cite{AW}, while the physical motivation has been discussed in \cite{GM}. It was shown that in the regularised Ostrovsky equation there is a rapid adjustment of the mass within the ``temporal boundary layer". The non-zero mass is transported to a large distance in the opposite direction to the propagation of the main wave which has zero mass. 

These arguments are not applicable if we need to model periodic solutions of our physical equations, or solutions on a finite interval. Therefore, in \cite{KMP} we by-passed the zero-mass contradiction by developing a systematic approach to the construction of the weakly-nonlinear solution of the initial-value problem for the Boussinesq-type equation with the Ostrovsky term (\ref{BousOst}) by considering  the deviation from the mean value. Rigorous estimates for the error terms  were obtained, similarly to \cite{PSW},  and convergence rates predicted by this derivation were confirmed by numerical experimentation. The results in \cite{KMP} were obtained in terms of spatial Fourier series for a periodic domain, where the initial condition for $u$ had non-zero mean value. 

In order to obtain a solution in a more explicit form, in this paper we aim to derive a weakly-nonlinear solution of ``d'Alembert's type''.
We develop a novel multiple-scales procedure, constructing the solution of the Boussinesq-type equation with variable coefficients describing the deviation from the oscillating mean value in the form of an asymptotic multiple-scales expansion in increasing powers of $\sqrt{\epsilon}$, using fast characteristic variables and,  importantly, two slow time variables. The procedure allows one to find explicitly the contributions at each order in terms of solutions of the leading-order equations, and still by-pass the zero-mass contradiction, similarly to our earlier work \cite{KMP}.

The paper is organised as follows. We construct a weakly-nonlinear d'Alembert-type solution of the Cauchy problem for a regularised Boussinesq-Klein-Gordon equation, for the case when the initial condition for $u$ may have non-zero mean, in Section \ref{sec:BOWNL}. Rigorous justification of the constructed solution follows from Theorem 2 of \cite{KMP}, where the error terms were controlled in appropriate function spaces. 
The constructed weakly-nonlinear solution is compared with direct numerical simulations in Section \ref{sec:Num}, for a number of periodic solutions on a finite interval. Results are shown for specific values of $\gamma$ and several choices for the mean value of the initial condition, and we also perform the detailed error analysis,
including an increasing number of terms in the weakly-nonlinear expansion. We then consider some examples for localised initial conditions with non-zero mass on a large interval  in Section \ref{sec:GHRes} and show that in this case our solution agrees with the results of previous studies by Grimshaw \cite{Grimshaw99} and Grimshaw and Helfrich \cite{GH} (a localised initial condition with zero mass was considered in  \cite{KMP}). We also illustrate that the Ostrovsky equation can not be used directly for initial conditions with non-zero mass on a finite periodic interval, while even the leading-order constructed solution 
agrees well with the exact (numerical) solution, and consider the KdV cnoidal wave initial conditions.
We conclude in Section \ref{sec:Conc}. We outline the extension of the constructed solution  to the general case, when both initial conditions, for $u$ and $u_t$, may have non-zero mean, in Appendix A. Numerical methods used in our examples are described in Appendix B.

\section{Weakly-nonlinear d'Alembert-type solution}
\label{sec:BOWNL}
We consider the following Cauchy problem for a BKG equation on the domain $\Omega = [-L, L] \times [0, T]$:
\begin{equation}
u_{tt} - c^2 u_{xx} = \epsilon \lsq \frac{\alpha}{2} \lb u^2 \rb_{xx} + \beta u_{ttxx} - \gamma u \rsq,
\label{BousOstOld}
\end{equation}
\begin{equation}
u |_{t=0} = F(x), \quad u_t |_{t=0} = V(x),
\label{BousOstIC}
\end{equation}
where $F$ and $V$ are assumed to be sufficiently smooth $(2L)$-periodic functions. 

As in the earlier work \cite{KMP} (see also \cite{Grimshaw99}), we integrate (\ref{BousOstOld}) in $x$ over the period $2L$ and obtain an evolution equation for the mean value of the form
\begin{equation}
\diffn{2}{ }{t} \int_{-L}^{L} u(x,t) \dd{x} = - \epsilon \gamma \int_{-L}^{L} u(x,t) \dd{x}.
\label{MeanValEq}
\end{equation}
Solving this equation we have the formula for the mean value
\begin{equation}
\langle u \rangle (t) := \frac{1}{2L} \int_{-L}^{L} u(x,t) \dd{x} = A \cos{\lb \sqrt{\epsilon \gamma} t \rb} + B \sin{\lb \sqrt{\epsilon \gamma} t \rb}.
\label{MeanValGS}
\end{equation}
Then, taking the mean value of the initial conditions we obtain
\begin{equation}
\langle u \rangle (t) := \frac{1}{2L} \int_{-L}^{L} u(x,t) \dd{x} = F_{0} \cos{\lb \sqrt{\epsilon \gamma} t \rb} + V_{0} \frac{\sin{\lb \sqrt{\epsilon \gamma} t \rb}}{\sqrt{\epsilon \gamma}},
\label{MeanVal}
\end{equation}
where we have
\begin{equation}
F_{0} = \frac{1}{2L} \int_{-L}^{L} F(x) \dd{x} \eqtext{and} V_{0} = \frac{1}{2L} \int_{-L}^{L} V(x) \dd{x}.
\label{MeanValIC}
\end{equation}
To eliminate $O\lb \frac{1}{\sqrt{\epsilon}} \rb$ oscillations
in the mean value $\langle u \rangle (t)$ we require that
\begin{equation}
V_{0} = \frac{1}{2L} \int_{-L}^{L} V(x) \dd{x} = 0.
\label{utcondition}
\end{equation}
The condition $V_0 = 0$ appears naturally in many physical applications and it is imposed in all cases considered in the main part of this paper in order to simplify the derivations. However, we would like to note that this condition can be removed, and the developed method can be extended to the general case with both $F_0 \ne 0$ and $V_0 \ne 0$ (see Appendix A).

The mean value is subtracted from the original solution to obtain an equation with zero mean value. Thus, we take $\tilde{u} = u - F_{0} \cos{\lb \omega t \rb}$, where $\omega = \sqrt{\epsilon \gamma}$, so we obtain
\begin{equation}
\tilde u_{tt} - c^2 \tilde u_{xx} = \epsilon \lsq \alpha F_0 \cos{\lb \omega t \rb} \tilde u_{xx} + \frac{\alpha}{2} \lb \tilde u^2 \rb_{xx} + \beta \tilde u_{ttxx} - \gamma \tilde u \rsq
\label{BousOstEq}
\end{equation}
and
\begin{equation}
\tilde u |_{t=0} = F(x) - F_0, \quad \tilde u_t |_{t=0} = V(x).
\label{BousOstICnew}
\end{equation}

We now look for a weakly-nonlinear solution of the form
\begin{align}
 \tilde u \lb x, t \rb &= f^{+} \lb \xi_{+}, \tau, T \rb +  f^{-} \lb \xi_{-}, \tau, T \rb + \sqrt{\epsilon} P \lb \xi_{-}, \xi_{+}, \tau, T \rb + \epsilon Q \lb \xi_{-}, \xi_{+}, \tau, T \rb \notag \\
&~~~+ \epsilon^{3/2} R \lb \xi_{-}, \xi_{+}, \tau, T \rb  + \epsilon^2 S \lb \xi_{-}, \xi_{+}, \tau, T \rb + \O{\epsilon^{5/2}},
\label{WNLSol}
\end{align}
where we introduce the following fast characteristic and slow time variables
\begin{equation*}
\xi_{\pm} = x \pm c t, \quad \tau = \sqrt{\epsilon} t, \quad T = \epsilon t.
\end{equation*}
Note that, unlike \cite{KMP}, we introduce two slow time scales and look for a solution of d'Alembert-type, similar to \cite{KM, KM12}, but on a periodic domain instead of the infinite interval. We now substitute (\ref{WNLSol}) into (\ref{BousOstEq}) and (\ref{BousOstICnew}) and collect the terms at equal powers of $\sqrt{\epsilon}$ to find expressions for all functions in the expansion. We noted earlier that the function $u$ is $2L$-periodic in $x$, therefore we require that $f^{-}$ and $f^{+}$ are also $2L$-periodic in $\xi_{-}$ and $\xi_{+}$, respectively. Moreover, it is natural to assume that all terms in the asymptotic expansion for $u$ are products of the functions $f^{-}$, $f^{+}$, and their derivatives. This assumption then implies that all terms are periodic in $\xi_{-}/\xi_{+}$, at fixed $\xi_{+}/\xi_{-}$. Furthermore, as the functions $f^{-}$ and $f^{+}$ have zero mean i.e.
\begin{equation}
\frac{1}{2L} \int_{-L}^{L} f^{\pm} \dd{\xi_{\pm}} = 0,
\end{equation}
then all terms in (\ref{WNLSol}) will have zero mean in $\xi_{-}/\xi_{+}$, at fixed $\xi_{+}/\xi_{-}$.

The equation is satisfied at leading order, therefore we move to $\O{\sqrt{\epsilon}}$.  At each stage we will also satisfy the initial condition for the previous order, as the functions at a given order are introduced by comparing terms at the previous order of $\sqrt{\epsilon}$.

At $\O{\sqrt{\epsilon}}$ the right-hand side of (\ref{BousOstEq}) does not contribute to the equation, so we have
\begin{equation}
- 4 c^2 P_{\xi_{-} \xi_{+}} - 2 c f_{\xi_{-} \tau}^{-} + 2 c f_{\xi_{+} \tau}^{+} = 0.
\label{Osqrteps}
\end{equation}
We average (\ref{Osqrteps}) with respect to the fast spatial variable $x$ at constant $\xi_{-}$ or $\xi_{+}$ i.e. in the reference frame moving with the linear speed of right- or left-propagating waves, respectively. Therefore, at constant $\xi_{-}$ for example, we have
\begin{equation}
\frac{1}{2L} \int_{-L}^{L} P_{\xi_{-} \xi_{+}} \dd{x} = \frac{1}{4L} \int_{-2L - \xi_{-}}^{2L - \xi_{-}} P_{\xi_{-} \xi_{+}} \dd{\xi_{+}} = \frac{1}{4L} \lsq P_{\xi_{-}} \rsq_{-2L - \xi_{-}}^{2L - \xi_{-}} = 0.
\label{PAvg}
\end{equation}
A similar result can be obtained for $\xi_{+}$ and we see that under the averaging $P_{\xi_{-} \xi_{+}} = 0$. Averaging (\ref{Osqrteps}) at constant $\xi_{-}$ and $\xi_{+}$ therefore gives
\begin{equation}
f_{\xi_{-} \tau}^{-} = 0 \eqtext{and} f_{\xi_{+} \tau}^{+} = 0,
\label{OsqrtepsAvgtemp}
\end{equation}
implying that
\begin{equation}
f^{-} = \tilde{f}^{-} \lb \xi_{-}, T \rb + B^{-} \lb \tau, T \rb \eqtext{and} f^{+} = \tilde{f}^{+} \lb \xi_{+}, T \rb + B^{+} \lb \tau, T \rb.
\label{OsqrtepsAvg}
\end{equation}
Noting that we have zero mean of all functions in the expansion, we require that $B^{\pm} = 0$. We will apply this rule at all orders to eliminate any functions of only $\tau$ and $T$. Substituting (\ref{OsqrtepsAvg}) into (\ref{Osqrteps}) gives
\begin{equation}
P_{\xi_{-} \xi_{+}} = 0 \RA P = g^{-} \lb \xi_{-}, \tau, T \rb + g^{+} \lb \xi_{+}, \tau, T \rb.
\label{PExp}
\end{equation}
We rewrite our weakly-nonlinear solution to accommodate these changes, so we have (omitting tildes for $f^{\pm}$)
\begin{align}
\tilde u \lb x, t \rb  &= f^{+} \lb \xi_{+}, T \rb + f^{-} \lb \xi_{-}, T \rb + \sqrt{\epsilon} \lsq g^{+} \lb \xi_{+}, \tau, T \rb + g^{-} \lb \xi_{-}, \tau, T \rb \rsq + \epsilon Q \lb \xi_{-}, \xi_{+}, \tau, T \rb   \notag \\
&~~~ + \epsilon^{3/2} R \lb \xi_{-}, \xi_{+}, \tau, T \rb + \epsilon^2 S \lb \xi_{-}, \xi_{+}, \tau, T \rb + \O{\epsilon^{5/2}}.
\label{WNLOsqrteps}
\end{align}

Substituting (\ref{WNLOsqrteps}) into the initial conditions (\ref{BousOstICnew}) and collecting terms at $\O{1}$ we obtain d'Alembert's formulae for the initial conditions for $f^{\pm}$:
\begin{equation}
	\lset
	\begin{aligned}
		\left. f^{-} + f^{+} \right|_{T=0} = F \lb x \rb - F_0, \\
		\left. -cf_{\xi_{-}}^{-} + cf_{\xi_{+}}^{+} \right|_{T=0} = V \lb x \rb
	\end{aligned}\right.
	\RA  f^{\pm}|_{T=0} = \frac{1}{2c} \lb c [F \lb \xi_{\pm} \rb -  F_0] \pm \int_{-L}^{\xi_{\pm}} V \lb \sigma \rb \dd{\sigma} \rb.
	\label{fIC}
\end{equation}

We now consider the equation at $\O{\epsilon}$, using the results from the previous order:
\begin{align}
-4 c^2 Q_{\xi_{-} \xi_{+}} &= 2cg_{\xi_{-} \tau}^{-} - 2cg_{\xi_{+} \tau}^{+} + \lb 2 c f_T^{-} + \alpha f^{-} f_{\xi_{-}}^{-} + \beta c^2 f_{\xi_{-} \xi_{-} \xi_{-}}^{-} \rb_{\xi_{-}} - \gamma f^{-} \notag \\
&~~~+ \alpha F_0 \cos{\lb \sqrt{\gamma} \tau \rb} \lb f_{\xi_{-} \xi_{-}}^{-} + f_{\xi_{+} \xi_{+}}^{+} \rb + \alpha \lb f_{\xi_{-} \xi_{-}}^{-} f^{+} + 2 f_{\xi_{-}}^{-} f_{\xi_{+}}^{+} + f^{-} f_{\xi_{+} \xi_{+}}^{+} \rb \notag \\
&~~~+ \lb - 2 c f_T^{+} + \alpha f^{+} f_{\xi_{+}}^{+} + \beta c^2 f_{\xi_{+} \xi_{+} \xi_{+}}^{+} \rb_{\xi_{+}} - \gamma f^{+}.
\label{Oeps}
\end{align}
Averaging (\ref{Oeps}) with respect to $x$ at constant $\xi_{-}$ or constant $\xi_{+}$ yields
\begin{equation}
\pm 2 c g_{\xi_{\pm} \tau}^{\pm} =  \alpha F_0  \cos{\lb \sqrt{\gamma} \tau \rb} f_{\xi_{\pm} \xi_{\pm}}^{\pm} + A^{\pm}  \lb \xi_{\pm}, T \rb,
\label{OepsAvgSec}
\end{equation}
where 
\begin{equation}
A^{\pm}  \lb \xi_{\pm}, T \rb = \lb \mp 2 c f_{T}^{\pm} + \alpha f^{\pm} f_{\xi_{\pm}}^{\pm} + \beta c^2 f_{\xi_{\pm} \xi_{\pm} \xi_{\pm}}^{\pm} \rb_{\xi_{\pm}} - \gamma f^{\pm},
\end{equation}
and to avoid secular terms we require $A^{\pm} = 0$. Therefore we have the following equations for $f^{\pm}$ and $g^{\pm}$:
\begin{equation}
\lb \mp 2 c f_{T}^{\pm} + \alpha f^{\pm} f_{\xi_{\pm}}^{\pm} + \beta c^2 f_{\xi_{\pm} \xi_{\pm} \xi_{\pm}}^{\pm} \rb_{\xi_{\pm}} = \gamma f^{\pm},
\label{feq}
\end{equation}
and
\begin{equation}
g^{\pm} = \pm \frac{\alpha F_0}{2 c \sqrt{\gamma}} \sin{\lb \sqrt{\gamma} \tau \rb} f_{\xi_{\pm}}^{\pm} + G^{\pm} \lb \xi_{\pm}, T \rb = \pm \theta f_{\xi_{\pm}}^{\pm} + G^{\pm} \lb \xi_{\pm}, T \rb,
\label{geq}
\end{equation}
where we have introduced the coefficient
\begin{equation}
\theta = \frac{\alpha F_0}{2 c \sqrt{\gamma}} \sin{\lb \sqrt{\gamma} \tau \rb},
\label{theta}
\end{equation}
and the functions $G^{\pm}$ are to be found at the next order. Substituting (\ref{feq}) and (\ref{geq}) into (\ref{Oeps}) and integrating with respect to the characteristic variables we obtain
\begin{equation}
Q \lb \xi_{-}, \xi_{+}, \tau, T \rb = h^{+} \lb \xi_{+}, \tau, T \rb + h^{-} \lb \xi_{-}, \tau, T \rb + h_{c} \lb \xi_{-}, \xi_{+}, T \rb,
\label{Qeq}
\end{equation}
where
\begin{equation}
h_{c} = -\frac{\alpha}{4c^2} \lb 2 f^{-} f^{+} + f_{\xi_{-}}^{-} \int_{-L}^{\xi_{+}} f^{+} (\sigma) \dd{\sigma} +  f_{\xi_{+}}^{+} \int_{-L}^{\xi_{-}} f^{-}(\sigma) \dd{\sigma} \rb.
\label{hceq}
\end{equation}
We again update the weakly-nonlinear solution to reflect the new results for $Q$, derived in (\ref{Qeq}), so we have
\begin{align}
\tilde u \lb x, t \rb &= f^{+} \lb \xi_{+}, T \rb + f^{-} \lb \xi_{-}, T \rb + \sqrt{\epsilon} \left [g^{+} \lb \xi_{+}, \tau, T \rb + g^{-} \lb \xi_{-}, \tau, T \rb \right ] \notag \\
&~~~ + \epsilon \left [ h^{+} \lb \xi_{+}, \tau, T \rb + h^{-} \lb \xi_{-}, \tau, T \rb + h_{c} \lb \xi_{-}, \xi_{+}, T \rb \right ] \notag \\
&~~~+ \epsilon^{3/2} R \lb \xi_{-}, \xi_{+}, \tau, T \rb + \epsilon^2 S \lb \xi_{-}, \xi_{+}, \tau, T \rb + \O{\epsilon^{5/2}}.
\label{WNLOeps}
\end{align}

Substituting (\ref{WNLOeps}) into (\ref{BousOstICnew}) and now collecting terms at $\O{\sqrt{\epsilon}}$ we obtain 
\begin{equation*}
	\lset
	\begin{aligned}
		\left. g^{-} + g^{+} \right|_{T=0} &= 0, \\
		\left. -cg_{\xi_{-}}^{-} + cg_{\xi_{+}}^{+} \right|_{T=0} &= 0
	\end{aligned}\right.
	\RA 
	\lset
	\begin{aligned}
		\left. -\theta f_{\xi_{-}}^{-} + G^{-} + \theta f_{\xi_{+}}^{+} + G^{+} \right|_{T=0} &= 0, \\
		\left. -c\theta f_{\xi_{-} \xi_{-}}^{-} - cG_{\xi_{-}}^{-} + c\theta f_{\xi_{+} \xi_{+}}^{+} + cG_{\xi_{+}}^{+} \right|_{T=0} &= 0.
	\end{aligned}\right.
\end{equation*}
From (\ref{theta}) we see that, at $T=0$, $\theta = 0$ and therefore we have
\begin{equation}
	\lset
	\begin{aligned}
		\left. G^{-} + G^{+} \right|_{T=0} &= 0, \\
		\left. -cG_{\xi_{-}}^{-} + cG_{\xi_{+}}^{+} \right|_{T=0} &= 0
	\end{aligned}\right.
	\RA G^{\pm}|_{T=0} = 0.
\label{GIC}
\end{equation}

Next, at  $\O{\epsilon^{3/2}}$ we obtain
\begin{align}
-4 c^2 R_{\xi_{-} \xi_{+}} &= 2 c h_{\xi_{-} \tau}^{-} - 2 c h_{\xi_{+} \tau}^{+} + \lb 2 c g_T^{-} + \alpha \lb f^{-} g^{-} \rb_{\xi_{-}} + \beta c^2 g_{\xi_{-} \xi_{-} \xi_{-}}^{-} \rb_{\xi_{-}} - \gamma g^{-} \notag \\
&~~~- g_{\tau \tau}^{-} - g_{\tau \tau}^{+} + \lb - 2 c g_T^{+} + \alpha \lb f^{+} g^{+} \rb_{\xi_{+}} + \beta c^2 g_{\xi_{+} \xi_{+} \xi_{+}}^{+} \rb_{\xi_{+}} - \gamma g^{+} \notag \\
&~~~+ \alpha \lb f_{\xi_{-} \xi_{-}}^{-} g^{+} + 2 f_{\xi_{-}}^{-} g_{\xi_{+}}^{+} + f^{-} g_{\xi_{+} \xi_{+}}^{+} + g_{\xi_{-} \xi_{-}}^{-} f^{+} + 2 g_{\xi_{-}}^{-} f_{\xi_{+}}^{+} + g^{-} f_{\xi_{+} \xi_{+}}^{+} \rb \notag \\
&~~~+ \alpha F_0 \cos{\lb \sqrt{\gamma} \tau \rb} \lb g_{\xi_{-} \xi_{-}}^{-} + g_{\xi_{+} \xi_{+}}^{+} \rb.
\label{Oeps32}
\end{align}
Substituting (\ref{geq}) into (\ref{Oeps32}) and averaging with respect to $x$ at constant $\xi_{-}$ or constant $\xi_{+}$ yields
\begin{align}
\pm 2 c h_{\xi_{\pm} \tau}^{\pm} &= \theta \lb \mp 2 c f_{T}^{\pm} + \alpha f^{\pm} f_{\xi_{\pm}}^{\pm} + \beta c^2 f_{\xi_{\pm} \xi_{\pm} \xi_{\pm}}^{\pm} \rb_{\xi_{\pm} \xi_{\pm}} - \gamma \theta f_{\xi_{\pm}}^{\pm} \notag \\
&~~~+ \lb \mp 2 c G_{T}^{\pm} + \alpha \lb f^{\pm} G^{\pm} \rb_{\xi_{\pm}} + \beta c^2 G_{\xi_{\pm} \xi_{\pm} \xi_{\pm}}^{\pm} \rb_{\xi_{\pm}} - \gamma G^{\pm} \notag \\
&~~~ \pm \gamma \theta f_{\xi_{\pm}}^{\pm} \pm  \alpha F_0 \theta \cos{\lb \sqrt{\gamma} \tau \rb} f_{\xi_{\pm} \xi_{\pm} \xi_{\pm}}^{\pm}.
\label{Oeps32Avg}
\end{align}
Differentiating (\ref{feq}) with respect to the appropriate characteristic variable, we can eliminate the first line from (\ref{Oeps32Avg}) and therefore we have an expression for $h_{\xi_{\pm} \tau}^{\pm}$ of the form
\begin{equation}
h_{\xi_{\pm} \tau}^{\pm} = \frac{\theta \gamma}{2c} f_{\xi}^{\pm} + \frac{\alpha F_0 \theta}{2c} \cos{\lb \sqrt{\gamma} \tau \rb} f_{\xi_{\pm} \xi_{\pm} \xi_{\pm}}^{\pm} + \tilde{G}^{\pm} \lb \xi_{\pm}, T \rb, 
\label{heqdiff}
\end{equation}
where
\begin{equation}
\tilde{G}^{\pm} \lb \xi_{\pm}, T \rb = \lb \mp 2 c G_{T}^{\pm} + \alpha \lb f^{\pm} G^{\pm} \rb_{\xi_{\pm}} + \beta c^2 G_{\xi_{\pm} \xi_{\pm} \xi_{\pm}}^{\pm} \rb_{\xi_{\pm}} - \gamma G^{\pm}.
\end{equation}
To avoid secular terms we require that $\tilde{G}^{\pm} = 0$. Therefore we have an equation for $G^{\pm}$ of the form
\begin{equation}
\lb \mp 2 c G_{T}^{\pm} + \alpha \lb f^{\pm} G^{\pm} \rb_{\xi_{\pm}} + \beta c^2 G_{\xi_{\pm} \xi_{\pm} \xi_{\pm}}^{\pm} \rb_{\xi_{\pm}} = \gamma G^{\pm}.
\label{Geq}
\end{equation}
At this stage we note that the initial condition for $G^{\pm}$, as derived in (\ref{GIC}), is $G^{\pm}|_{T=0} = 0$ and therefore we see that $G^{\pm} = 0$ for all times. Integrating (\ref{heqdiff}) we obtain
\begin{equation}
h^{\pm} = \frac{\gamma \rho}{2c} f^{\pm} - \frac{\gamma \rho^2}{2} f_{\xi_{\pm} \xi_{\pm}}^{\pm} + \phi^{\pm} \lb \xi_{\pm}, T \rb,
\label{heq}
\end{equation}
where $ \displaystyle \rho =  \pinv{\tau} \theta = - \frac{\alpha F_0}{2 c \gamma} \cos \lb \sqrt{\gamma} \tau \rb$.
Substituting (\ref{heq}) into (\ref{Oeps32}) and integrating with respect to the characteristic variables we have
\begin{equation} 
R \lb \xi_{-}, \xi_{+}, \tau, T \rb = \psi^{+} \lb \xi_{+}, \tau, T \rb+  \psi^{-} \lb \xi_{-}, \tau, T \rb + \psi_{c} \lb \xi_{-}, \xi_{+}, \tau, T \rb,
\label{Req}
\end{equation}
where
\begin{equation}
\psi_{c} = -\frac{\alpha}{4c^2} \lb f^{-} f_{\xi_{+}}^{+} - f_{\xi_{-}}^{-} f^{+} + f_{\xi_{+} \xi_{+}} \int_{-L}^{\xi_{-}} f^{-}(\sigma) \dd{\sigma} - f_{\xi_{-} \xi_{-}}^{-} \int_{-L}^{\xi_{+}} f^{+}(\sigma) \dd{\sigma} \rb.
\label{psiceq}
\end{equation}
We once again update our weakly-nonlinear solution:
\begin{align}
\tilde u \lb x, t \rb &= f^{+} \lb \xi_{+}, T \rb + f^{-} \lb \xi_{-}, T \rb + \sqrt{\epsilon} \left [g^{+} \lb \xi_{+}, \tau, T \rb + g^{-} \lb \xi_{-}, \tau, T \rb \right ] \notag \\
&~~~ + \epsilon \left [ h^{+} \lb \xi_{+}, \tau, T \rb + h^{-} \lb \xi_{-}, \tau, T \rb + h_{c} \lb \xi_{-}, \xi_{+}, T \rb \right ] \notag \\
&~~~+ \epsilon^{3/2} \lsq \psi^{+} \lb \xi_{+}, \tau, T \rb + \psi^{-} \lb \xi_{-}, \tau, T \rb+ \psi_{c} \lb \xi_{-}, \xi_{+}, \tau, T \rb \rsq \notag \\
&~~~+ \epsilon^2 S \lb \xi_{-}, \xi_{+}, \tau, T \rb + \O{\epsilon^{5/2}}.
\label{WNLOeps32}
\end{align}

Substituting (\ref{WNLOeps32}) into the initial conditions (\ref{BousOstICnew}) and now collecting terms at $\O{\epsilon}$ we obtain 
\begin{align*}
	&\lset
	\begin{aligned}
		\left. h^{-} + h^{+} + h_{c} \right|_{T=0} &= 0, \\
		\left. f_{T}^{-} + f_{T}^{+} + g_{\tau}^{-} + g_{\tau}^{+} - c h_{\xi_{-}}^{-} + c h_{\xi_{+}}^{+} - c h_{c\, \xi_{-}} + c h_{c\, \xi_{+}} \right|_{T=0} &= 0
	\end{aligned}\right. \\
	&\RA   \phi^{\pm}|_{T=0} = \frac{1}{2c}  \lb c J \lb \xi_{\pm} \rb \mp \int_{-L}^{x \pm t} K \lb \sigma \rb \dd{\sigma} \rb ,
\end{align*}
where we define
\begin{align}
J &= -h_{c} - \frac{\gamma \rho}{2c} f^{-} + \frac{\gamma \rho^2}{2} f_{\xi_{-} \xi_{-}}^{-} - \frac{\gamma \rho}{2c} f^{+} + \left . \frac{\gamma \rho^2}{2} f_{\xi_{+} \xi_{+}}^{+} \right |_{T=0}, \notag \\
K &= f_{T}^{-} + f_{T}^{+} + \frac{\gamma \rho}{2c} f_{\xi_{-}}^{-} - \frac{\gamma \rho}{2c} f_{\xi_{+}}^{+} + \frac{\gamma \rho^2}{2} f_{\xi_{-} \xi_{-} \xi_{-}}^{-}  \notag \\
&~~~   \left . - \frac{\gamma \rho^2}{2} f_{\xi_{+} \xi_{+} \xi_{+}}^{+} - h_{c\, \xi_{-}} + h_{c\, \xi_{+}}\right |_{T=0}.
\label{phiIC}
\end{align}

Finally, at $\O{\epsilon^2}$ we have
\begin{align}
-4 c^2 S_{\xi_{-} \xi_{+}} &= -2 g_{\tau T}^{-} - 2 g_{\tau T}^{+} - f_{TT}^{-} - f_{TT}^{+} - h_{\tau \tau}^{-} - h_{\tau \tau}^{+} + 2 c h_{\xi_{-} \tau}^{-} - 2 c h_{\xi_{+} \tau}^{+} + 2 c h_{c_{\xi_{-} T}}^{-} \notag \\
&~~~- 2 c h_{c_{\xi_{+} T}}^{+} + 2c \psi_{\xi_{-} \tau}^{-} - 2c \psi_{\xi_{+} \tau}^{+} + \alpha \lb f^{-} h^{-} \rb_{\xi_{-} \xi_{-}} + \alpha \lb f^{+} h^{+} \rb_{\xi_{+} \xi_{+}} \notag \\
&~~~+ \frac{\alpha}{2} \lb g^{-^2} \rb_{\xi_{-} \xi_{-}} + \frac{\alpha}{2} \lb g^{+^2} \rb_{\xi_{+} \xi_{+}} + \beta c^2 h_{\xi_{-} \xi_{-} \xi_{-} \xi_{-}}^{-} + \beta c^2 h_{\xi_{+} \xi_{+} \xi_{+} \xi_{+}}^{+} - \gamma h^{-} - \gamma h^{+} \notag \\
&~~~+  \alpha F_0 \cos{\lb \sqrt{\gamma} \tau \rb} \lb h_{\xi_{-} \xi_{-}}^{-} + h_{\xi_{+} \xi_{+}}^{+} \rb - 2 c \beta g_{\xi_{-} \xi_{-} \xi_{-} \tau}^{-} + 2 c \beta g_{\xi_{+} \xi_{+} \xi_{+} \tau}^{+} \notag \\
&~~~- 2 c \beta f_{\xi_{-} \xi_{-} \xi_{-} T}^{-} + 2 c \beta f_{\xi_{+} \xi_{+} \xi_{+} T}^{+} - 4 \mu_{c},
\label{Oeps2}
\end{align}
where the last term $\mu_{c}$ contains all the coupling terms between $f^{\pm}$, $g^{\pm}$ and $h^{\pm}$. 
When averaging (\ref{Oeps2}) with respect to $x$ at constant $\xi_{-}$ or constant $\xi_{+}$, the coupling terms are averaged out and therefore the averaging yields
\begin{align}
\pm 2 c \psi_{\xi_{\pm} \tau}^{\pm} &= H_1^{\pm} \lb \xi_{\pm}, \tau, T \rb + H_2^{\pm} \lb \xi_{\pm}, T \rb,
\label{Oeps2Avg}
\end{align}
where the functions $H_{1}^{\pm}$, $H_{2}^{\pm}$ are found from (\ref{Oeps2}). Integrating (\ref{Oeps2Avg}) with respect to the relevant characteristic variables, we see that to avoid secular terms we require $H_{2}^{\pm} = 0$, implying
\begin{align}
\lb \mp 2 c \phi_{T}^{\pm} + \alpha \lb f^{\pm} \phi^{\pm} \rb_{\xi_{\pm}} + \beta c^2 \phi_{\xi_{\pm} \xi_{\pm} \xi_{\pm}}^{\pm} \rb_{\xi_{\pm}} &= \gamma \phi^{\pm} + f_{TT}^{\pm} \mp 2 c \beta f_{\xi_{\pm} \xi_{\pm} \xi_{\pm} T}^{\pm} \notag \\
&~~~ + \frac{\gamma \tilde{\theta}^2}{2} f_{\xi_{\pm} \xi_{\pm}}^{\pm} - \frac{\alpha \tilde{\theta}^2}{2} \lb f_{\xi_{\pm}}^{\pm^2} \rb_{\xi_{\pm} \xi_{\pm}},
\label{phieq}
\end{align}
where
\begin{equation}
\tilde{\theta} = \frac{\theta}{\sin{\lb \sqrt{\gamma} \tau \rb}} = \frac{\alpha F_0}{2 c \sqrt{\gamma}}.
\label{thetatilde}
\end{equation}
At this stage we have fully defined all functions present at $\O{\epsilon}$, but the procedure can be continued to any order.

To summarise, returning to the original variable $u(x,t)$, the weakly-nonlinear d'Alembert-type solution of the original Cauchy problem (\ref{BousOstOld}), (\ref{BousOstIC}) for the case when $V_0 = 0$ (see Appendix A for the extension to the general case)  takes the form
\begin{align}
u \lb x, t \rb &= F_0 \cos (\sqrt{\gamma} \tau) + f^{+} \lb \xi_{+}, T \rb +  f^{-} \lb \xi_{-}, T \rb  +\sqrt{\epsilon}  \lsq g^{+} \lb \xi_{+}, \tau, T \rb + g^{-} \lb \xi_{-}, \tau, T \rb \rsq \notag \\
&~~~+ \epsilon \lsq h^{+} \lb \xi_{+}, \tau, T \rb + h^{-} \lb \xi_{-}, \tau, T \rb + h_{c} \lb \xi_{-}, \xi_{+}, T \rb \rsq + \O{\epsilon^{3/2}},
\label{WNLFinal}
\end{align}
where the leading order function $f^+(\xi_+, T)$ and $f^-(\xi_-, T)$ satisfy two independent Ostrovsky equations
\begin{equation}
\lb \mp 2 c f_{T}^{\pm} + \alpha f^{\pm} f_{\xi_{\pm}}^{\pm} + \beta c^2 f_{\xi_{\pm} \xi_{\pm} \xi_{\pm}}^{\pm} \rb_{\xi_{\pm}} = \gamma f^{\pm},
\label{feq1}
\end{equation}
which should be solved subject to the following initial conditions
\begin{equation}
f^{\pm}|_{T=0} = \frac{1}{2c} \lb c [F \lb \xi_{\pm} \rb - F_0] \pm \int_{-L}^{\xi_{\pm}} V \lb \sigma \rb \dd{\sigma} \rb.
\end{equation}
The accuracy of this leading order solution is easily improved by adding terms of $\O{\sqrt{\epsilon}}$, where
\begin{equation}
g^{\pm} = \pm \frac{\alpha F_0}{2 c \sqrt{\gamma}} \sin{\lb \sqrt{\gamma} \tau \rb} f_{\xi_{\pm}}^{\pm}  = \pm \theta f_{\xi_{\pm}}^{\pm},
\label{geq1}
\end{equation}
and we have introduced the coefficient
\begin{equation}
\theta = \frac{\alpha F_0}{2 c \sqrt{\gamma}} \sin{\lb \sqrt{\gamma} \tau \rb} = \tilde \theta \sin \lb \sqrt{\gamma} \tau \rb.
\label{theta1}
\end{equation}
To improve the accuracy of the solution even further, we need to add the $\O{\epsilon}$ terms, where
\begin{align}
h^{\pm} &= \frac{\gamma \rho}{2c} f^{\pm} - \frac{\gamma \rho^2}{2} f_{\xi_{\pm} \xi_{\pm}}^{\pm} + \phi^{\pm} \lb \xi_{\pm}, T \rb,
\label{heq1} \\
h_{c} &= -\frac{\alpha}{4c^2} \lb 2 f^{-} f^{+} + f_{\xi_{-}}^{-} \int_{-L}^{\xi_{+}} f^{+} (\sigma) \dd{\sigma} +  f_{\xi_{+}}^{+} \int_{-L}^{\xi_{-}} f^{-}(\sigma) \dd{\sigma} \rb,
\label{hceq1} 
\end{align}
and the functions $ \phi^{\pm} \lb \xi_{\pm}, T \rb$ are found by solving the linearised Ostrovsky equations
\begin{align}
\lb \mp 2 c \phi_{T}^{\pm} + \alpha \lb f^{\pm} \phi^{\pm} \rb_{\xi_{\pm}} + \beta c^2 \phi_{\xi_{\pm} \xi_{\pm} \xi_{\pm}}^{\pm} \rb_{\xi_{\pm}} &= \gamma \phi^{\pm} + f_{TT}^{\pm} \mp 2 c \beta f_{\xi_{\pm} \xi_{\pm} \xi_{\pm} T}^{\pm} \notag \\
&~~~ + \frac{\gamma \tilde{\theta}^2}{2} f_{\xi_{\pm} \xi_{\pm}}^{\pm} - \frac{\alpha \tilde{\theta}^2}{2} \lb f_{\xi_{\pm}}^{\pm^2} \rb_{\xi_{\pm} \xi_{\pm}},
\end{align}
subject to the initial conditions
\begin{equation}
\phi^{\pm}|_{T = 0} = \frac{1}{2c} \left. \lb c J \lb \xi_{\pm} \rb \mp \int_{-L}^{\xi_{\pm}} K \lb \sigma \rb \dd{\sigma} \rb \right |_{T = 0},
\label{phipm} 
\end{equation}
where the functions $J$ and $K$ are given by the formulae (\ref{phiIC}).

Validity of the asymptotic expansion (\ref{WNLFinal}) follows from Theorem 2 of \cite{KMP}, where rigorous estimates for the error terms have been obtained in appropriate function spaces. We would like to emphasise that the Ostrovsky equations (\ref{feq1}) derived for the deviation from the mean have the same form as the usual Ostrovsky equations with constant coefficients, which can be derived from the original problem formulation. However, they are now solved for initial conditions which by construction have zero mass. Thus, the zero-mass contradiction has been by-passed.

\section{Periodic solutions on a finite interval}
\label{sec:Num}

In this section we compare the weakly-nonlinear solution (\ref{WNLFinal}) to the ``exact" solution of the Cauchy problem (\ref{BousOstOld}), (\ref{BousOstIC}),  obtained by direct numerical simulations. We compare several approximations with an increasing number of terms and perform the error analysis. Let us denote the direct numerical solution to (\ref{BousOstOld}), (\ref{BousOstIC}) as $u_{\text{num}}$, weakly-nonlinear solution (\ref{WNLFinal}) with only the leading order terms included as $u_{1}$, with terms up to and including $\O{\sqrt{\epsilon}}$ terms as $u_{2}$ and with terms up to and including $\O{\epsilon}$ as $u_{3}$. We consider the maximum absolute error over $x$, defined as
\begin{equation}
e_{i} = \max_{-L \leq x \leq L} \abs{u_{\text{num}} \lb x, t \rb - u_{i} \lb x, t \rb}, \quad i = 1, 2, 3,
\label{MaxErr}
\end{equation}
and use a least-squares power fit to determine how the maximum absolute error varies with the small parameter $\epsilon$. Therefore we write the errors in the form
\begin{equation}
\mathrm{exp} \lsq e_{i} \rsq = C_{i} \epsilon^{\alpha_{i}},
\label{Err}
\end{equation}
and take the logarithm of both sides to form the error plot (the exponential factor is included so that we have $e_{i}$ as the plotting variable). The values of $C_{i}$ and $\alpha_{i}$ are found using the MATLAB function \textit{polyfit}.

To determine the initial conditions, we note that the leading-order weakly-nonlinear solution is governed by the Ostrovsky equations (\ref{feq1}) which, if we take $\gamma = 0$, will reduce to KdV equations. We choose the first initial condition, defining $u(x, 0)$, to be the solitary wave solution of the KdV equation constructed in the usual way (see, for example, \cite{Whitham, Johnson97}). We add a constant to increase the mean value of the initial condition (and therefore the value of $F_0$), and to initiate a non-trivially evolving solution. We choose the second initial condition, defining $u_t (x,0)$  in such a way that there is no leading-order left-propagating wave. Explicitly we take
\begin{align}
u(x,0) &= A \sechn{2}{\frac{x}{\Lambda}} + d, \\
u_t (x,0) &= \frac{2cA}{\Lambda} \sechn{2}{\frac{x}{\Lambda}} \tanhn{ }{\frac{x}{\Lambda}},
\label{IC}
\end{align}
where $d$ is a constant and we have
\begin{equation}
A = \frac{6ck^2}{\alpha}, \quad \Lambda = \frac{\sqrt{2c \beta}}{k}.
\label{BOstCoefs}
\end{equation}
Here, $k$ is a parameter, and we choose $k = \sqrt{\alpha/3c}$.
The mean value term $F_0$ is given by
\begin{equation}
F_0 = d + \frac{A \Lambda}{2 L} \lb \tanh{(L)} - \tanh{(-L)} \rb \approx d + \frac{A \Lambda}{L} \eqtext{for sufficiently large $L$.}
\label{c0val}
\end{equation}
The initial condition for $f^{-}$ is chosen using (\ref{fIC}), and we have $f^{+} = 0$. The initial conditions for the functions $\phi^{\pm}$ are chosen using the formulae (\ref{phiIC}).

\subsection{\texorpdfstring{Example 1: $c = \alpha = \beta = 1$, $\gamma = 0.1$ and $\gamma = 0.5$}{Example 1: c = alpha = beta = 1, gamma = 0.1 and gamma = 0.5}}
\label{sec:Resc1}
In this section we solve (\ref{BousOstOld}) with $c = \alpha = \beta = 1$. 
We compute the results for various values of $\gamma$ and $d$ (corresponding to different values of $F_0$). In Figure \ref{fig:g01} we show that, for $\gamma = 0.1$, an increase in $d$ from $d=1$ to $d=7$ increases the error in the solution (the weakly-nonlinear solution is less accurate). The same behaviour occurs in Figure \ref{fig:g05}, and by comparison to Figure \ref{fig:g01} we see that an increase in $\gamma$ from $\gamma = 0.1$ to $\gamma = 0.5$ also increases the error. 
\begin{figure}[!htbp]
	\subfigure[$\gamma = 0.1$ and $d = 1$.]{\includegraphics[width=0.48\textwidth]{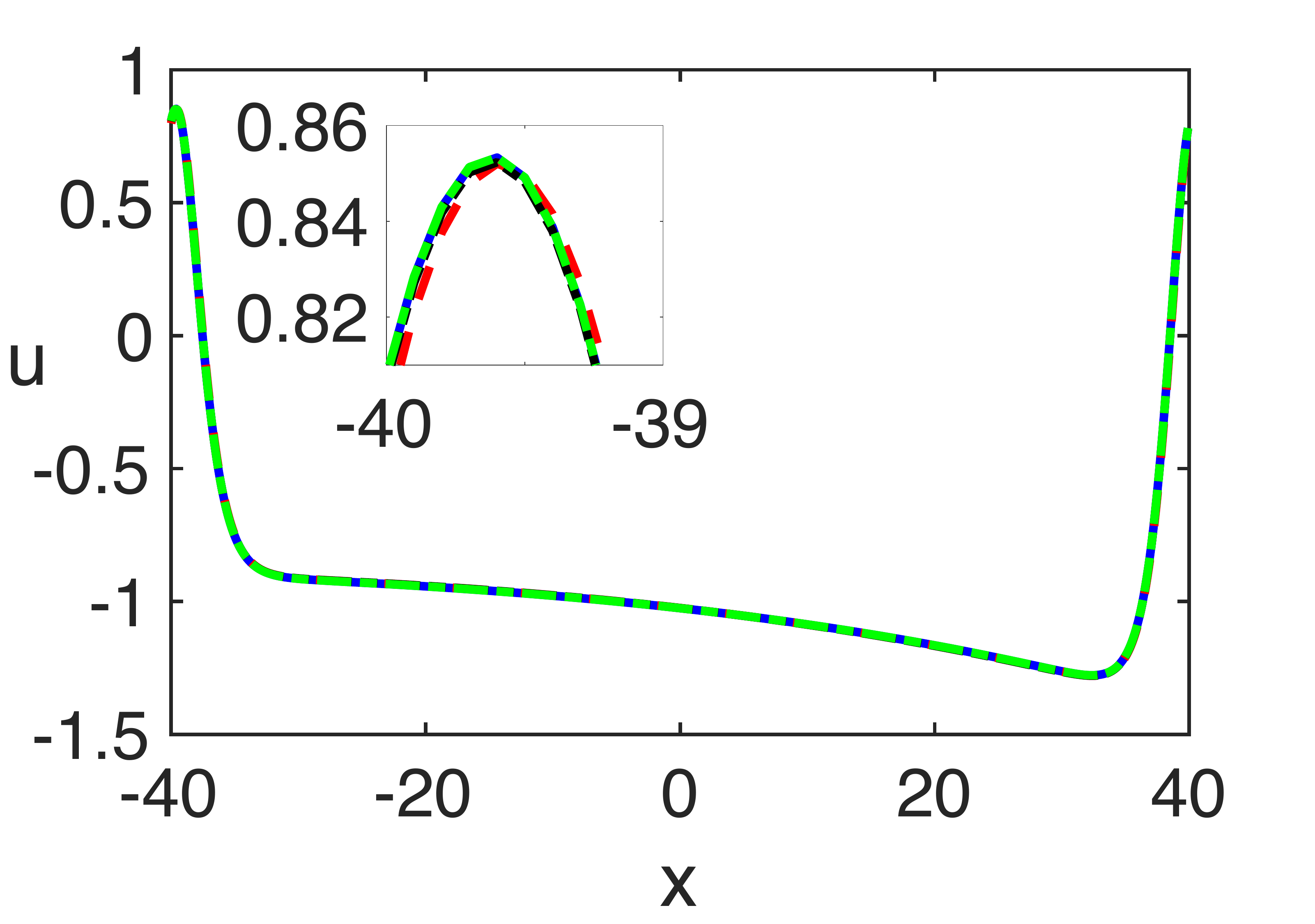}}~
	\subfigure[$\gamma = 0.1$ and $d = 7$.]{\includegraphics[width=0.48\textwidth]{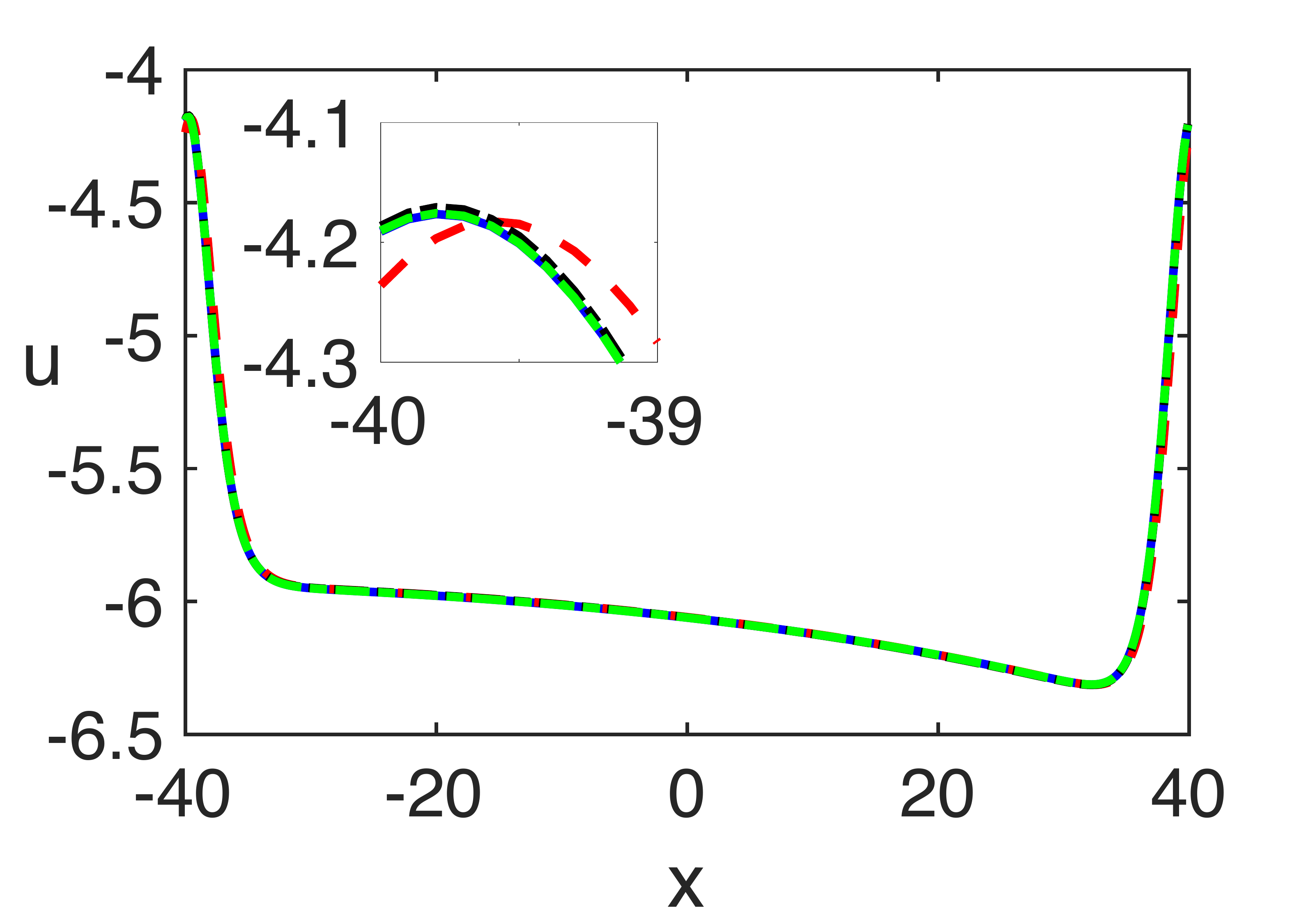}}
	\caption{\small A comparison of the numerical solution (solid, blue) at $t=1/\epsilon$ and the weakly-nonlinear solution including leading-order (dashed, red), $\O{\sqrt{\epsilon}}$ (dash-dot, black) and $\O{\epsilon}$ (dotted, green) corrections, for (a) $d=1$ and (b) $d=7$. Parameters are $L=40$, $N=800$, $k = 1/\sqrt{3}$, $\alpha = \beta = c = 1$, $\gamma = 0.1$, $\epsilon = 0.001$, $\Delta t = 0.01$ and $\Delta T = \epsilon \Delta t$. The solution agrees reasonably well to leading order, and this agreement is improved with the addition of higher-order corrections.}
	\label{fig:g01}
	\vspace{2em}
	\subfigure[$\gamma = 0.5$ and $d = 1$.]{\includegraphics[width=0.48\textwidth]{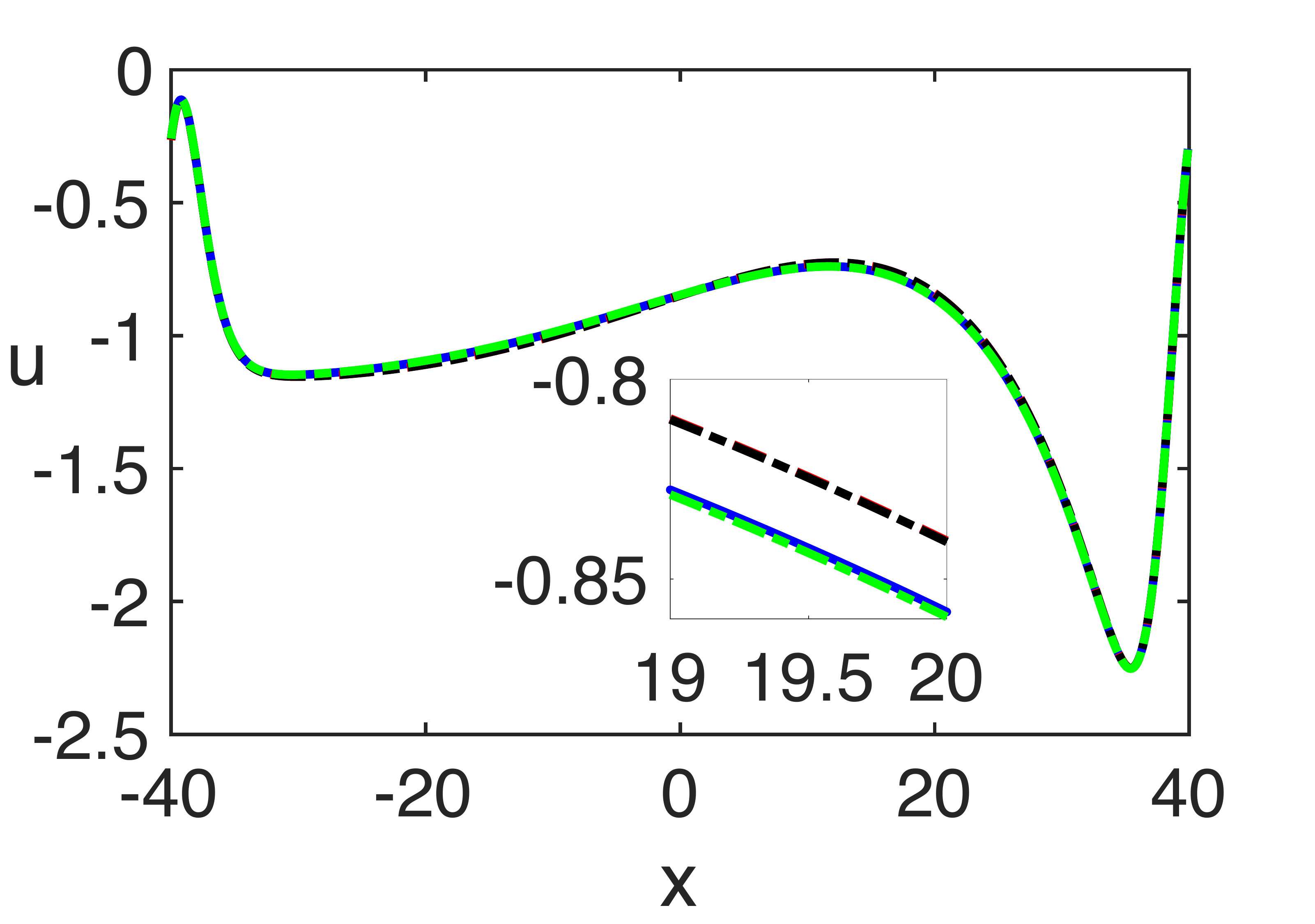}}~
	\subfigure[$\gamma = 0.5$ and $d = 7$.]{\includegraphics[width=0.48\textwidth]{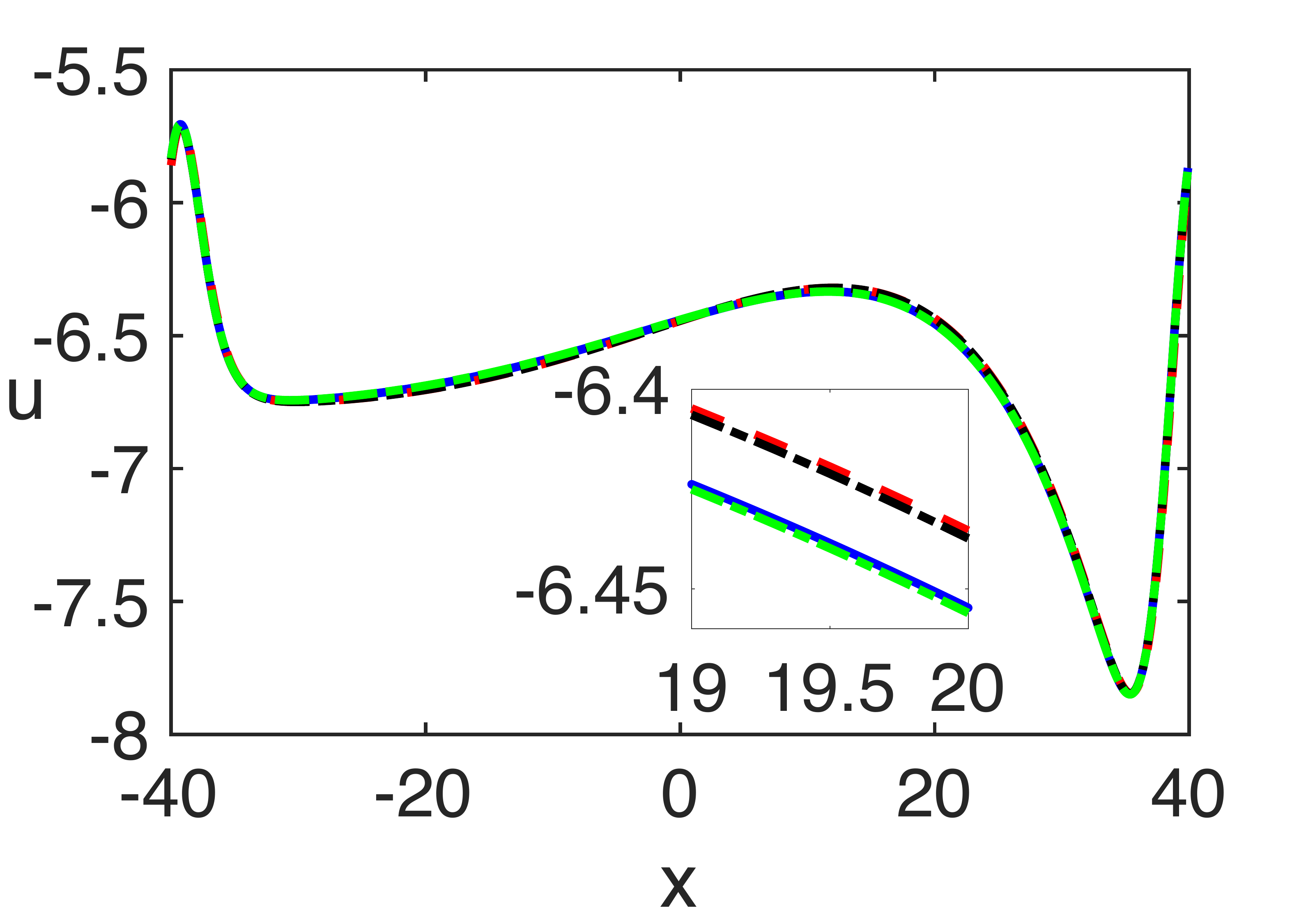}}
	\caption{\small A comparison of the numerical solution (solid, blue) at $t=1/\epsilon$ and  the weakly-nonlinear solution including leading-order (dashed, red), $\O{\sqrt{\epsilon}}$ (dash-dot, black) and $\O{\epsilon}$ (dotted, green) corrections, for (a) $d=1$ and (b) $d=7$. Parameters are $L=40$, $N=800$, $k = 1/\sqrt{3}$, $\alpha = \beta = c = 1$, $\gamma = 0.5$, $\epsilon = 0.001$, $\Delta t = 0.01$ and $\Delta T = \epsilon \Delta t$. The solution agrees well to leading order, and this agreement is improved with the addition of higher-order corrections.}
	\label{fig:g05}
\end{figure}

To further understand the behaviour of the errors, we plot the corresponding error curves for the cases shown in Figures \ref{fig:g01} and \ref{fig:g05}. These results are presented in Figure \ref{fig:Errg01} and Figure \ref{fig:Errg05} for $\gamma = 0.1$ and $\gamma = 0.5$, respectively. We see that the error curves in Figure \ref{fig:Errg01} have slope 0.5, 1 and 1.5, corresponding to errors at $\O{\sqrt{\epsilon}}$, $\O{\epsilon}$ and $\O{\epsilon^{3/2}}$ respectively. This can be understood from (\ref{WNLFinal}), as the inclusion of terms at a given order of the expansion will result in errors at the next order.
\begin{figure}[!htbp]
	\subfigure[$\gamma = 0.1$ and $d = 1$.]{\includegraphics[width=0.48\textwidth]{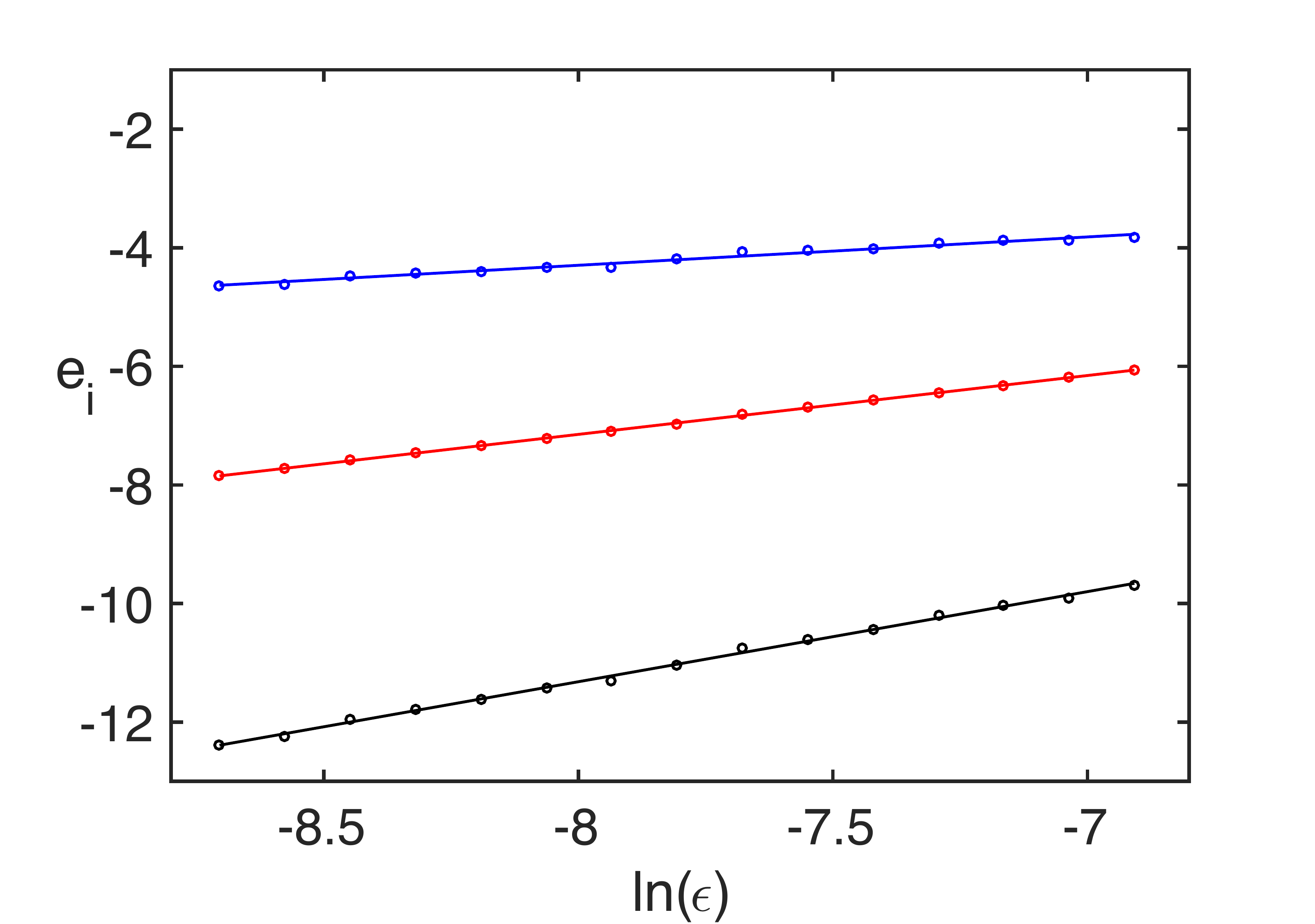}}~
	\subfigure[$\gamma = 0.1$ and $d = 7$.]{\includegraphics[width=0.48\textwidth]{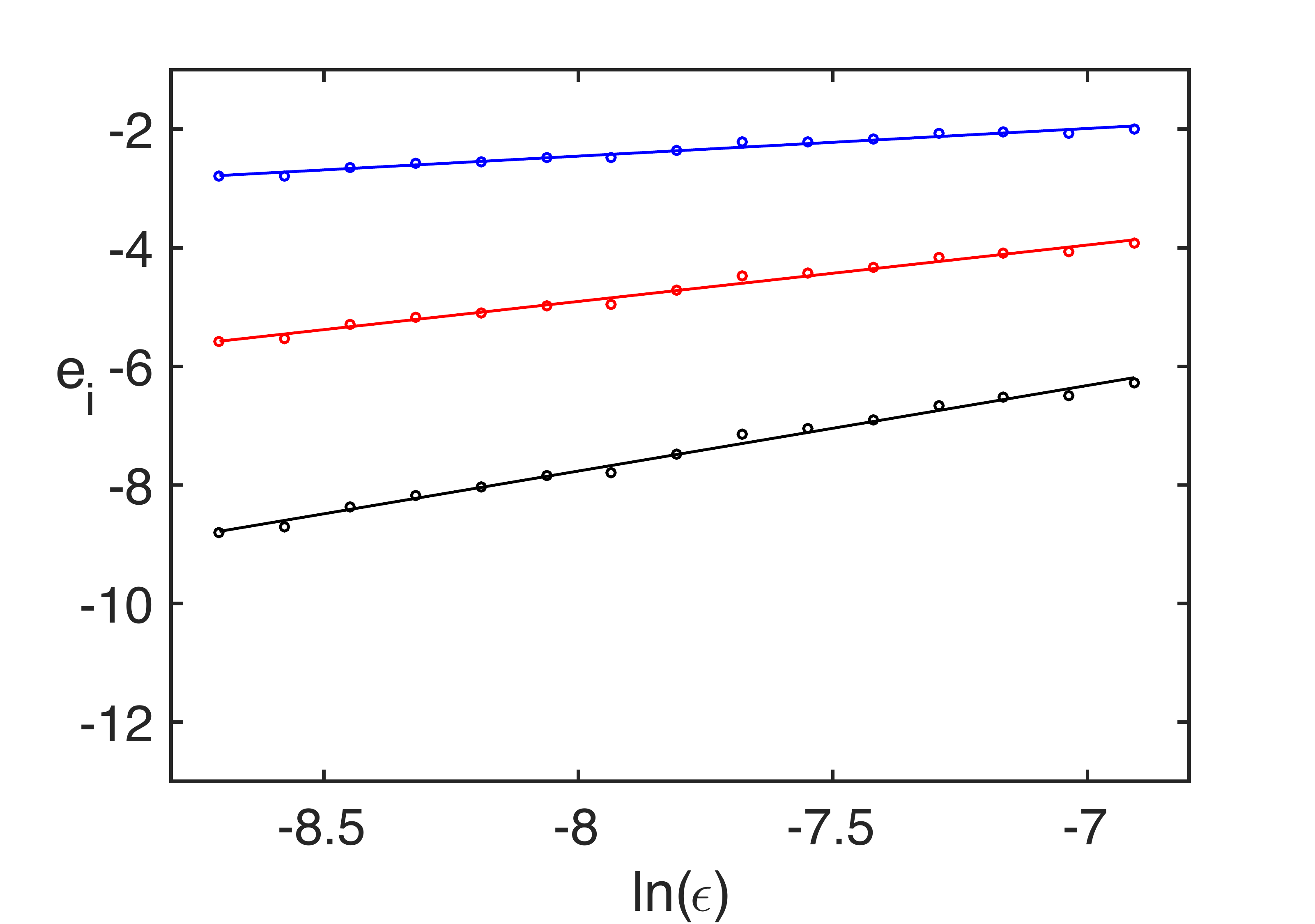}}
	\caption{\small A comparison of error curves for varying values of $\epsilon$, at $t=1/\epsilon$, for the weakly-nonlinear solution including leading-order (upper, blue), $\O{\sqrt{\epsilon}}$ (middle, red) and $\O{\epsilon}$ (lower, black) corrections, for (a) $d=1$ and (b) $d=7$. Parameters are $L=40$, $N=800$, $k = 1/\sqrt{3}$, $\alpha = \beta = c = 1$, $\gamma = 0.1$, $\Delta t = 0.01$ and $\Delta T = \epsilon \Delta t$. The inclusion of more terms in the expansion increases the accuracy, and the errors increase for larger values of $d$.}
	\label{fig:Errg01}
	\vspace{2em}
	\subfigure[$\gamma = 0.5$ and $d = 1$.]{\includegraphics[width=0.48\textwidth]{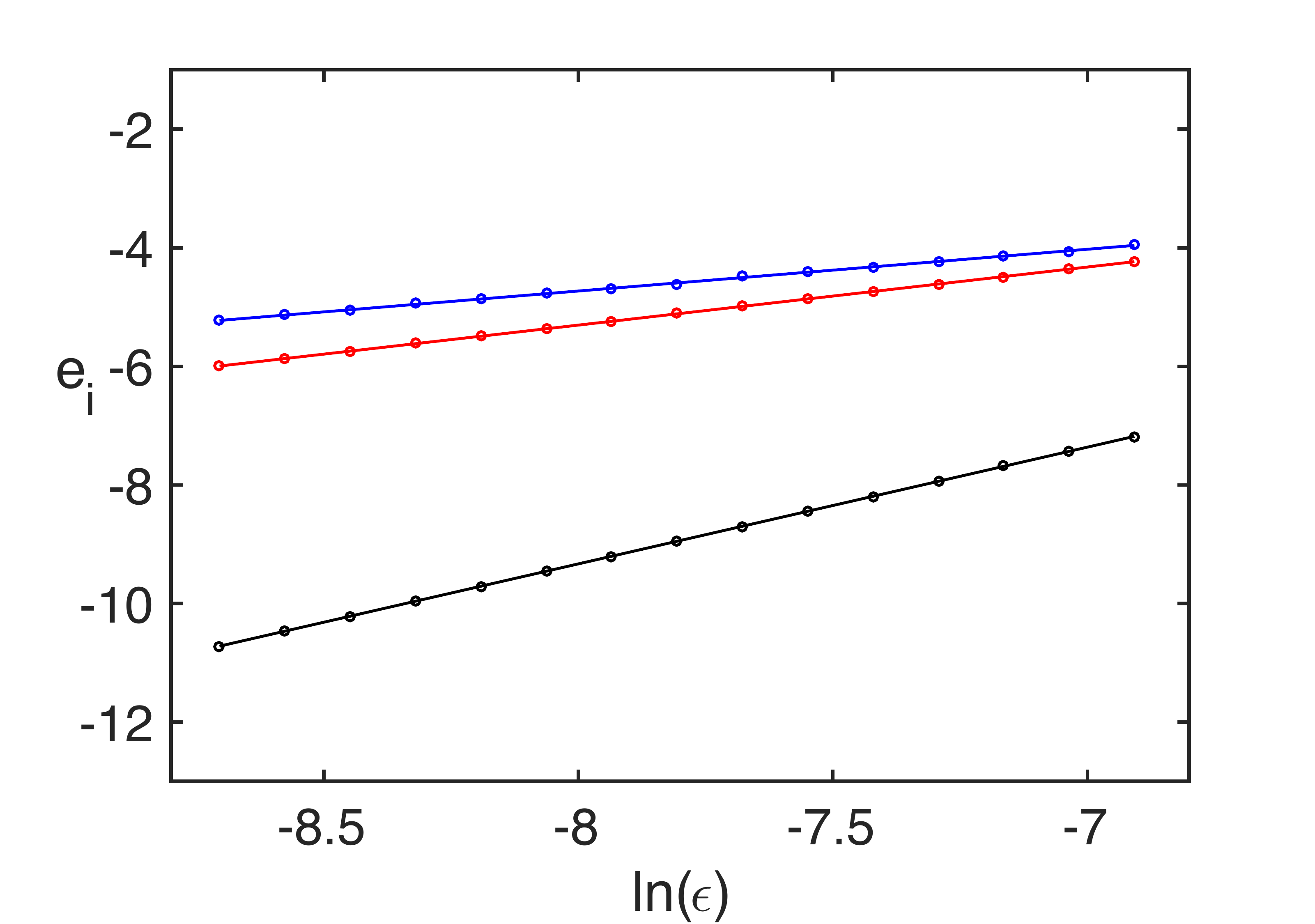}}~
	\subfigure[$\gamma = 0.5$ and $d = 7$.]{\includegraphics[width=0.48\textwidth]{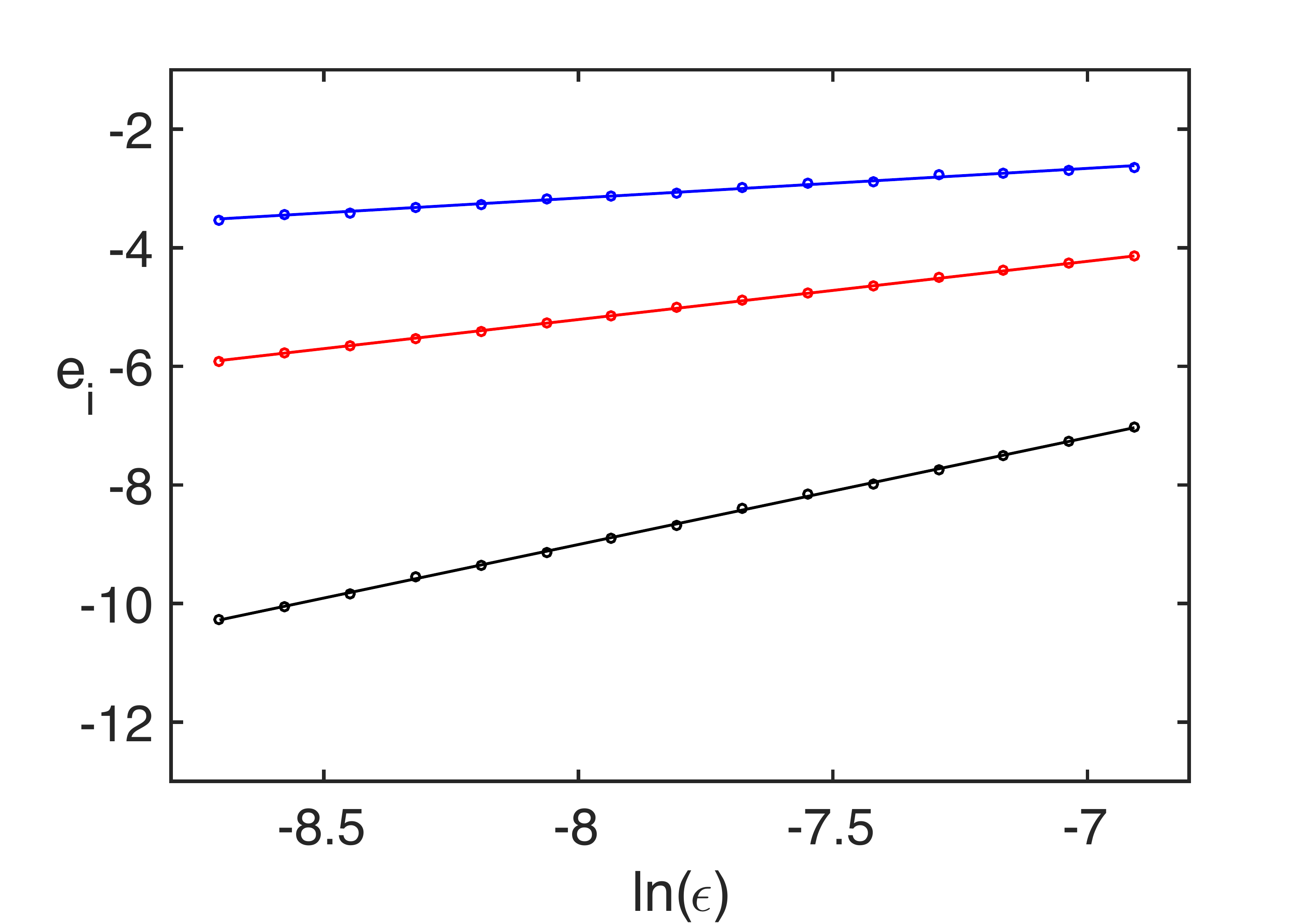}}
	\caption{\small A comparison of error curves for varying values of $\epsilon$, at $t=1/\epsilon$, for the weakly-nonlinear solution including leading-order (upper, blue), $\O{\sqrt{\epsilon}}$ (middle, red) and $\O{\epsilon}$ (lower, black) corrections, for (a) $d=1$ and (b) $d=7$. Parameters are $L=40$, $N=800$, $k = 1/\sqrt{3}$, $\alpha = \beta = c = 1$, $\gamma = 0.5$, $\Delta t = 0.01$ and $\Delta T = \epsilon \Delta t$. The inclusion of more terms in the expansion increases the accuracy, and the errors increase for larger values of $d$. The upper and lower curves are steeper for smaller values of $d$, and tend to their theoretical values as $d$ increases.}
	\label{fig:Errg05}
\end{figure}
An interesting observation is that, as $\gamma$ increases for a fixed value of $d$, the value of $e_{1}$ tends to $e_{2}$ and similarly $e_{3}$ tends to the value expected with the inclusion of the next order of terms, i.e. from 1.5 to 2. Analysing the form of equation (\ref{geq}) shows that, as $\gamma$ increases, the magnitude of these terms decreases. However, from the initial condition for $\phi$ as given in (\ref{phiIC}), the magnitude of $\phi$ will increase as $\gamma$ increases. Therefore the gradient of the error curves will tend to integer powers of epsilon, so the fractional powers will tend to the next largest integer power.

A further observation is that the increase of $d$ will result in an increase in the errors, as we saw in Figure \ref{fig:g01} and Figure \ref{fig:g05}. This can be seen by comparing the two images in Figure \ref{fig:Errg01} and again in Figure \ref{fig:Errg05}. Furthermore, the gradient of the error curves tends towards the expected theoretical values as the value of $F_0$ increases. This is expected as the magnitude of the terms in (\ref{geq}) increase as $d$ increases.

To identify this behaviour more clearly, we tabulate the values of $e_{i}, i=\overline{1,3}$ for a range of values of $d$ and $\gamma$. These results are shown in Tables \ref{tab:e1}, \ref{tab:e2} and \ref{tab:e3}. From these Tables we clearly see that the error values are close to the theoretical values for $\gamma = 0.1$, while for a larger value of $\gamma = 0.5$ they tend towards the theoretical values with the increase of $d$. We also confirm that, as $\gamma$ increases, the values of $e_{1}$ and $e_{3}$ tend to the next largest integer value.
\begin{table}
\centering
\begin{tabularx}{0.9\textwidth}{| Y | Y | Y | Y | Y | Y | Y |}
\hline
\multirow{2}{*}{Value of $\gamma$} &\multicolumn{2}{c|}{$d=1$} & \multicolumn{2}{c|}{$d=4$} & \multicolumn{2}{c|}{$d=7$}\\ \cline{2-7}
& $\alpha_{1}$ & $C_{1}$ & $\alpha_{1}$ & $C_{1}$ & $\alpha_{1}$ & $C_{1}$ \\ \hhline{|=|=|=|=|=|=|=|}
$\gamma = 0.1$ & 0.478 & -0.471 & 0.468 & 0.742 & 0.466 & 1.274 \\ \hline
$\gamma = 0.3$ & 0.559 & -0.260 & 0.504 & 0.552 & 0.500 & 1.070 \\ \hline
$\gamma = 0.5$ & 0.704 & 0.904 & 0.519 & 0.457 & 0.500 & 0.838 \\ \hline
\end{tabularx}
\caption{Maximum absolute error scaling parameters for the leading-order weakly-nonlinear solution for the initial condition in (\ref{IC}). The domain lengths and parameters are $\alpha = \beta = c = 1$, $L=40$ and $k = 1/\sqrt{3}$.}
\label{tab:e1}
\end{table}
\begin{table}
\centering
\begin{tabularx}{0.9\textwidth}{| Y | Y | Y | Y | Y | Y | Y |}
\hline
\multirow{2}{*}{Value of $\gamma$} &\multicolumn{2}{c|}{$d=1$} & \multicolumn{2}{c|}{$d=4$} & \multicolumn{2}{c|}{$d=7$}\\ \cline{2-7}
& $\alpha_{2}$ & $C_{2}$ & $\alpha_{2}$ & $C_{2}$ & $\alpha_{2}$ & $C_{2}$ \\ \hhline{|=|=|=|=|=|=|=|}
$\gamma = 0.1$ & 0.993 & 0.795 & 0.962 & 1.848 & 0.952 & 2.707 \\ \hline
$\gamma = 0.3$ & 0.988 & 1.819 & 0.989 & 1.947 & 0.991 & 2.268 \\ \hline
$\gamma = 0.5$ & 0.979 & 2.529 & 0.980 & 2.565 & 0.982 & 2.645 \\ \hline
\end{tabularx}
\caption{Maximum absolute error scaling parameters for the weakly-nonlinear solution including $\O{\sqrt{\epsilon}}$ terms for the initial condition in (\ref{IC}). The domain lengths and parameters are $\alpha = \beta = c = 1$, $L=40$ and $k = 1/\sqrt{3}$.}
\label{tab:e2}
\end{table}
\begin{table}
\centering
\begin{tabularx}{0.9\textwidth}{| Y | Y | Y | Y | Y | Y | Y |}
\hline
\multirow{2}{*}{Value of $\gamma$} &\multicolumn{2}{c|}{$d=1$} & \multicolumn{2}{c|}{$d=4$} & \multicolumn{2}{c|}{$d=7$}\\ \cline{2-7}
& $\alpha_{3}$ & $C_{3}$ & $\alpha_{3}$ & $C_{3}$ & $\alpha_{3}$ & $C_{3}$ \\ \hhline{|=|=|=|=|=|=|=|}
$\gamma = 0.1$ & 1.519 & 0.835 & 1.455 & 2.505 & 1.443 & 3.776 \\ \hline
$\gamma = 0.3$ & 1.920 & 4.755 & 1.643 & 3.290 & 1.528 & 3.310 \\ \hline
$\gamma = 0.5$ & 1.969 & 6.418 & 1.913 & 6.071 & 1.805 & 5.434 \\ \hline
\end{tabularx}
\caption{Maximum absolute error scaling parameters for the weakly-nonlinear solution including $\O{\epsilon}$ terms for the initial condition in (\ref{IC}). The domain lengths and parameters are $\alpha = \beta = c = 1$, $L=40$ and $k = 1/\sqrt{3}$.}
\label{tab:e3}
\end{table}

A further interesting point arises from Figures \ref{fig:Errg01} and \ref{fig:Errg05}. There will be a value of $\epsilon$ where the error curves intercept, suggesting that, for applications of the weakly-nonlinear solution, the inclusion of higher-order terms only improves the solution when $\epsilon$ is below the values of the intercepts. We introduce the following notations:
\begin{equation}
\epsilon_{1} = \text{Intercept of $e_{1}$ and $e_{2}$ curves}, \quad \epsilon_{2} = \text{Intercept of  $e_{2}$ and $e_{3}$ curves}.
\label{ErrIntNotation}
\end{equation}
More preceisely, $\epsilon_{1}$ is the intercept of the leading-order and $\O{\sqrt{\epsilon}}$ error curves, and $\epsilon_{2}$ is the intercept of the $\O{\sqrt{\epsilon}}$ and $\O{\epsilon}$ error curves. We can calculate this using the values from Tables \ref{tab:e1} - \ref{tab:e3} and the results are shown in Table \ref{tab:WNLInt}. We can see that, in general, an increase in $\gamma$ results in a smaller value of $\epsilon_{1}$ or $\epsilon_{2}$ i.e. as $\gamma$ increases we require a smaller value of $\epsilon$ to improve the accuracy of the solution with the inclusion of higher-order terms.
\begin{table}
\centering
\begin{tabularx}{0.9\textwidth}{| Y | Y | Y | Y | Y | Y | Y |}
\hline
\multirow{2}{*}{Value of $\gamma$} &\multicolumn{3}{c|}{Intercept value $\epsilon_1$} & \multicolumn{3}{c|}{Intercept value $\epsilon_2$} \\ \cline{2-7}
& $d=1$ & $d=4$ & $d=7$ & $d=1$ & $d=4$ & $d=7$ \\ \hhline{|=|=|=|=|=|=|=|}
$\gamma = 0.1$ & 0.0856 & 0.1066 & 0.0524 & 0.9268 & 0.2638 & 0.1134 \\ \hline
$\gamma = 0.3$ & 0.0079 & 0.0563 & 0.0872 & 0.0428 & 0.1283 & 0.1436 \\ \hline
$\gamma = 0.5$ & 0.0027 & 0.0103 & 0.0235 & 0.0197 & 0.0233 & 0.0337 \\ \hline
\end{tabularx}
\caption{Intercept point of error curves in Figures \ref{fig:Errg01} and \ref{fig:Errg05}, in terms of $\epsilon$, representing the maximum value of $\epsilon$ at which the inclusion of $\O{\sqrt{\epsilon}}$ or $\O{\epsilon}$ terms will decrease the error.}
\label{tab:WNLInt}
\end{table}

Another observation is that, for $\gamma > 0.1$, as $d$ increases we see a corresponding increase in $\epsilon_{1}$ or $\epsilon_{2}$. This suggests that for large $\gamma$, when a small value of $\epsilon_{1}$ and $\epsilon_{2}$ is required to validate the inclusion of higher-order terms from the weakly-nonlinear expansion, the threshold is increased with an increase in $d$. More importantly, in all cases here we have $\epsilon_{1} < \epsilon_{2}$ and therefore, if the inclusion of $\O{\sqrt{\epsilon}}$ terms improves the accuracy of the solution for the choice of $\epsilon$, then the inclusion of $\O{\epsilon}$ terms will also improve the accuracy of the solution, without further restriction on $\epsilon$.

\subsection{\texorpdfstring{Example 2: $c = \alpha = \beta = 2$, $\gamma = 0.1$ and $\gamma = 0.5$}{Example 1: c = alpha = beta = 2, gamma = 0.1 and gamma = 0.5}}
\label{sec:Resc2}
We now consider the system with a higher characteristic speed, namely that $c = 2$ and we also vary the values of $\alpha$ and $\beta$, so we take $\alpha = \beta = 2$. As was done in Section \ref{sec:Resc1}, we compare the solution to the initial-value problem (\ref{BousOstOld}), (\ref{BousOstIC}) to the weakly-nonlinear solution with an increasing number of terms included in the expansion. The initial conditions take the same form as Section \ref{sec:Resc1} with the new coefficients, so for $u(x,0)$ we have (\ref{IC}), the initial condition for $f^{\pm}$ is found from (\ref{fIC}) and the initial conditions for $\phi^{\pm}$ are given by (\ref{phiIC}).

We plot the results for $\gamma = 0.1$ and $\gamma = 0.5$, with $d = 1$ and $d = 7$, for direct comparison with the results for $c = 1$, in Figures \ref{fig:c2g01} and \ref{fig:c2g05}. We can see that there is a larger phase shift in these cases; comparing directly between the cases for $\gamma = 0.5$ when $c = 1$ and $c = 2$, we see that the phase shift of the leading-order solution (red, dashed line) is distinctly larger in the latter case than the former. Furthermore, the difference between the cases including terms up to $\O{\sqrt{\epsilon}}$ and terms up to $\O{\epsilon}$ are more clearly highlighted in this case than the previous results for $c = 1$, as can again be seen clearly from Figure \ref{fig:c2g01} and \ref{fig:c2g05} for the enhanced inserts in each image.

\begin{figure}[!htbp]
	\subfigure[$\gamma = 0.1$ and $d = 1$.]{\includegraphics[width=0.48\textwidth]{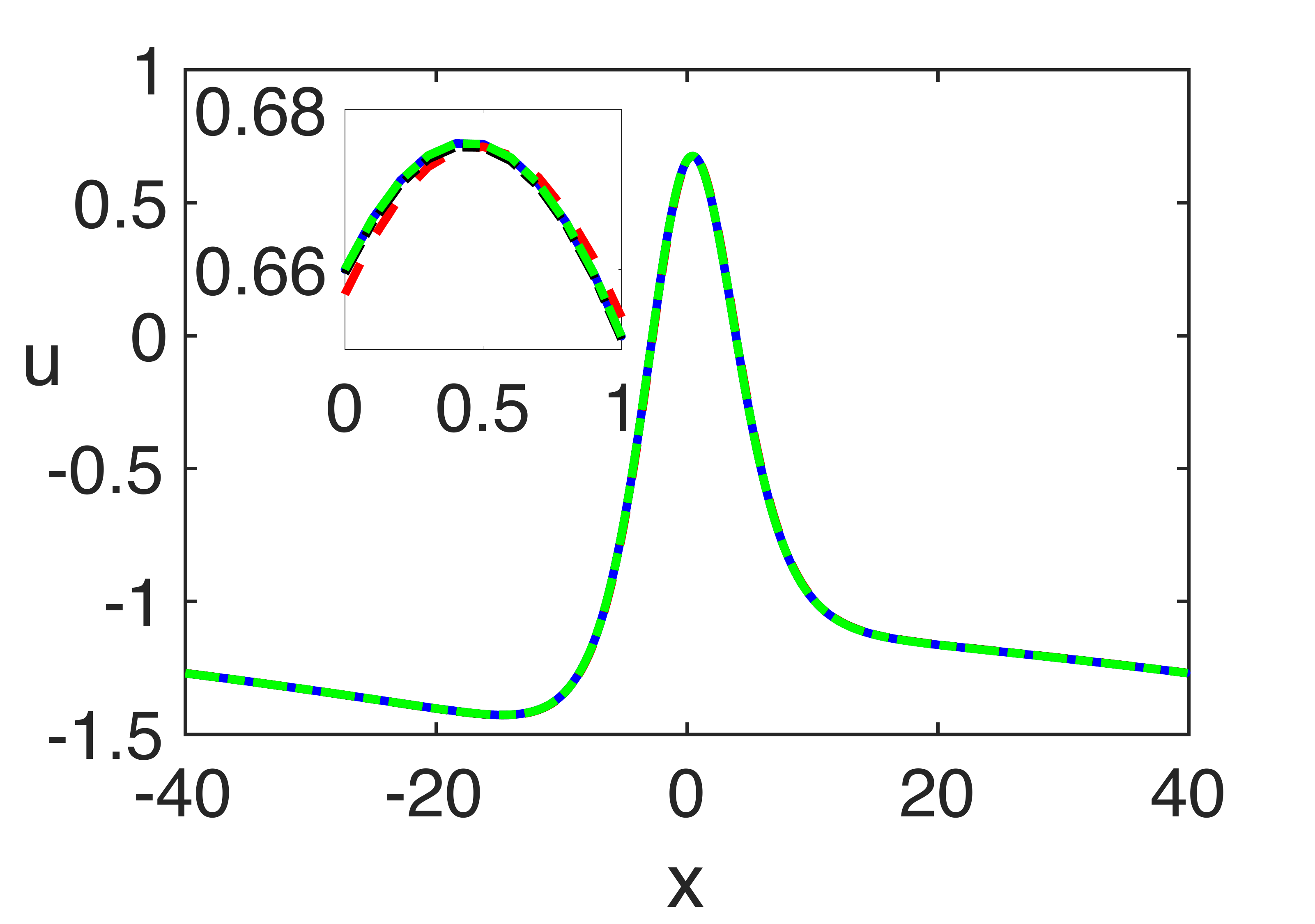}}~
	\subfigure[$\gamma = 0.1$ and $d = 7$.]{\includegraphics[width=0.48\textwidth]{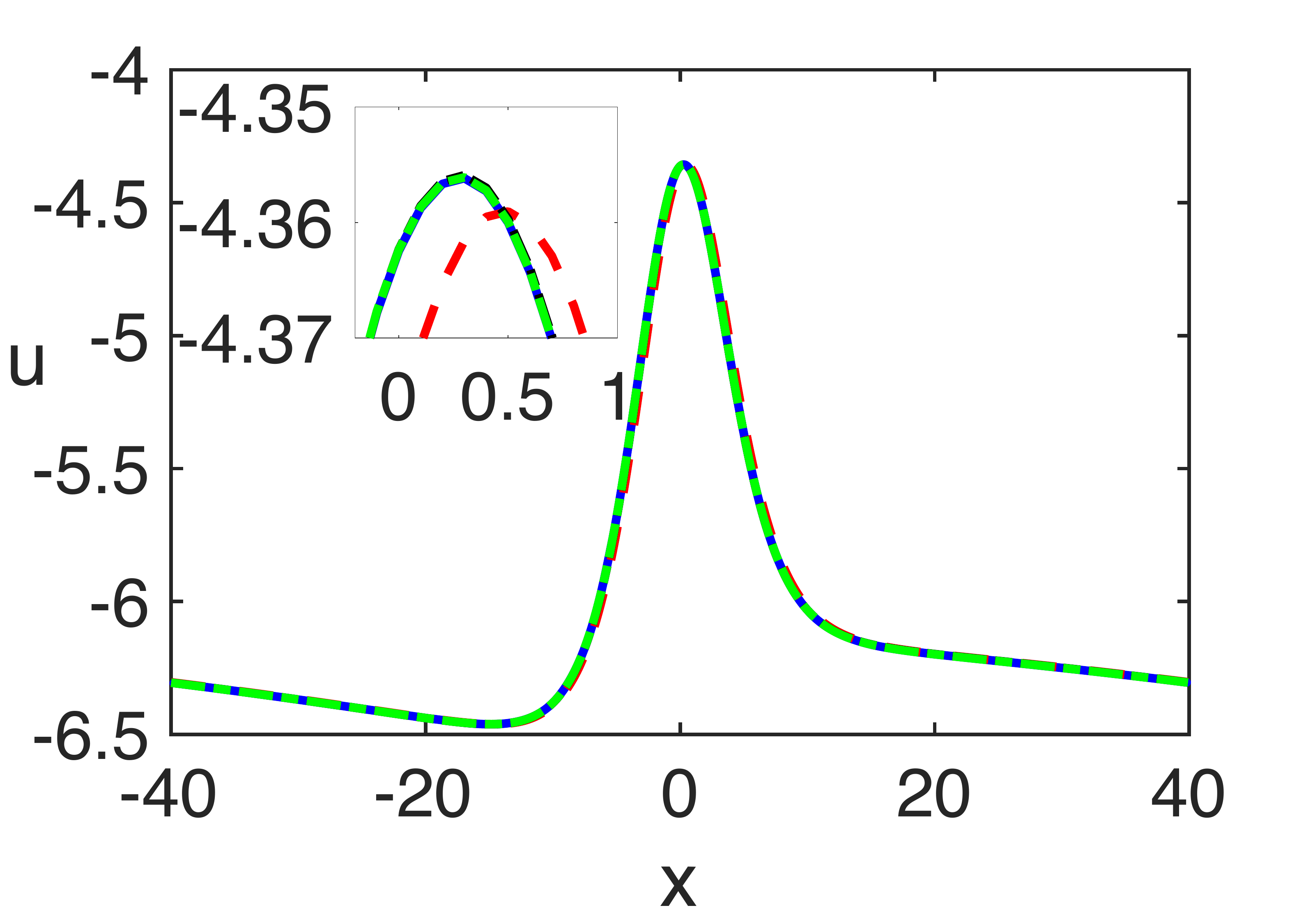}}
	\caption{\small A comparison of the numerical solution (solid, blue) at $t=1/\epsilon$ and the weakly-nonlinear solution including leading-order (dashed, red), $\O{\sqrt{\epsilon}}$ (dash-dot, black) and $\O{\epsilon}$ (dotted, green) corrections, for (a) $d=1$ and (b) $d=7$. Parameters are $L=40$, $N=800$, $k = 1/\sqrt{3}$, $\alpha = \beta = c = 2$, $\gamma = 0.1$, $\epsilon = 0.001$, $\Delta t = 0.01$ and $\Delta T = \epsilon \Delta t$. The solution agrees reasonably well to leading order, and this agreement is improved with the addition of higher-order corrections.}
	\label{fig:c2g01}
	\vspace{2em}
	\subfigure[$\gamma = 0.5$ and $d = 1$.]{\includegraphics[width=0.48\textwidth]{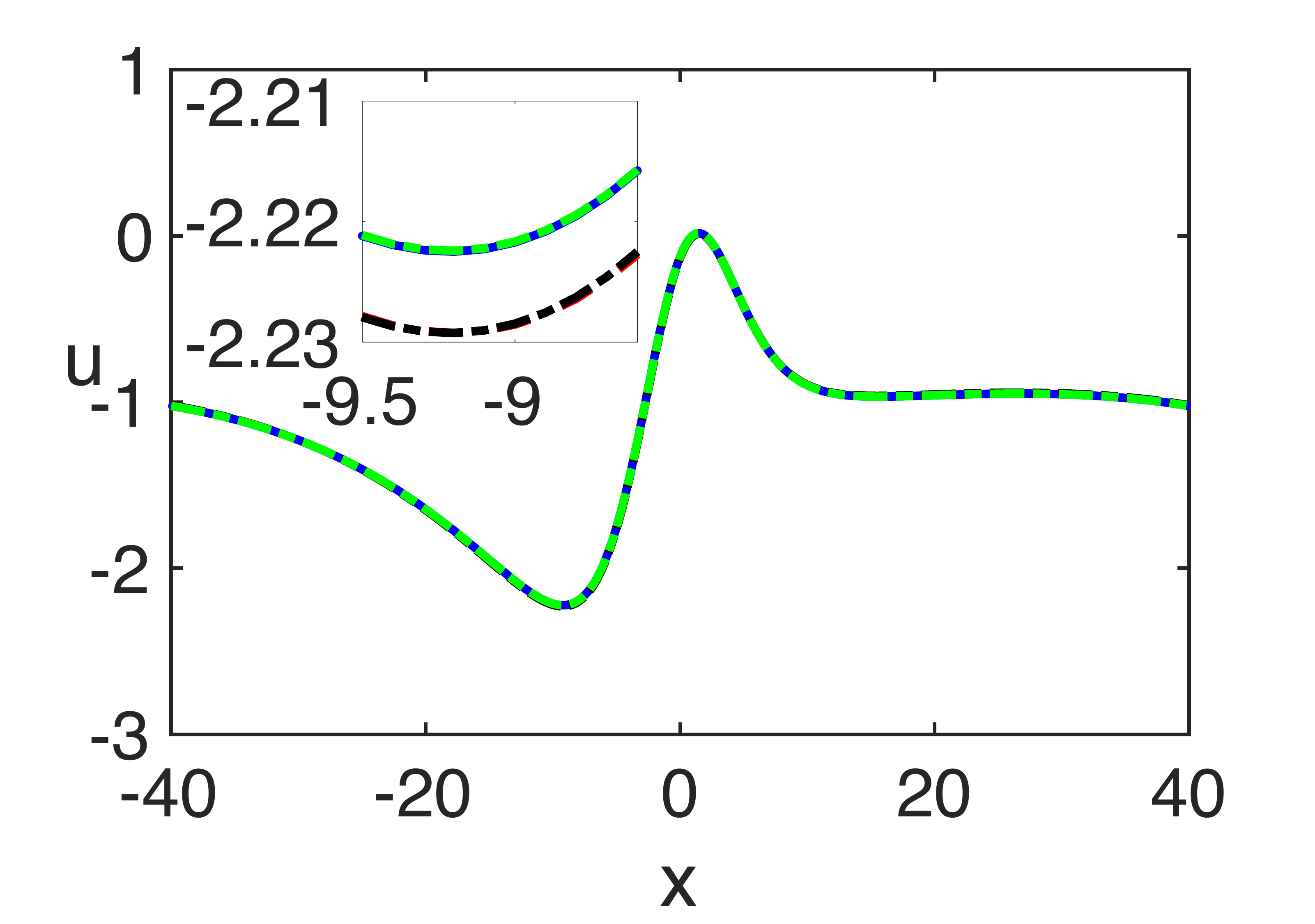}}~
	\subfigure[$\gamma = 0.5$ and $d = 7$.]{\includegraphics[width=0.48\textwidth]{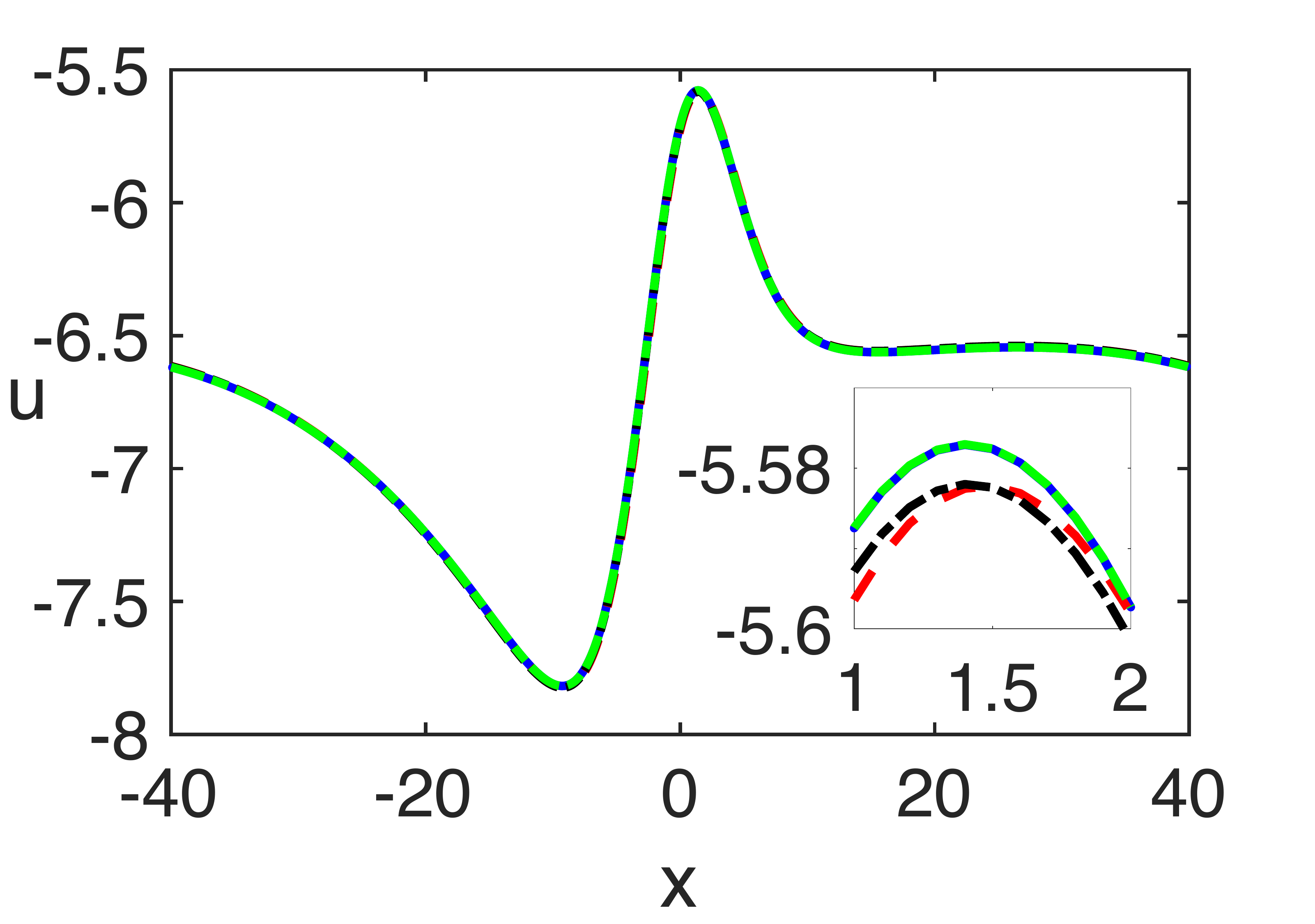}}
	\caption{\small A comparison of the numerical solution (solid, blue) at $t=1/\epsilon$ and the weakly-nonlinear solution including leading-order (dashed, red), $\O{\sqrt{\epsilon}}$ (dash-dot, black) and $\O{\epsilon}$ (dotted, green) corrections, for (a) $d=1$ and (b) $d=7$. Parameters are $L=40$, $N=800$, $k = 1/\sqrt{3}$, $\alpha = \beta = c = 2$, $\gamma = 0.5$, $\epsilon = 0.001$, $\Delta t = 0.01$ and $\Delta T = \epsilon \Delta t$. The solution agrees reasonably well to leading order and this agreement is improved with the addition of higher-order corrections.}
	\label{fig:c2g05}
\end{figure}
As before we plot the corresponding error curves for the cases in Figures \ref{fig:c2g01} and \ref{fig:c2g05}. These results are presented in Figure \ref{fig:Errc2g01} and Figure \ref{fig:Errc2g05} for $\gamma = 0.1$ and $\gamma = 0.5$, respectively. We again see that the error curves in Figure \ref{fig:Errc2g01} have slope 0.5, 1 and 1.5, as expected. It is worth noting that the errors are similar to their previous cases for $c = 1$. However, as $\gamma$ increases, while the curves for the leading-order and the inclusion of $\O{\sqrt{\epsilon}}$ terms do tend towards each other as in the case for $c = 1$, the rate at which this occurs is slower.
\begin{figure}[!htbp]
	\subfigure[$\gamma = 0.1$ and $d = 1$.]{\includegraphics[width=0.48\textwidth]{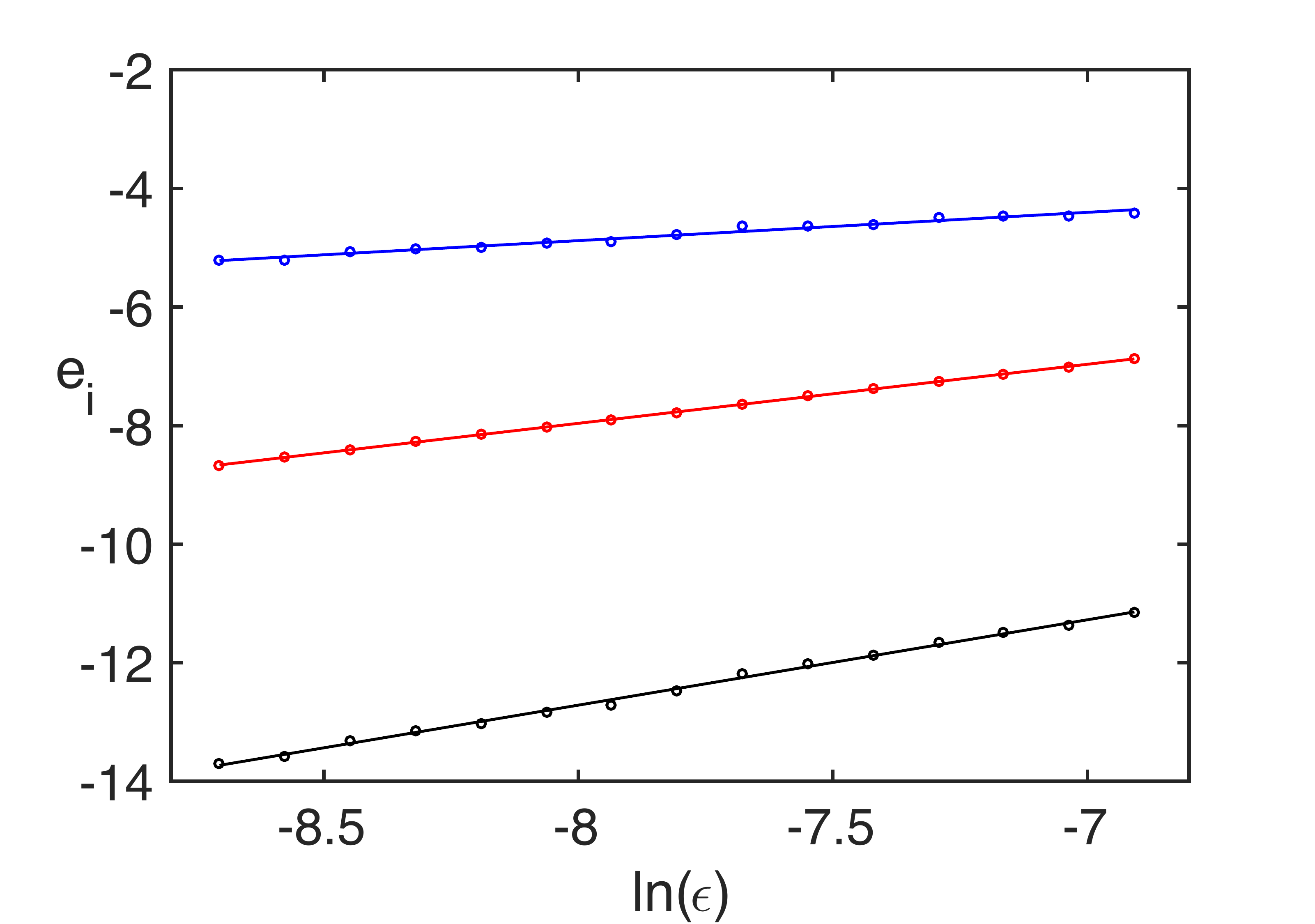}}~
	\subfigure[$\gamma = 0.1$ and $d = 7$.]{\includegraphics[width=0.48\textwidth]{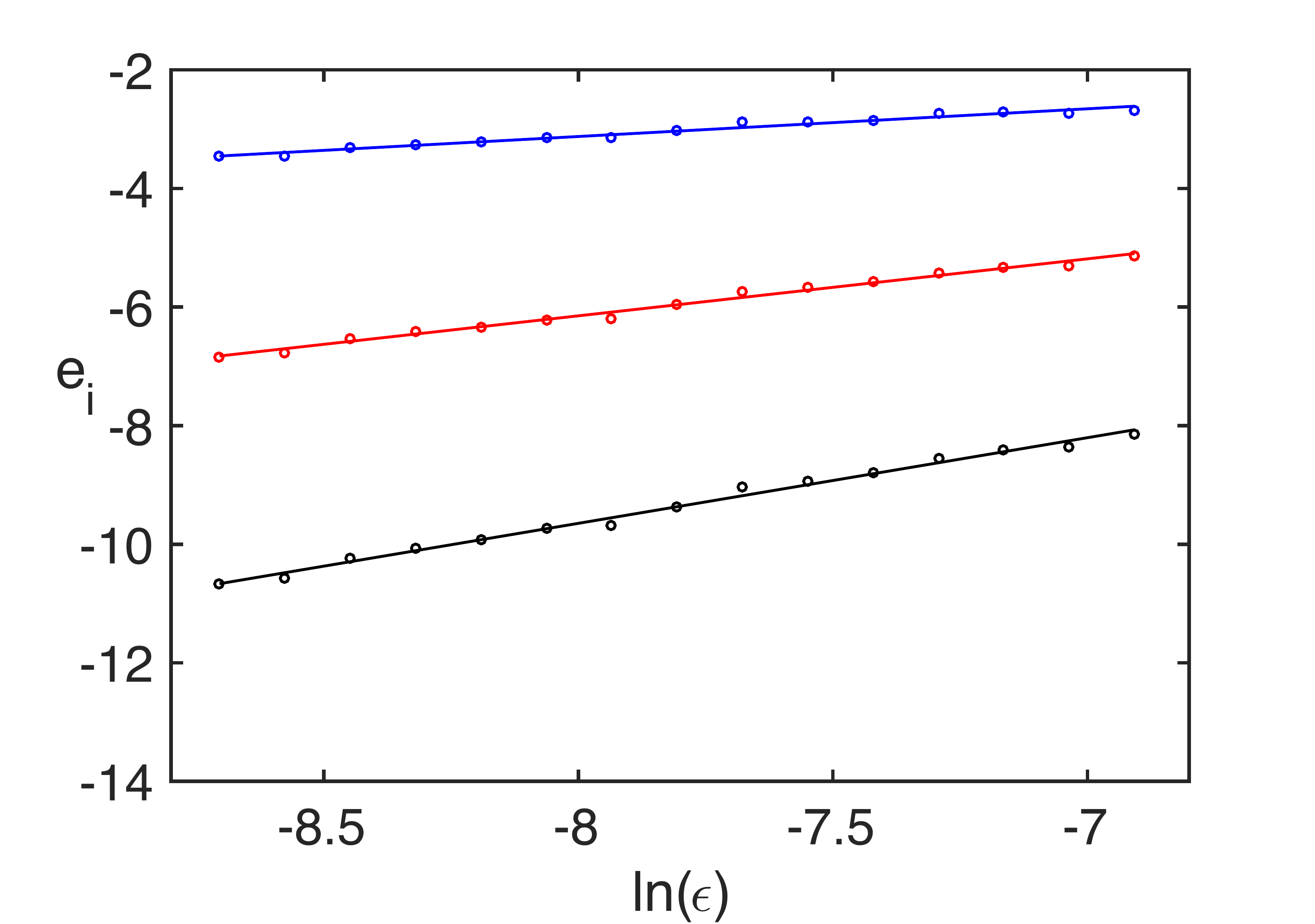}}
	\caption{\small A comparison of error curves for varying values of $\epsilon$, at $t=1/\epsilon$, for the weakly-nonlinear solution including leading-order (upper, blue), $\O{\sqrt{\epsilon}}$ (middle, red) and $\O{\epsilon}$ (lower, black) corrections, for (a) $d=1$ and (b) $d=7$. Parameters are $L=40$, $N=800$, $k = 1/\sqrt{3}$, $\alpha = \beta = c = 2$, $\gamma = 0.1$, $\Delta t = 0.01$ and $\Delta T = \epsilon \Delta t$. The inclusion of more terms in the expansion increases the accuracy, and the errors increase for larger values of $d$.}
	\label{fig:Errc2g01}
	\vspace{2em}
	\subfigure[$\gamma = 0.5$ and $d = 1$.]{\includegraphics[width=0.48\textwidth]{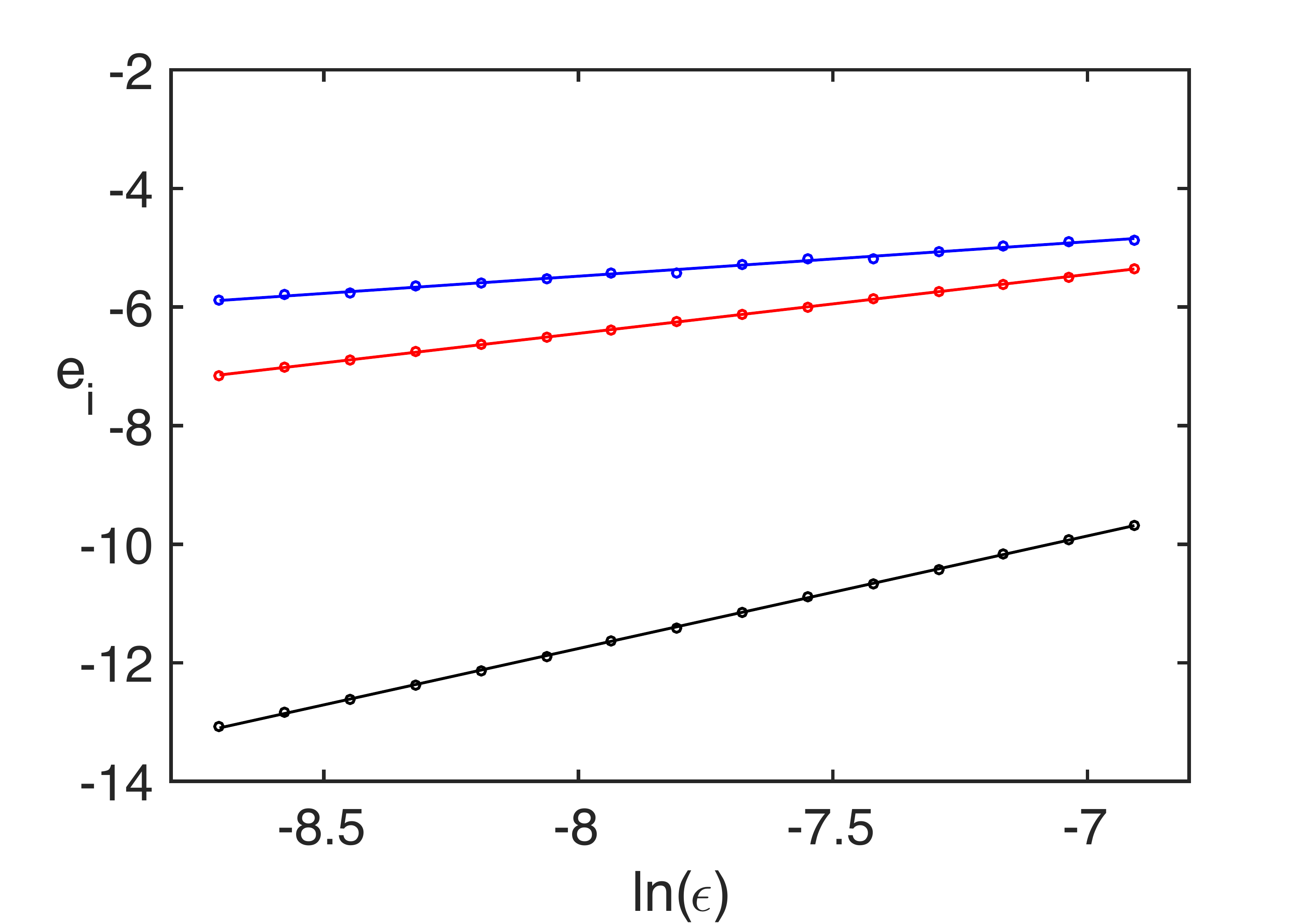}}~
	\subfigure[$\gamma = 0.5$ and $d = 7$.]{\includegraphics[width=0.48\textwidth]{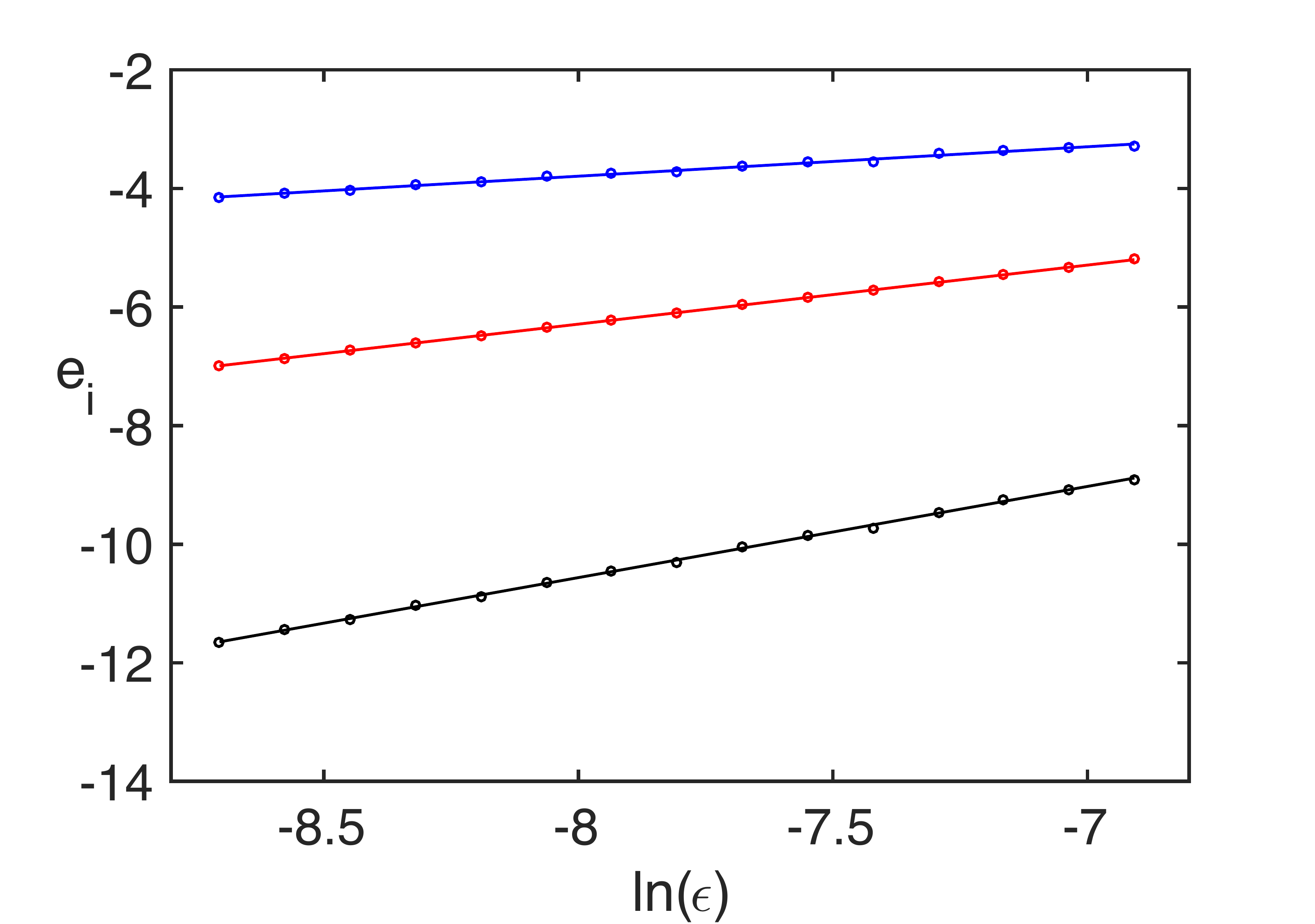}}
	\caption{\small A comparison of error curves for varying values of $\epsilon$, at $t=1/\epsilon$, for the weakly-nonlinear solution including leading-order (upper, blue), $\O{\sqrt{\epsilon}}$ (middle, red) and $\O{\epsilon}$ (lower, black) corrections, for (a) $d=1$ and (b) $d=7$. Parameters are $L=40$, $N=800$, $k = 1/\sqrt{3}$, $\alpha = \beta = c = 2$, $\gamma = 0.5$, $\Delta t = 0.01$ and $\Delta T = \epsilon \Delta t$. The inclusion of more terms in the expansion increases the accuracy, and the errors increase for larger values of $d$. The upper and lower curves are steeper for smaller values of $d$, and tend to their theoretical values as $d$ increases.}
	\label{fig:Errc2g05}
\end{figure}

To inspect this behaviour further, we again tabulate the errors in Tables \ref{tab:c2e1}, \ref{tab:c2e2} and \ref{tab:c2e3} for the inclusion of leading-order terms, $\O{\sqrt{\epsilon}}$ and $\O{\epsilon}$ i.e. $e_{1}$, $e_{2}$ and $e_{3}$. The errors are close to the theoretical values as before, however the theoretical slope values are obtained for a wider range of $\gamma$ and $d$ values in contrast to the previous case for $c = 1$. If we analyse the form of the expressions in (\ref{WNLFinal}) we can see that, for $\O{\sqrt{\epsilon}}$, the term is identical to the previous case if $c = \alpha$, as we have here. However, for $\O{\epsilon}$ (as can be seen from (\ref{heq}) the first term in this expression is smaller due to the divisor of $2c$. This is further reflected in the initial condition for $\phi$, as the terms from (\ref{heq}) are present here as well. Therefore, the values present at $\O{\epsilon}$ are likely to be smaller and therefore more distinct from the previous value at $\O{\sqrt{\epsilon}}$, resulting in the estimates being obtained more distinctly for the same set of $\epsilon$ values as before. This behaviour is replicated for several values of $\gamma$ and $d$, and indeed while for large $\gamma$ we see the slope values differing from their theoretical estimates, the magnitude of the divergence is less than the previous case.
\begin{table}
\centering
\begin{tabularx}{0.9\textwidth}{| Y | Y | Y | Y | Y | Y | Y |}
\hline
\multirow{2}{*}{Value of $\gamma$} &\multicolumn{2}{c|}{$d=1$} & \multicolumn{2}{c|}{$d=4$} & \multicolumn{2}{c|}{$d=7$}\\ \cline{2-7}
& $\alpha_{1}$ & $C_{1}$ & $\alpha_{1}$ & $C_{1}$ & $\alpha_{1}$ & $C_{1}$ \\ \hhline{|=|=|=|=|=|=|=|}
$\gamma = 0.1$ & 0.477 & -1.061 & 0.469 & 0.091 & 0.467 & 0.615 \\ \hline
$\gamma = 0.3$ & 0.522 & -1.154 & 0.498 & -0.136 & 0.497 & 0.389 \\ \hline
$\gamma = 0.5$ & 0.581 & -0.828 & 0.501 & -0.314 & 0.496 & 0.178 \\ \hline
\end{tabularx}
\caption{Maximum absolute error scaling parameters for the leading-order weakly-nonlinear solution for the initial condition in (\ref{IC}). The domain lengths and parameters are $\alpha = \beta = c = 2$, $L=40$ and $k = 1/\sqrt{3}$.}
\label{tab:c2e1}
\end{table}

\begin{table}
\centering
\begin{tabularx}{0.9\textwidth}{| Y | Y | Y | Y | Y | Y | Y |}
\hline
\multirow{2}{*}{Value of $\gamma$} &\multicolumn{2}{c|}{$d=1$} & \multicolumn{2}{c|}{$d=4$} & \multicolumn{2}{c|}{$d=7$}\\ \cline{2-7}
& $\alpha_{2}$ & $C_{2}$ & $\alpha_{2}$ & $C_{2}$ & $\alpha_{2}$ & $C_{2}$ \\ \hhline{|=|=|=|=|=|=|=|}
$\gamma = 0.1$ & 0.996 & 0.007 & 0.974 & 0.825 & 0.961 & 1.540 \\ \hline
$\gamma = 0.3$ & 0.996 & 0.893 & 0.996 & 1.063 & 0.996 & 1.348 \\ \hline
$\gamma = 0.5$ & 0.995 & 1.514 & 0.995 & 1.576 & 0.996 & 1.679 \\ \hline
\end{tabularx}
\caption{Maximum absolute error scaling parameters for the weakly-nonlinear solution including  $\O{\sqrt{\epsilon}}$ terms for the initial condition in (\ref{IC}). The domain lengths and parameters are $\alpha = \beta = c = 2$, $L=40$ and $k = 1/\sqrt{3}$.}
\label{tab:c2e2}
\end{table}

\begin{table}
\centering
\begin{tabularx}{0.9\textwidth}{| Y | Y | Y | Y | Y | Y | Y |}
\hline
\multirow{2}{*}{Value of $\gamma$} &\multicolumn{2}{c|}{$d=1$} & \multicolumn{2}{c|}{$d=4$} & \multicolumn{2}{c|}{$d=7$}\\ \cline{2-7}
& $\alpha_{3}$ & $C_{3}$ & $\alpha_{3}$ & $C_{3}$ & $\alpha_{3}$ & $C_{3}$ \\ \hhline{|=|=|=|=|=|=|=|}
$\gamma = 0.1$ & 1.440 & -1.197 & 1.449 & 0.978 & 1.444 & 3.433 \\ \hline
$\gamma = 0.3$ & 1.684 & 0.723 & 1.515 & 0.808 & 1.498 & 2.003 \\ \hline
$\gamma = 0.5$ & 1.899 & 1.907 & 1.646 & 1.507 & 1.538 & 1.738 \\ \hline
\end{tabularx}
\caption{Maximum absolute error scaling parameters for the weakly-nonlinear solution including $\O{\epsilon}$ terms for the initial condition in (\ref{IC}). The domain lengths and parameters are $\alpha = \beta = c = 2$, $L=40$ and $k = 1/\sqrt{3}$.}
\label{tab:c2e3}
\end{table}

As was seen in the case for $c = 1$, from Figures \ref{fig:Errg01} and \ref{fig:Errg05} we notice that there will be a value of $\epsilon$ where the error curves intercept, suggesting that the inclusion of higher-order terms only improves the solution when $\epsilon$ is below the values of the intercepts. We again calculate this limit using the values from Tables \ref{tab:c2e1} - \ref{tab:c2e3} for each of the cases considered in Figures \ref{fig:Errg01} and \ref{fig:Errg05}. The results are shown in Table \ref{tab:WNLIntc2}.
\begin{table}
\centering
\begin{tabularx}{0.9\textwidth}{| Y | Y | Y | Y | Y | Y | Y |}
\hline
\multirow{2}{*}{Value of $\gamma$} &\multicolumn{3}{c|}{Intercept value $\epsilon_{1}$} & \multicolumn{3}{c|}{Intercept value $\epsilon_{2}$} \\ \cline{2-7}
& $d=1$ & $d=4$ & $d=7$ & $d=1$ & $d=4$ & $d=7$ \\ \hhline{|=|=|=|=|=|=|=|}
$\gamma = 0.1$ & 0.1277 & 0.2338 & 0.1537 & 15.0550 & 0.7246 & 0.0199 \\ \hline
$\gamma = 0.3$ & 0.0133 & 0.0900 & 0.1463 & 1.2803 & 1.6345 & 0.2712 \\ \hline
$\gamma = 0.5$ & 0.0035 & 0.0218 & 0.0497 & 0.6474 & 1.1118 & 0.8969 \\ \hline
\end{tabularx}
\caption{Intercept point of error curves in Figures \ref{fig:Errc2g01} and \ref{fig:Errc2g05}, in terms of $\epsilon$, representing the maximum value of $\epsilon$ at which the inclusion of $\O{\sqrt{\epsilon}}$ or $\O{\epsilon}$ terms will decrease the error.}
\label{tab:WNLIntc2}
\end{table}

Let us analyse the conclusions we drew for the previous case, referring to the notation in (\ref{ErrIntNotation}). As $\gamma$ increases, we see that $\epsilon_{1}$ decreases for all values of $d$ whereas for $\epsilon_{2}$, in contrast to the previous case, as $d$ increases this behaviour is reversed i.e. for $d=1$ we have that $\epsilon_{2}$ decreases as $\gamma$ increases, but for $d=7$ we have that $\epsilon_{2}$ increases as $\gamma$ increases. Thus, the detailed behaviour is dependent upon the coefficients in the equation.

The behaviour observed for increasing $d$ is also different to the first example. For $\gamma > 0.1$ we see that $\epsilon_{1}$ increases as $d$ increases, but for $\epsilon_{2}$ there is no clear relation. This behaviour would need to be investigated further. As with the previous scenario, in almost all cases we see that $\epsilon_{1} < \epsilon_{2}$ and therefore, if the inclusion of $\O{\sqrt{\epsilon}}$ terms improves the accuracy of the solution for the choice of $\epsilon$, then the inclusion of $\O{\epsilon}$ terms will also improve the accuracy of the solution, without further restriction on $\epsilon$.

There is one further conclusion that can be drawn by comparing the values obtained for the first example. In almost every case (the only exception being $\gamma = 0.1$ and $d = 7$) we have that $\epsilon_{i}$ is smaller for the first example  $c = \alpha = \beta = 1$ than for the second case when $c = \alpha = \beta = 2$. This suggests that the threshold value of $\epsilon$ increases when the coefficients increase and therefore an accurate weakly-nonlinear solution will be applicable for larger values of $\epsilon$.

\section{Comparison of localised and periodic solutions}
\label{sec:GHRes}
In this section we aim to reproduce the well-known scenario for a localised initial condition with non-zero mass on a large (``infinite") interval, using our constructed solution. Previously, a similar comparison has been made within the scope of an initial-value problem for the regularised Boussinesq equation ($\gamma = 0$) in \cite{KM, KMP}, and for zero-mass initial condition in the BKG equation (\ref{BousOst}) in \cite{KMP}.  

As discussed in \cite{Grimshaw99} within the scope of a regularised Ostrovsky equation, a localised initial condition  of a soliton type (with non-zero mass) evolves into a right-propagating wave packet with zero mass, while a fast moving left-propagating wave carries the ``mass" away from this wave packet. This result was highlighted in numerical studies in \cite{GH}, describing the evolution of  the Ostrovsky equation with the KdV solitary wave initial condition, for various amplitudes of the soliton. 

Firstly, we look at the differences in the ``exact" (numerical) solutions of the BKG equation (\ref{BousOst}) when the initial condition is taken either as an exact soliton solution of the respective Boussinesq equation ($\gamma = 0$), or its approximation by the soliton solution of the Korteweg - de Vries (KdV) equation. Indeed, this is an approximation of the type used, for example, in the studies of the effect of rotation on an internal solitary wave in the ocean. The latter is approximated by the solution of the KdV equation.  

Thus, the initial condition takes the form of either a right-propagating KdV soliton,
\begin{equation}
u(x,0) = A\ \sechn{2}{\frac{x}{\Lambda}}, \quad u_t(x,0) = \frac{2cA}{\Lambda} \sechn{2}{\frac{x}{\Lambda}} \tanhn{ }{\frac{x}{\Lambda}},
\label{GHKdVSol}
\end{equation}
where $\displaystyle \Lambda = \sqrt{\frac{12 c^2 \beta}{\alpha A}}$, or a right-propagating Boussinesq soliton,
\begin{equation}
u(x,0) = A\ \sechn{2}{\frac{x}{\tilde{\Lambda}}}, \quad u_t(x,0) = \frac{2 v A}{\tilde{\Lambda}} \sechn{2}{\frac{x}{\tilde{\Lambda}}} \tanhn{ }{\frac{x}{\tilde{\Lambda}}},
\label{GHBousSol}
\end{equation}
where $\displaystyle \tilde{\Lambda} = \frac{2 v \sqrt{\epsilon \beta }}{\sqrt{v^2 - c^2}}$ and  $\displaystyle v = \sqrt{c^2 +  \frac{\epsilon \alpha  A}{3}}$,
parametrised by the amplitude $A$. Both initial conditions have non-zero mass. The respective numerical solutions of the equation (\ref{BousOst}) are shown in Figure \ref{fig:BousKdVBousSol} in the characteristic reference frame, i.e. $\tilde{x} = x - t$ and we omit the tilde in the figures. We can see that, on the full scale, there is no visible difference between the solutions. In the enhanced image to the right we see that there is a small phase shift between the main wavepackets moving to the right. The solution for the KdV soliton initial condition (blue, solid line) is moving slightly slower than the solution for the Boussinesq soliton (red, dashed line). In both cases, in the full-scale image, we can also see the emergence of a fast-moving left-propagating wave packet. This wave packet is generated from the initial soliton and carries the mass to the left, leaving a wave packet with zero mass, agreeing with the scenario described in \cite{Grimshaw99}.
\begin{figure}[!htbp]
	\subfigure[Full image.]{\includegraphics[width=0.48\textwidth]{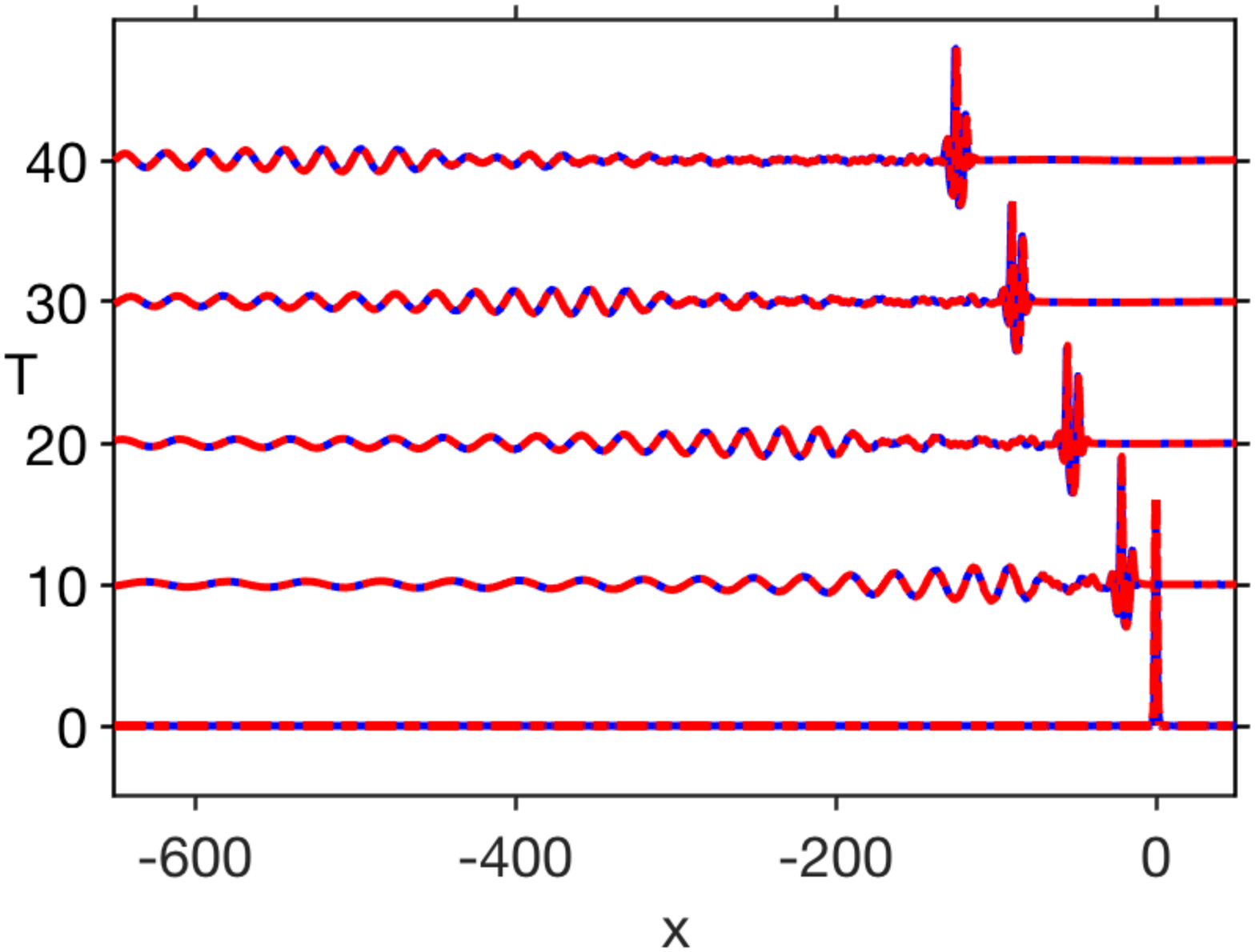}}~
	\subfigure[Enhanced image.]{\includegraphics[width=0.48\textwidth]{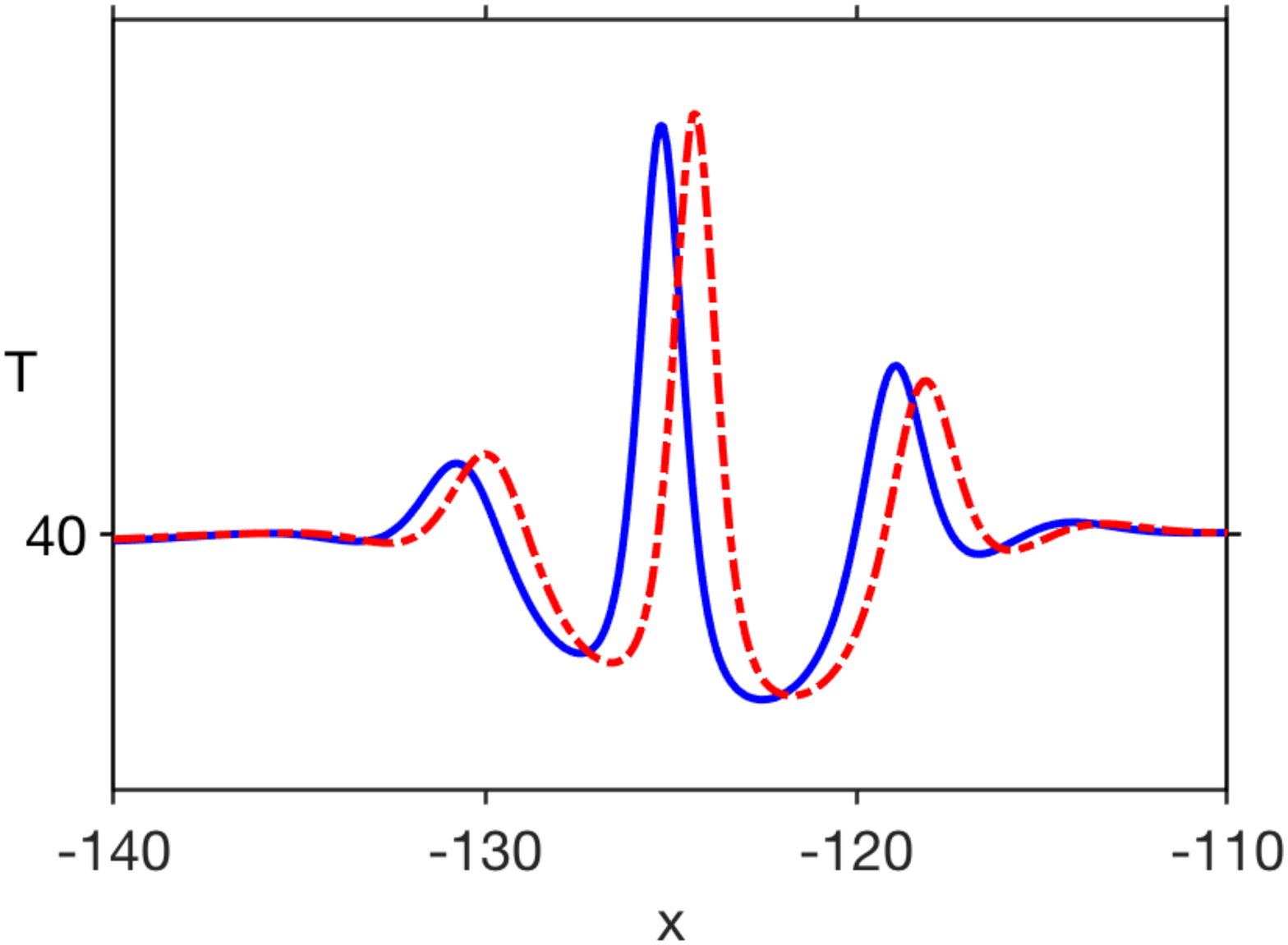}}
	\caption{\small A comparison of the numerical solution of the BKG equation presented at various times, for initial condition of a KdV soliton (blue, solid line) or a Boussinesq soliton (red, dashed line) in the moving reference frame, as presented for (a) the full domain, and (b) an enhanced domain. Parameters are $L=5,000$, $N=100,000$, $A = 32$, $c = 1$, $\alpha = \beta = \gamma = 2$, $\epsilon = 0.001$ and $\Delta t = 0.01$.}
	\label{fig:BousKdVBousSol}
\end{figure}
In the remainder of this section we are using the KdV soliton initial condition (\ref{GHKdVSol}).

Secondly, we compare our constructed solution (which by-passes the zero-mass contradiction) both with the ``exact" (numerical) solution of the BKG equation (\ref{BousOst}) and
with the corresponding direct solution of the Ostrovsky equation for non-zero mass initial conditions as modelled in \cite{GH}. 
Therefore, our comparisons in this section are restricted to the extended leading-order solution up to and including $\O{\sqrt{\epsilon}}$ terms:
\begin{align}
u \lb x, t \rb &= F_0 \cos \omega t +  f^{-}   - \sqrt{\epsilon} \  \frac{\alpha F_0}{2 c \sqrt{\gamma}} \sin (\sqrt{\gamma} \tau ) f^-_{\xi_{-}}   + \O{\epsilon},
\label{WNL_loc}
\end{align}
where the function $f^-$ solves the Ostrovsky equation
\begin{equation}
\lb 2c f^-_T + \alpha f^- f^-_{\xi_{-}} + \beta c^2 f^-_{\xi_{-} \xi_{-} \xi_{-}} \rb _ {\xi_{-}} = \gamma f^-
\label{Oeq}
\end{equation}
for the initial condition which has zero mass:
\begin{equation}
f^{-}|_{T=0} = A\ \sechn{2}{\frac{x}{\Lambda}}  - F_0,
\label{ICO}
\end{equation}
where
$
F_0 = 2 \tanh L.
$
(Validity of the constructed weakly-nonlinear solution up to $\O{\epsilon}$ has been illustrated in the previous section.) 
 
 We compare (\ref{WNL_loc}) - (\ref{ICO}) with the direct solution of the Ostrovsky equation (\ref{Oeq}) for the initial condition with non-zero mass:
 \begin{equation}
 f^{-}|_{T=0} = A\ \sechn{2}{\frac{x}{\Lambda}},
 \label{ICO1}
 \end{equation}
 and we compare both with the direct numerical solution of the BKG equation (\ref{BousOst}) with the initial conditions (\ref{GHKdVSol}). 
 
 We also make one comparison for a modified initial condition (\ref{IC}) on a finite periodic interval in order to better illustrate the difference between the behaviour of periodic solutions on a finite interval and localised solutions on a large interval. Indeed, solving the Ostrovsky equation directly  for a localised initial condition with non-zero mass defined on a large interval gives a good approximation to the exact solution of the Cauchy problem for the BKG equation (\ref{BousOst}), but using the same approach for periodic solutions with non-zero mass on a finite interval would lead to wrong results, as illustrated in Section \ref{sec:GHFinite}.

\subsection{\texorpdfstring{Example 1:  $c = \alpha = \beta = 1$, $\gamma = 0.1$}{Example 1: c = alpha = beta = 1, gamma = 0.1}}
\label{sec:GHFinite}
In this example we consider a modification of the initial condition used in Section 5.2 of \cite{KMP}, i.e.
use the initial condition  (\ref{GHKdVSol}) with the amplitude parameter $A$ chosen to be $A=1$.
The initial data is defined on a large domain, so we take $L = 80$ and $N = 1600$. We compare the exact numerical solution with our extended leading-order solution (\ref{WNL_loc}), and the direct solution of the Ostrovsky equation (\ref{Oeq}) for the KdV soliton initial condition (\ref{ICO1}) (with non-zero mass)
 at the time $t = 1/\epsilon$, where $\epsilon = 0.025$. Thus, this is essentially the same example as in \cite{KMP} but the initial condition has non-zero mass.

The results are plotted in Figure \ref{fig:KMPComparison}(a), where the exact numerical solution (blue, solid line) is plotted against the extended leading-order weakly-nonlinear solution (red, dashed line) and the direct solution of the Ostrovsky equation (black, dashed line). The mean value is very small, and it is subtracted from the initial condition, and added afterwards, which is a rough approximation, justifiable in the asymptotic sense on large domains \cite{GM}.  The errors are shown in Figure \ref{fig:KMPComparison}(b), where the error for the weakly-nonlinear solution is the blue, solid line, and the error in the direct solution of the Ostrovsky equation is the red, dashed line. We can see that the solutions are similar but with a vertical shift between the weakly-nonlinear solution and the direct solution of the Ostrovsky equation. We note that the constructed leading-order weakly-nonlinear solution (\ref{WNL_loc}) more accurately resolves the amplitude of the main wave-structure, but overall both solutions show good agreement with the direct numericial solution of the equation (\ref{BousOst}). The accuracy of the constructed weakly-nonlinear solution can be further improved by including higher-order terms (see the example in Section 5.2 in \cite{KMP}).
\begin{figure}[!htbp]
	\subfigure[Solutions in full domain.]{\includegraphics[width=0.48\textwidth]{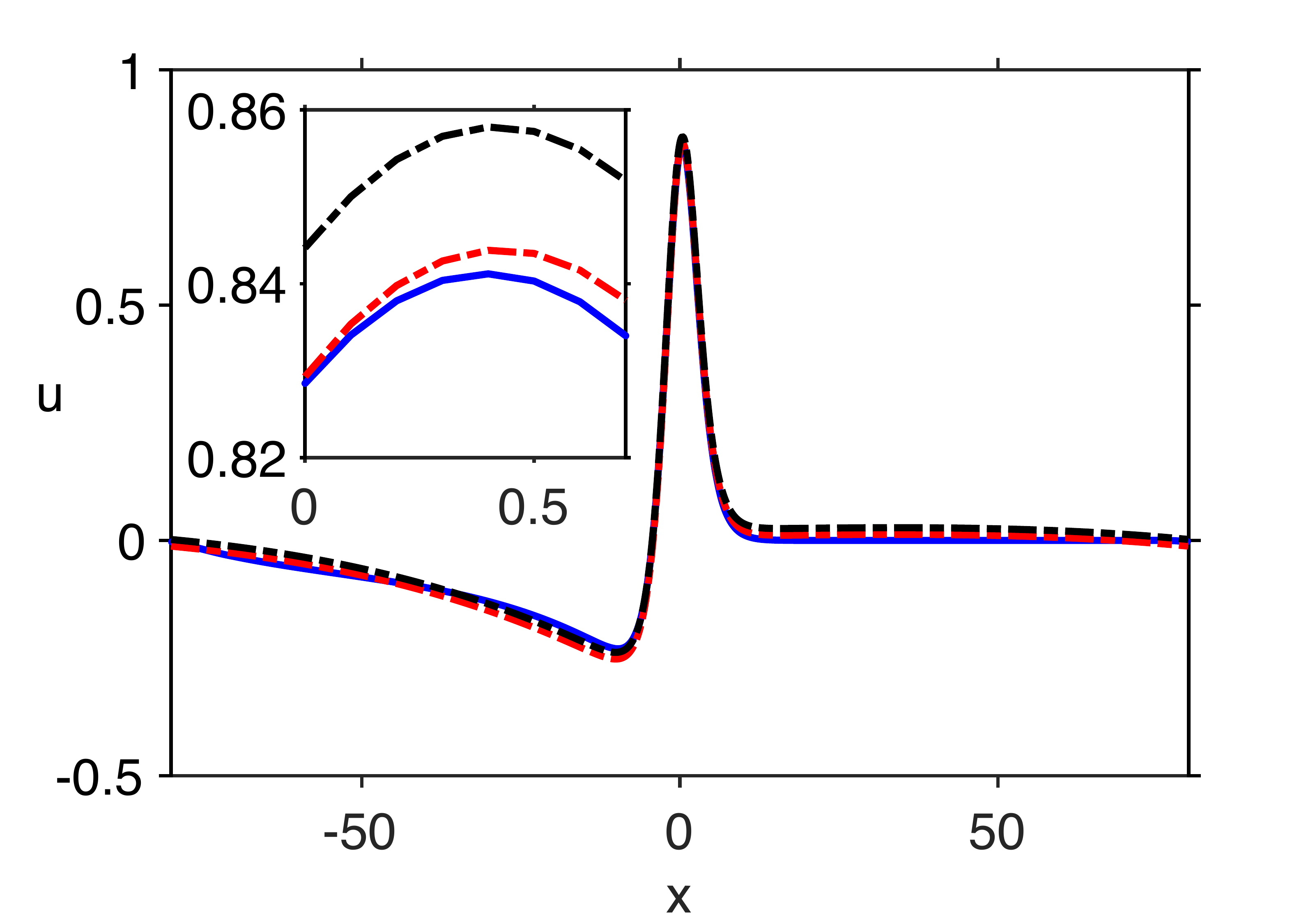}}~
	\subfigure[Difference between the numerical solution of Eq. (\ref{BousOst}) and the approximate solutions.]{\includegraphics[width=0.48\textwidth]{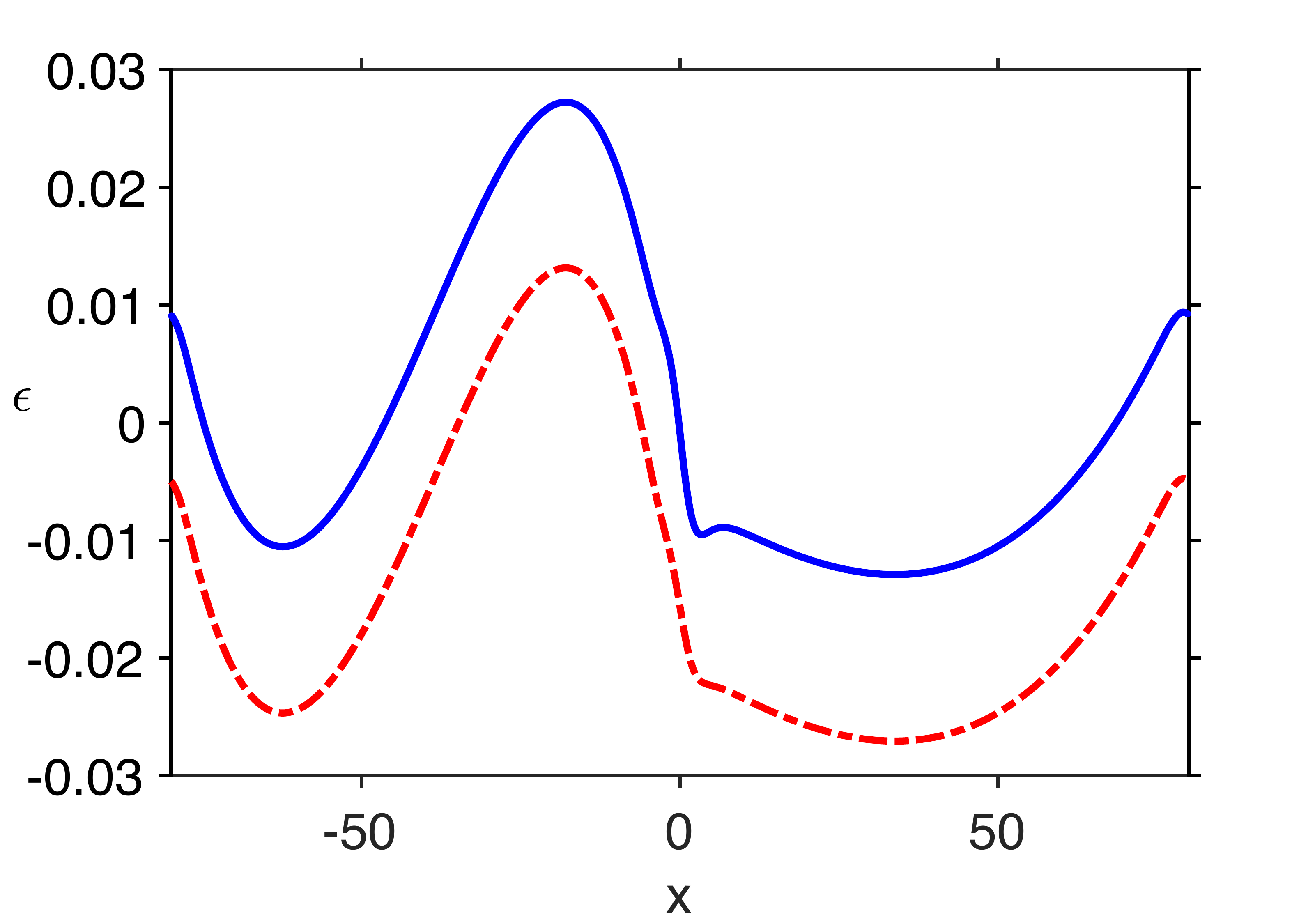}}
	\caption{\small A comparison of the numerical solution of the BKG equation (blue, solid line), extended leading-order weakly-nonlinear solution of the BKG equation (red, dashed line), and the solution of the Ostrovsky equation (black, dashed line), presented in (a) at $t = 1/\epsilon$ and the errors are plotted in (b) for the weakly-nonlinear solution (blue, solid line) and Ostrovsky equation (red, dashed line). Parameters are $L=80$, $N=1600$, $A=1$, $c = 1$, $\alpha = \beta = 1$, $\gamma = 0.1$, $\epsilon = 0.025$, $\Delta t = 0.01$ and $\Delta T = 0.000125$.}
	\label{fig:KMPComparison}
\end{figure}

In contrast to that behaviour, we also show the comparison between the constructed extended leading-order solution (\ref{WNL_loc}) and direct solution of the Ostrovsky equation for the initial condition with non-zero mass on a finite periodic interval, mirroring the case considered in the first example of Section (\ref{sec:Resc1}). In this case the initial condition has a larger mean, but we try to do the same as in the previous case of a large domain, i.e. we subtract the mean from the initial condition in order to do the numerics, and add it afterwards, and show that this is no longer a valid approximation.
We take the same parameters as used in Figure \ref{fig:g01}(a), namely $c = \alpha = \beta = 1$, $\gamma = 0.1$ and $\epsilon = 0.001$, with domain parameters $L=40$ and $N=800$. The initial condition is given by
\begin{align}
u(x,0) &= A \sechn{2}{\frac{x}{\Lambda}} + d,  \label{ICO2} \\
u_t (x,0) &= \frac{2cA}{\Lambda} \sechn{2}{\frac{x}{\Lambda}} \tanhn{ }{\frac{x}{\Lambda}},
\label{ComparisonIC}
\end{align}
where $A = 2$. This corresponds to the previous choice of $k = 1/\sqrt{3}$ in Section \ref{sec:Resc1}, and we have $d = 1$. 

The results are presented at $T = 0.5$ and $T = 1$ in Figure \ref{fig:Locald1g01}. We can see that the direct solution of the Ostrovsky equation 
for the initial condition (\ref{ICO2}) (black, dash-dotted line) has a large shift with respect to the exact numerical solution (blue, solid line),  while our constructed leading-order solution (red, dashed line) is very close to the exact solution. Comparing to the result at $T = 0.5$ we also notice that the direct solution of the Ostrovsky equation is oscillating, as a whole, at a different frequency to the exact solution. 
Thus, one can not use the Ostrovsky equation directly for initial conditions with non-zero mass on a finite periodic interval, and a valid approximation is instead provided by our constructed weakly-nonlinear solution.
\begin{figure}[!htbp]
	\subfigure[Result at $T = 0.5$.]{\includegraphics[width=0.48\textwidth]{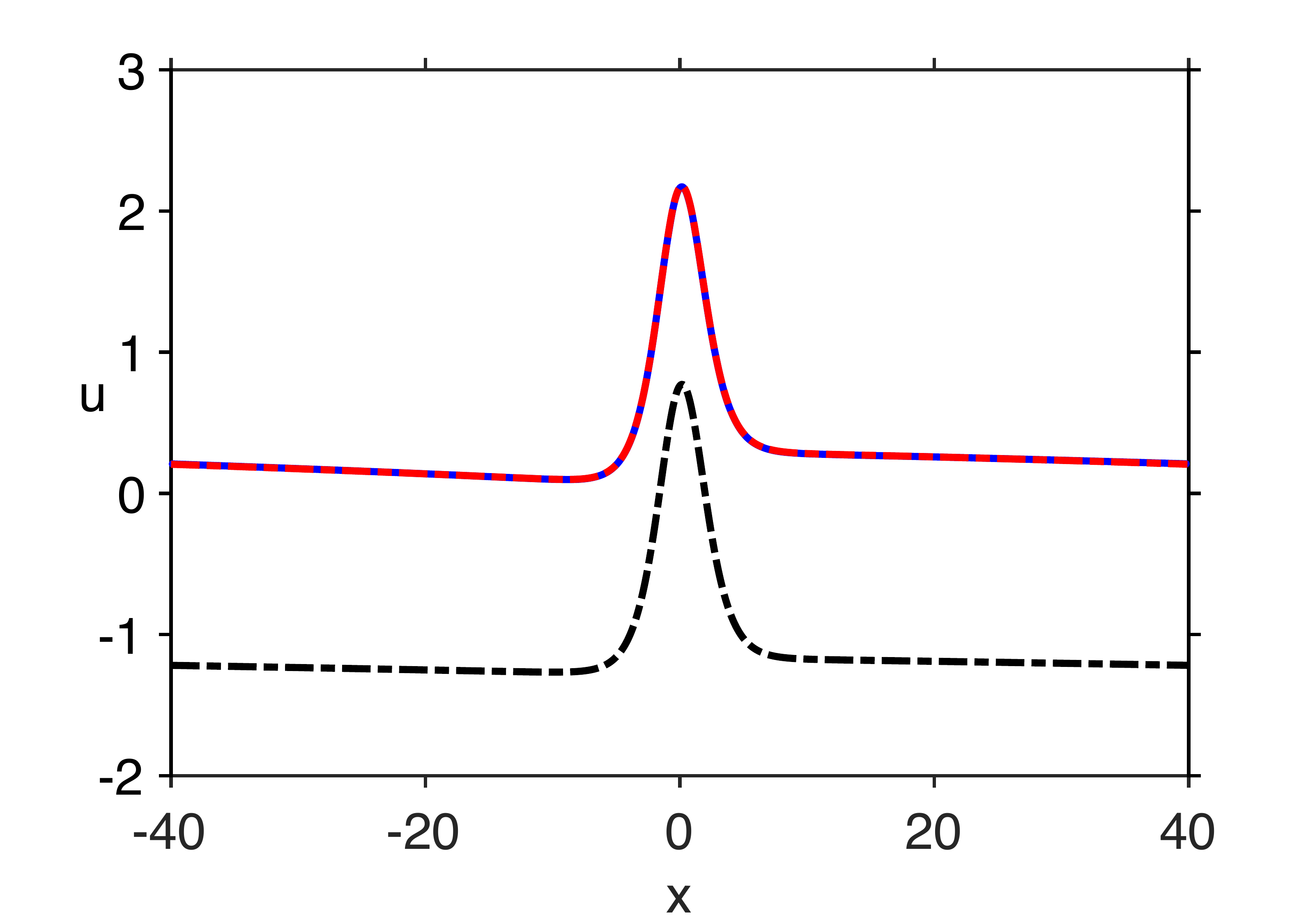}}~
	\subfigure[Result at $T = 1$.]{\includegraphics[width=0.48\textwidth]{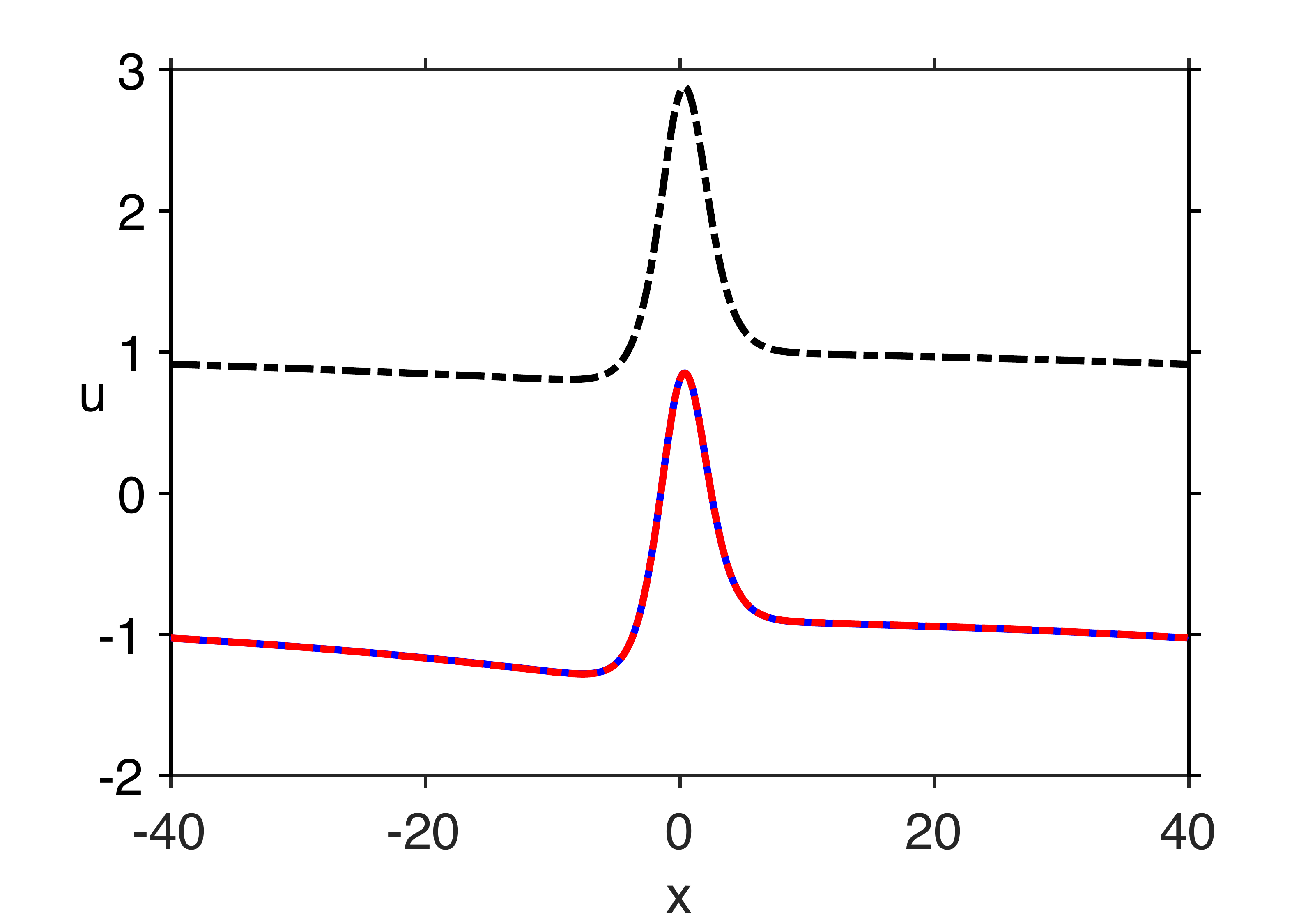}}
	\caption{\small A comparison of the numerical solution of the BKG equation (blue, solid line),  extended leading-order weakly-nonlinear solution of the BKG equation (red, dashed line), and the solution of the Ostrovsky equation (black, dashed line), presented at (a) $T = 0.5$, and (b) $T = 1$. Parameters are $L=40$, $N=800$, $A = 2$, $c = 1$, $\alpha = \beta = 1$, $\gamma = 0.1$, $\epsilon = 0.001$, $\Delta t = 0.01$ and $\Delta T = \epsilon \Delta t$.}
	\label{fig:Locald1g01}
\end{figure}

\subsection{\texorpdfstring{Example 2:  $c = 1,  \alpha = \beta = 2$, $\gamma = 2$}{Example 2: c = 1,  alpha = beta = 2, gamma = 2}}
Let us now consider the same example as in \cite{GH}, specifically we want the leading-order problem to be related to the Ostrovsky equation
\begin{equation}
\lb \eta_{t} + \eta \eta_{x} + \eta_{xxx} \rb_{x} = \eta.
\label{GHOstEq}
\end{equation}
Therefore, we consider the equation  (\ref{BousOstOld}) with the coefficients $c = 1$, $\alpha = 2$, $\beta = 2$, $\gamma = 2$, which yields the same coefficients in the leading-order Ostrovsky equation as in equation (\ref{GHOstEq}). The initial condition is again given by (\ref{GHKdVSol}) with $A = 32$ (to match the results in \cite{GH}), and the data is defined on a very large domain.

In this case the value of $F_0$ is close to zero and therefore the constructed weakly-nonlinear solution and the solution of the Ostrovsky equation are again in very good agreement. The comparison of these two solutions at $T = 40$ and the errors at this time are shown in Figure \ref{fig:WNLGHComparison}. We can see that the error is small and is caused by a slight phase shift. Therefore, as the agreement is very good between these two solutions, we will only use the constructed weakly-nonlinear solution to compare to the ``exact" (numerical) solution.

\begin{figure}[!htbp]
	\subfigure[Solution in enhanced domain.]{\includegraphics[width=0.48\textwidth]{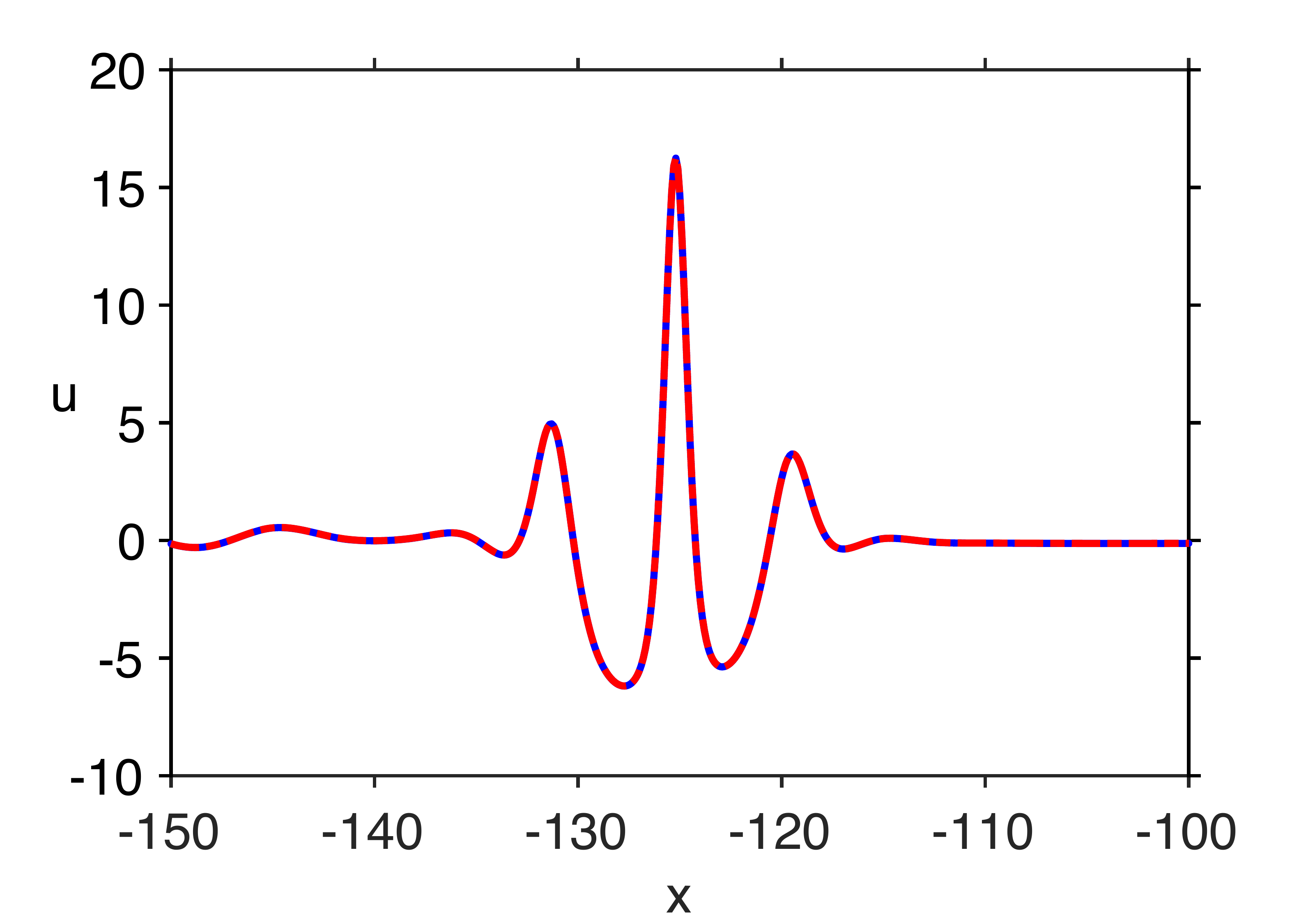}}~
	\subfigure[Difference between the numerical solution of Eq. (\ref{BousOst}) and solution of the Ostrovsky equation in enhanced domain.]{\includegraphics[width=0.48\textwidth]{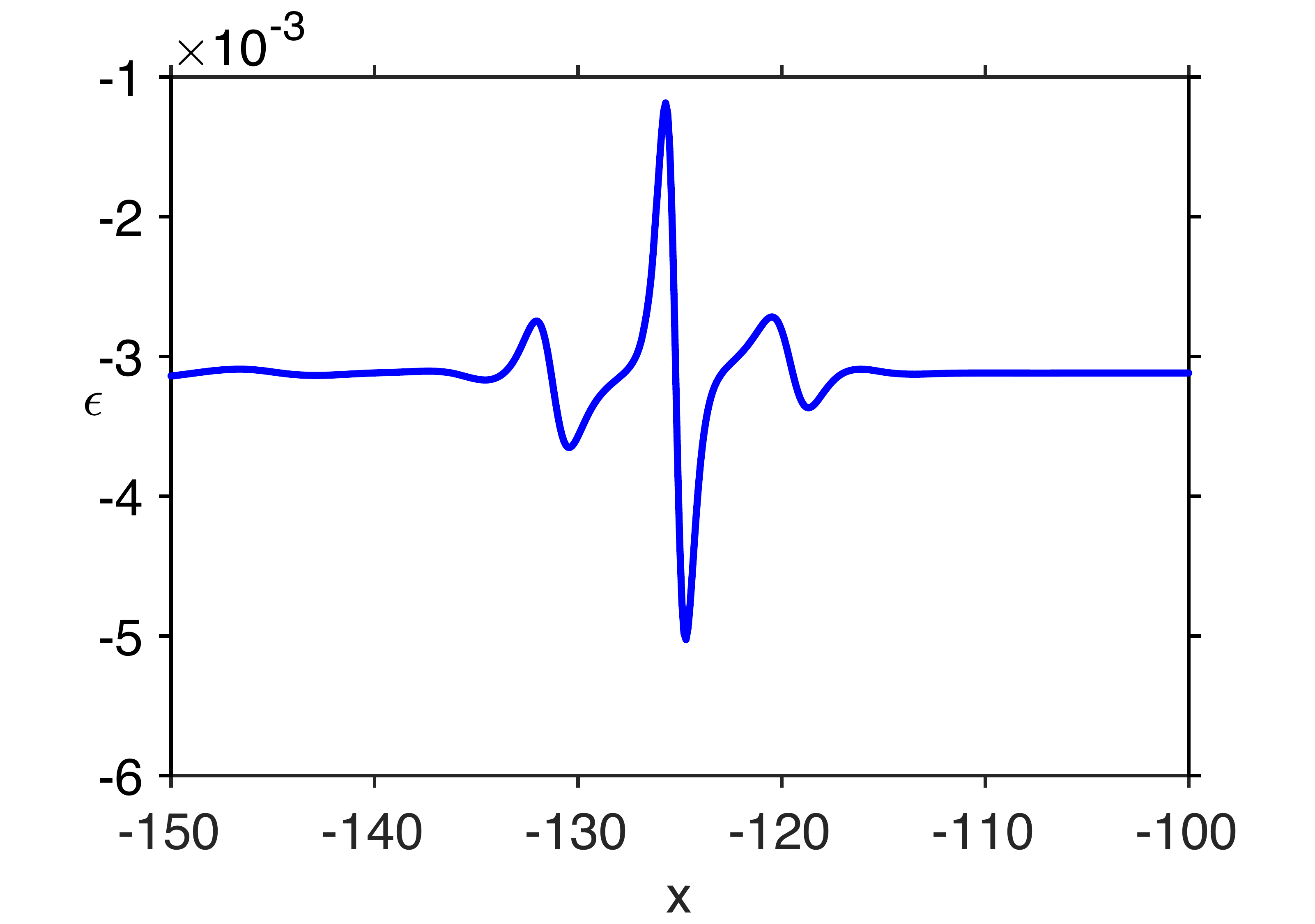}}
	\caption{\small A comparison of the extended leading-order weakly-nonlinear solution of the BKG equation (blue, solid line) and direct solution of the corresponding Ostrovsky equation (red, dashed line), in an enhanced domain, plotted in (a) at $T=40$, and the error between the solution is plotted in (b). Parameters are $L=5,000$, $N=100,00$, $A = 32$, $c = 1$, $\alpha = \beta = \gamma = 2$, $\epsilon = 0.001$, $\Delta t = 0.01$ and $\Delta T = \epsilon \Delta t$.}
	\label{fig:WNLGHComparison}
\end{figure}

To correspond with the results in \cite{GH}, the results are presented up to $T = 40$ and for $\epsilon = 0.001$ (i.e. for $t$ up to $t = 40,000$). We consider a large domain as was done in \cite{GH} and therefore we take $L = 5,000$ corresponding to $N = 100,000$. The step sizes are the same as in previous calculations. 
 
We compare our constructed weakly-nonlinear solution to the exact numerical solution in Figure \ref{fig:BousWNLComparison}. We note that the exact solution (blue, solid line) is in good qualitative agreement with the constructed solution (red, dashed line) even at such a large time ($T=40$).
\begin{figure}[!htbp]
	\centering
	\includegraphics[width=0.7\textwidth]{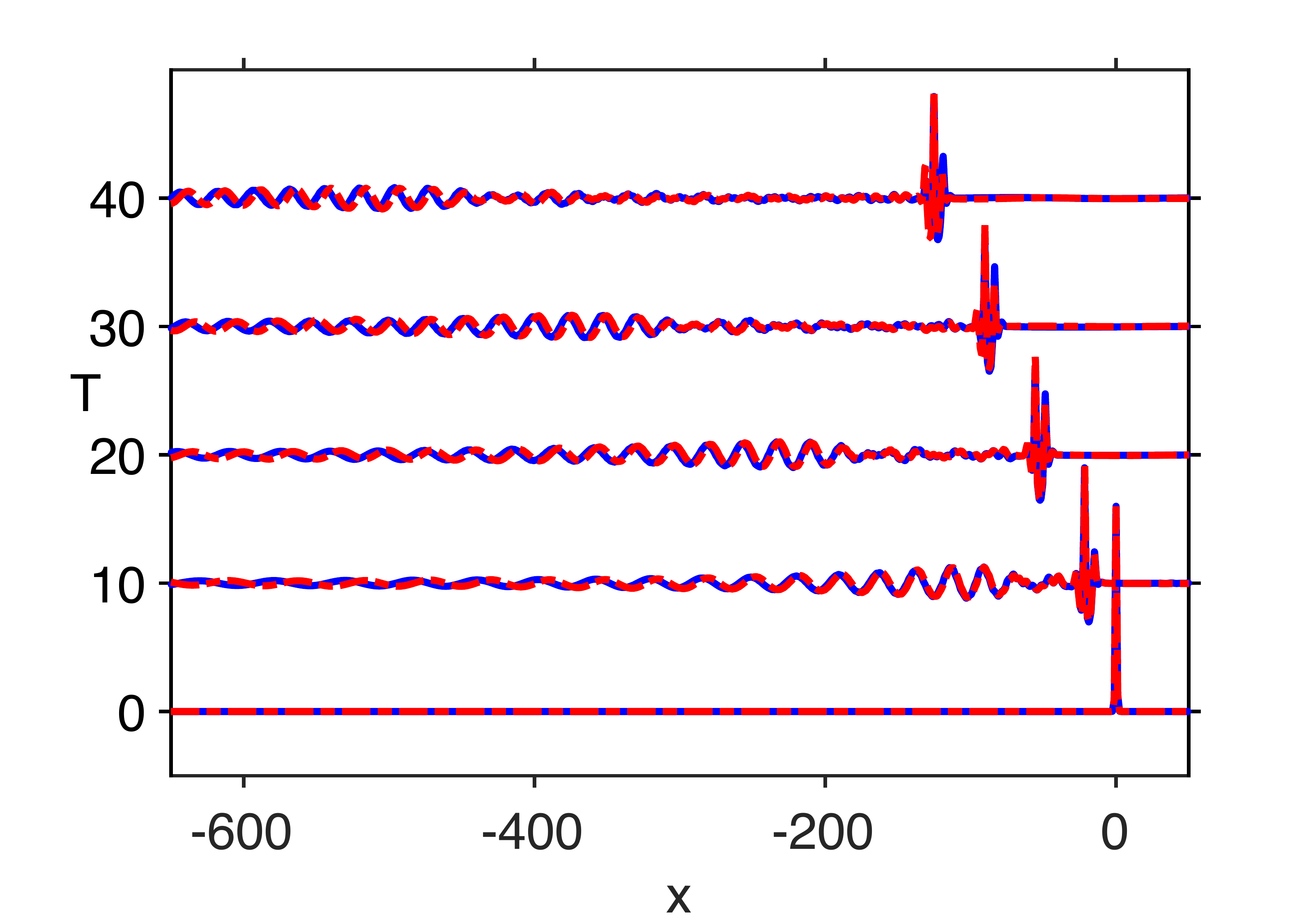}
	\caption{\small A comparison of the numerical solution of the BKG equation (blue, solid line) and the extended leading-order weakly-nonlinear solution (red, dashed line) in the moving reference frame, for the KdV soliton initial condition. Parameters are $L=5,000$, $N=100,000$, $A = 32$, $c = 1$, $\alpha = \beta = \gamma = 2$, $\epsilon = 0.001$, $\Delta t = 0.01$ and $\Delta T = \epsilon \Delta t$.}
	\label{fig:BousWNLComparison}
\end{figure}

\subsection{Example 3: cnoidal wave initial condition}
\label{sec:Cnoidal}
Finally in this section, we extend the previous examples to the initial condition in the form of a right-propagating KdV cnoidal wave of the equation (\ref{feq}) for $f^{-}$, i.e.
\begin{equation}
2 c f_{T}^{-} + \alpha f^{-} f_{\xi_{-}}^{-} + \beta c^2 f_{\xi_{-} \xi_{-} \xi_{-}}^{-}  = 0.
\label{feq-}
\end{equation}
The exact cnoidal wave solution can be written in terms of the Jacobi elliptic function as follows (e.g., \cite{Johnson97})
\begin{align}
&f^{-} =  - \frac{6 \beta c^3}{\alpha} \left (f_2 - (f_2 - f_3) {\rm cn}^2[(\xi^- + v T) \sqrt{\frac{f_1 - f_3}{2}} | m] \right ), \\
&\mbox{where} \quad v =  (f_1 + f_2 + f_3) \beta c^2, \quad m = \frac{f_2 - f_3}{f_1 - f_3}.
\end{align}
Here, the solution is parametrised by the constants $f_3 < f_2 < f_1$ such that the elliptic modulus $0 < m < 1$. The wave length is equal to
$$
L = 2 K(m)  \sqrt{\frac{2}{f_1 - f_3}},
$$
where $K(m)$ is the complete elliptic integral of the first kind.

Firstly we consider a comparison to the case in Section \ref{sec:Resc1}, which will be a limiting case of the cnoidal wave when $m \to 1$. Therefore we take the parameters $f_{1} = -1/6 + 1 \times 10^{-8}$, $f_{2} = -1/6$, $f_{3} = -1/2$, giving $m \approx 1 - 3 \times 10^{-8}$. This corresponds to the results in Figure \ref{fig:g05}(a) where we take the parameters $c = \alpha = \beta = 1$, $\gamma = 0.5$, $\epsilon = 0.001$ and the initial condition has a pedestal of amplitude $d = 1$. The results are compared in Figure \ref{fig:CnoidalComparison}. We can see that the constructed solution has high accuracy.
\begin{figure}[!htbp]
	\subfigure[Solution at $t=0$.]{\includegraphics[width=0.48\textwidth]{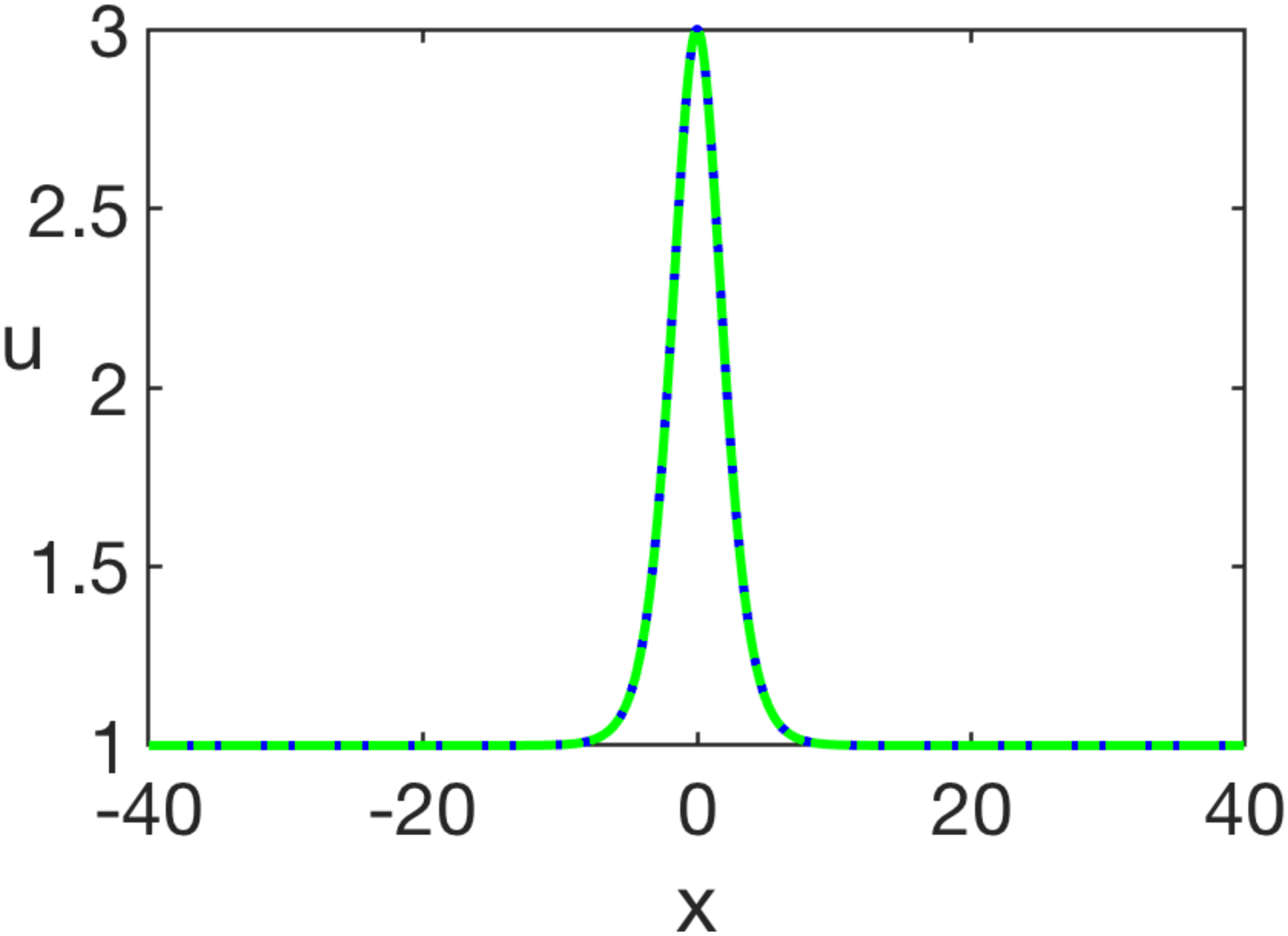}}~
	\subfigure[Solution at $t=1000$.]{\includegraphics[width=0.48\textwidth]{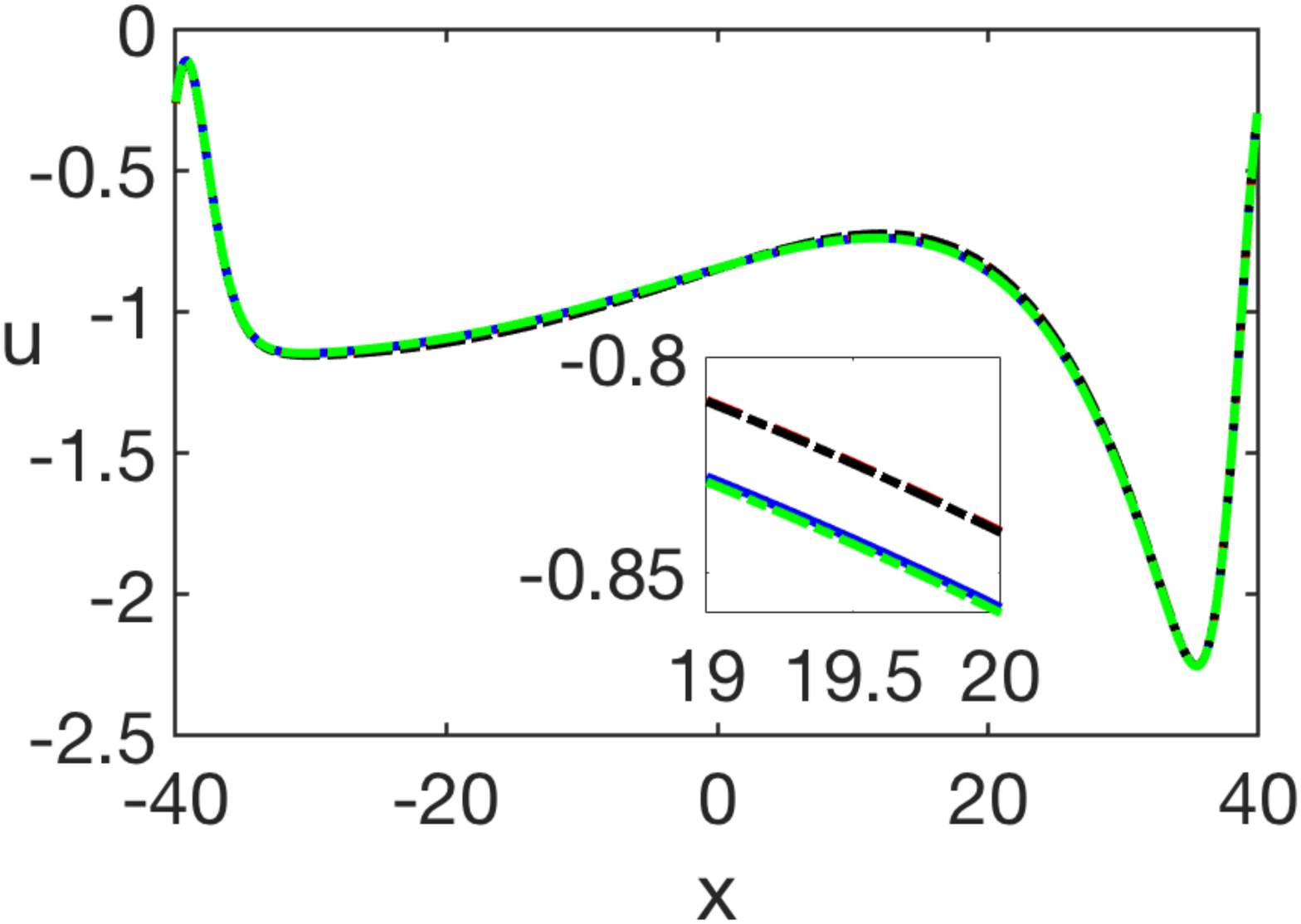}}
	\caption{\small A comparison of the numerical solution (solid, blue) and the weakly-nonlinear solution including leading-order (dashed, red), $\O{\sqrt{\epsilon}}$ (dash-dot, black) and $\O{\epsilon}$ (dotted, green) corrections, at (a) $t=0$ and (b) $t=1/\epsilon$. Parameters are $L=40$, $N=800$, $k = 1/\sqrt{3}$, $\alpha = \beta = c = 1$, $\gamma = 0.5$, $\epsilon = 0.001$, $\Delta t = 0.01$ and $\Delta T = \epsilon \Delta t$, with cnoidal wave parameters $f_{1} = -1/6 + 1 \times 10^{-8}$, $f_{2} = -1/6$, $f_{3} = -1/2$, giving $m = 1 - 3 \times 10^{-8}$. The solution agrees well to leading order, and this agreement is improved with the addition of higher-order corrections.}
	\label{fig:CnoidalComparison}
\end{figure}

We now consider two cases for a cnoidal wave with more than one period of the wave in the domain, one without a pedestal and a second with a pedestal term. As we have shown in Figure \ref{fig:CnoidalComparison} that the solution can be constructed up to $\O{\epsilon}$, we will only take terms up to and including $\O{\sqrt{\epsilon}}$ to be consistent with the results in Section \ref{sec:GHRes}. We take the parameters to be $c = 1$ and $\alpha = \beta = \gamma = 2$, with $\epsilon = 0.001$. For the first case, where we have no pedestal term, we take $f_{1} = 10^{-3}$, $f_{2} = 0$, $f_{3} = -1/6$, giving $m \approx 0.994$. The results for this case are presented at $t = 3/\epsilon$ in Figure \ref{fig:CnoidalC2d0}. The agreement between the numerical solution (solid, blue line) and constructed weakly-nonlinear solution at leading order (dashed, red) and including $\O{\sqrt{\epsilon}}$ terms (dash-dot, black) is good, and we note that the weakly-nonlinear solutions are indistinguishable at this scale.
\begin{figure}[!htbp]
	\subfigure[Solution at $t=0$.]{\includegraphics[width=0.48\textwidth]{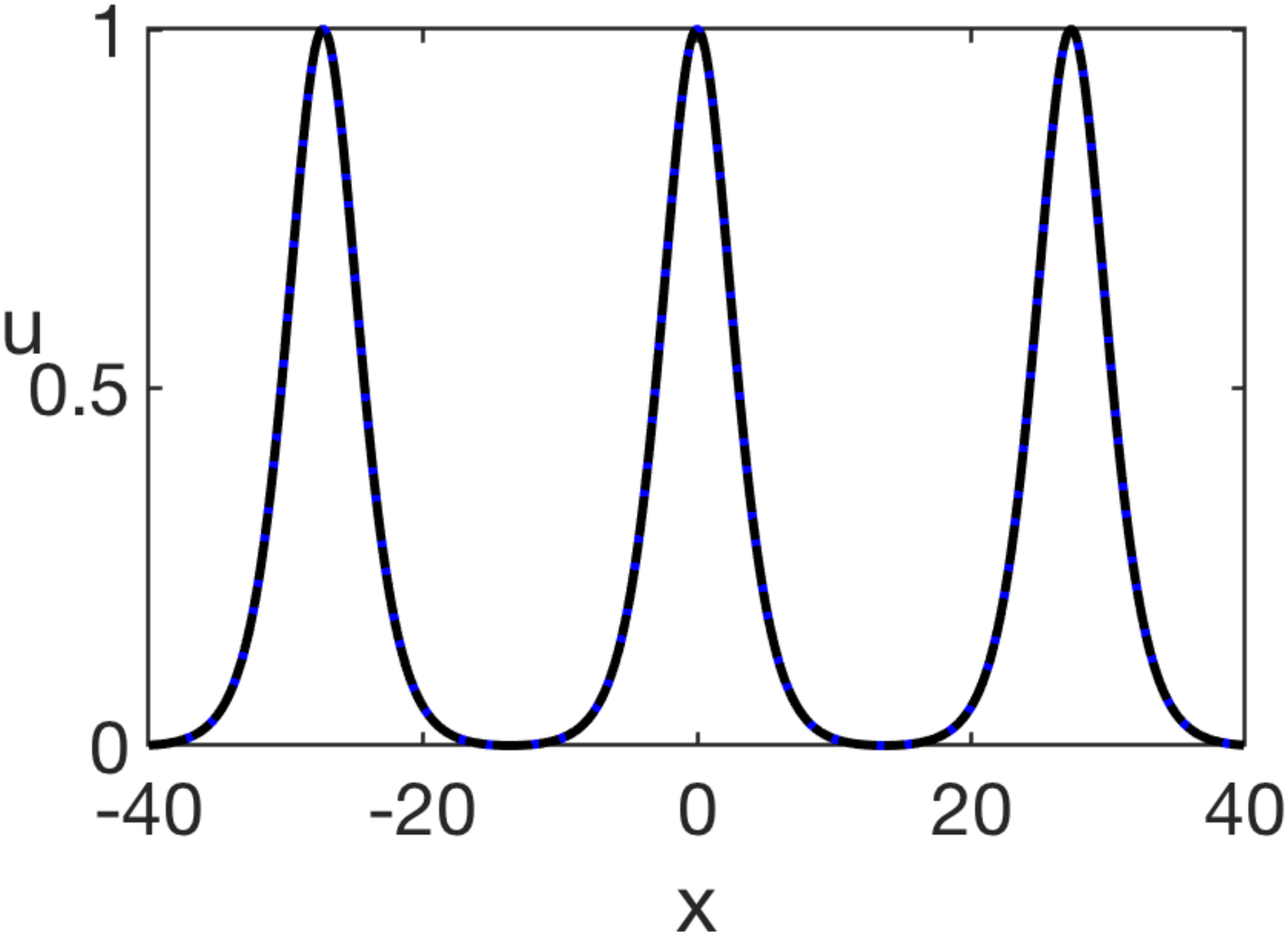}}~
	\subfigure[Solution at $t=3000$.]{\includegraphics[width=0.48\textwidth]{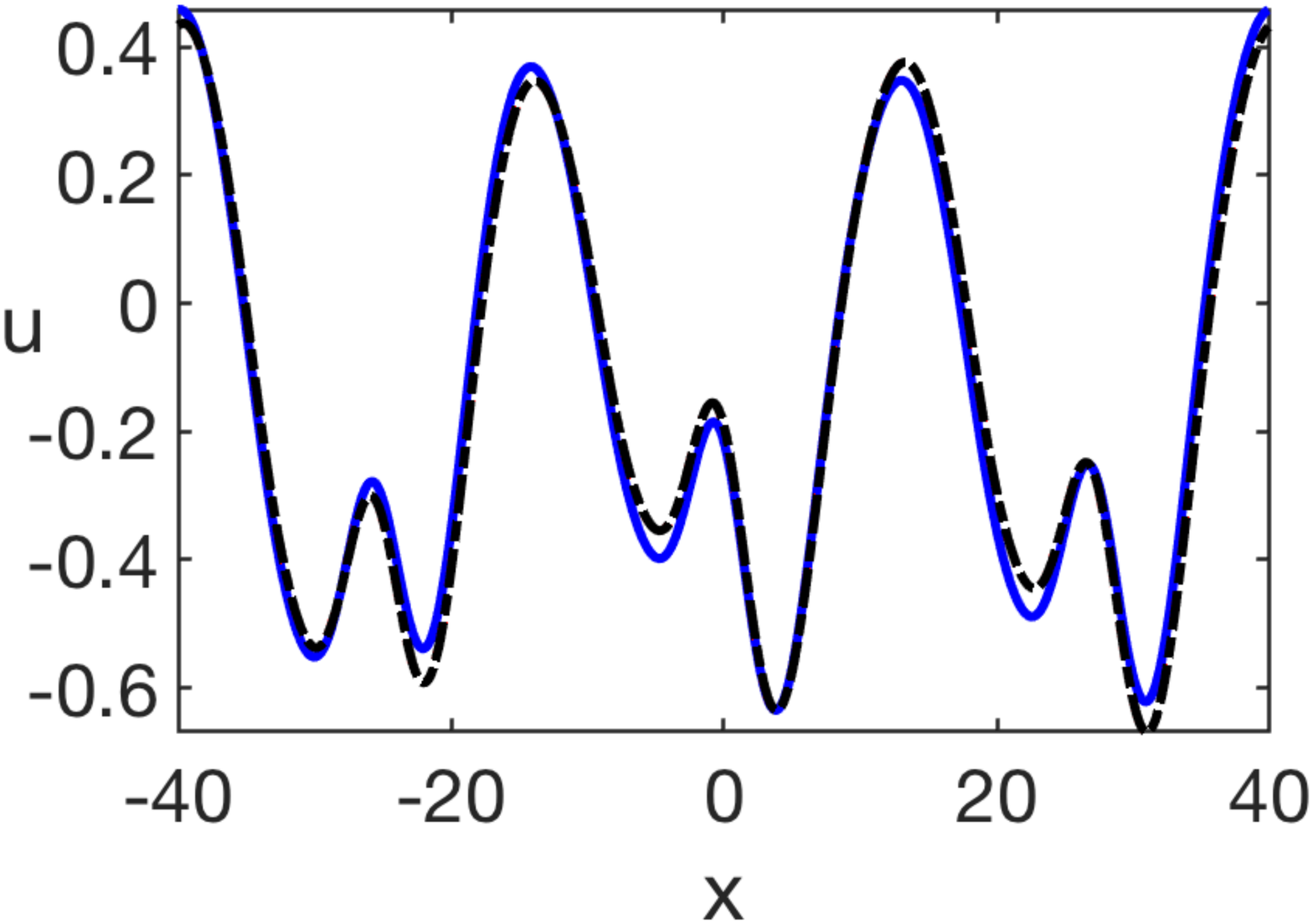}}
	\caption{\small A comparison of the numerical solution (solid, blue) and the weakly-nonlinear solution including leading-order (dashed, red) and $\O{\sqrt{\epsilon}}$ (dash-dot, black) terms, at (a) $t=0$ and (b) $t=3/\epsilon$. Parameters are $L=40$, $N=800$, $k = 1/\sqrt{3}$, $c = 1$, $\alpha = \beta = \gamma = 2$, $\epsilon = 0.001$, $\Delta t = 0.01$ and $\Delta T = \epsilon \Delta t$, with cnoidal wave parameters $f_{1} = 10^{-3}$, $f_{2} = 0$, $f_{3} = -1/6$, giving $m \approx 0.994$. The solutions are in good agreement.}
	\label{fig:CnoidalC2d0}
\end{figure}

The second case is for a pedestal of amplitude $d = 1$, with the same parameters as the previous case. As we have a pedestal, the cnoidal wave parameters are $f_{1} = -1/6 + 10^{-3}$, $f_{2} = -1/6$, $f_{3} = -1/3$, giving $m \approx 0.994$. The results are presented in Figure \ref{fig:CnoidalC2d1} and we see that there is again good agreement between the results, with no discernible difference  in accuracy of the constructed solution compared to the case with no pedestal.
\begin{figure}[!htbp]
	\subfigure[Solution at $t=0$.]{\includegraphics[width=0.48\textwidth]{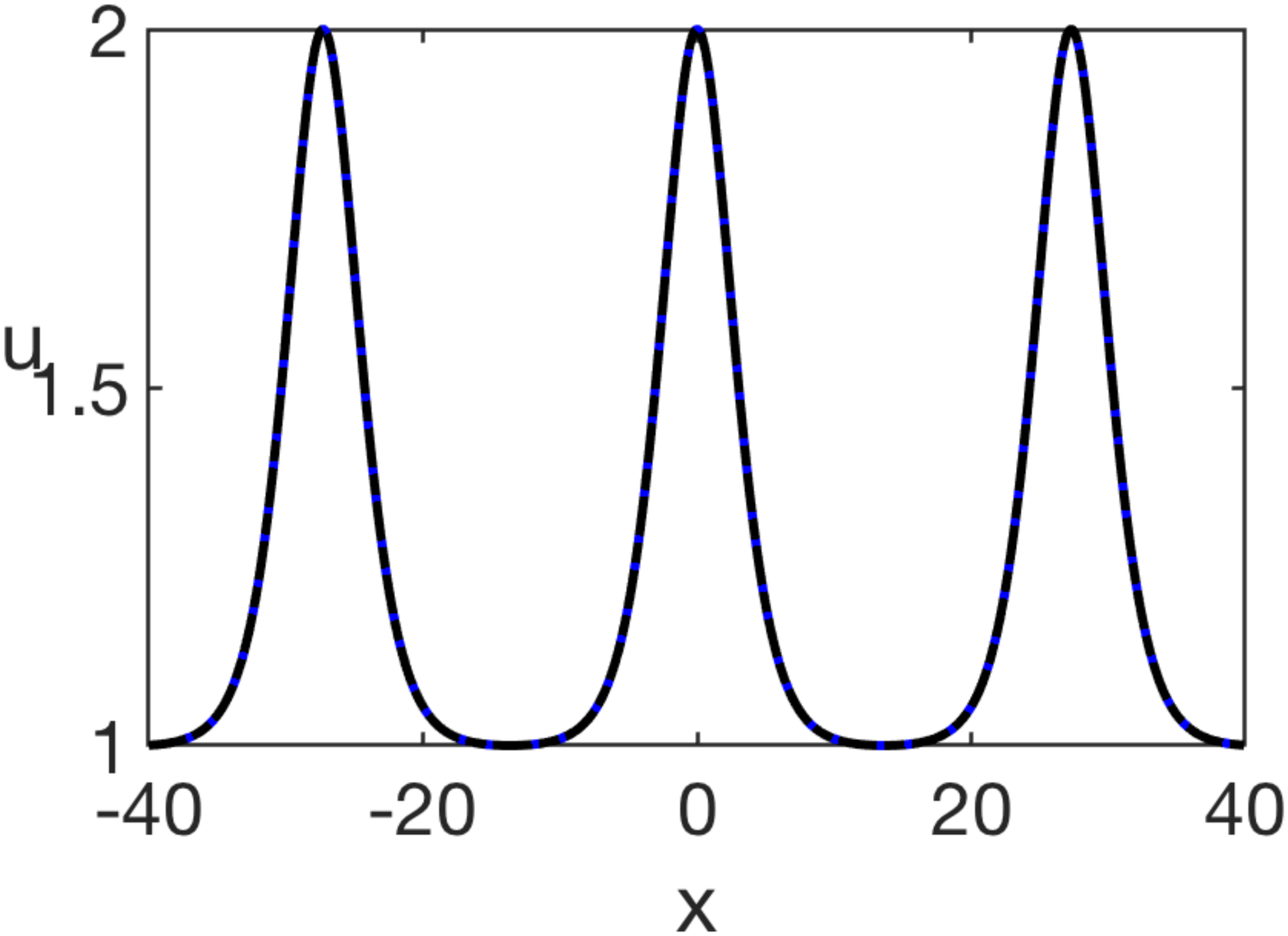}}~
	\subfigure[Solution at $t=3000$.]{\includegraphics[width=0.48\textwidth]{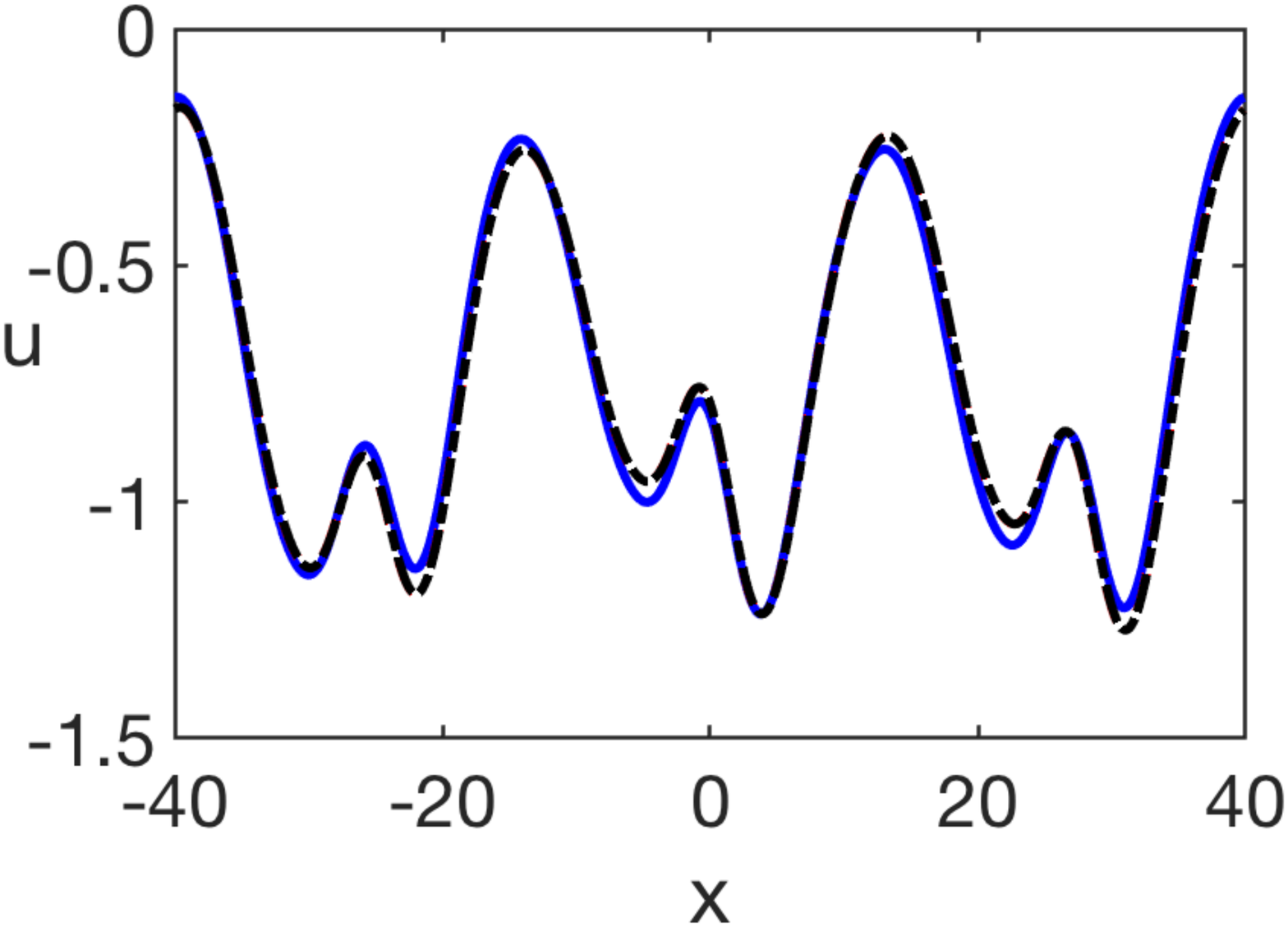}}
	\caption{\small A comparison of the numerical solution (solid, blue) and the weakly-nonlinear solution including leading-order (dashed, red) and $\O{\sqrt{\epsilon}}$ (dash-dot, black) terms, at (a) $t=0$ and (b) $t=3/\epsilon$. Parameters are $L=40$, $N=800$, $k = 1/\sqrt{3}$, $c = 1$, $\alpha = \beta = \gamma = 2$, $\epsilon = 0.001$, $\Delta t = 0.01$ and $\Delta T = \epsilon \Delta t$, with cnoidal wave parameters $f_{1} = -1/6 + 10^{-3}$, $f_{2} = -1/6$, $f_{3} = -1/3$, giving $m \approx 0.994$. The solutions are in good agreement.}
	\label{fig:CnoidalC2d1}
\end{figure}

The constructed solution could find useful applications in the studies of the modulational instability of periodic solutions in physical systems mentioned in the Introduction (e.g., \cite{WJ, WJ15}) similarly to our earlier work, in a different context, in \cite{GGK}.

\section{Conclusions}
\label{sec:Conc}
Our study has been mainly  devoted to the initial-value problem for the Boussinesq-Klein-Gordon (BKG) equation (\ref{BousOstOld}) in the class of periodic functions on a finite interval, where the first initial condition in (\ref{BousOstIC}) has non-zero mean (an extension of the weakly-nonlinear solution to the general case when both initial conditions, for $u$ and $u_t$, may have non-zero mean, is outline in Appendix A).  Such problem formulation is relevant to the studies of the evolution of periodic waves in the oceanic context, since the developed methodology can be extended to rotation-modified strongly-nonlinear and Boussinesq-type systems discussed in \cite{O, Shrira81, Shrira86, Helfrich2007} (see also \cite{GOSS}), as well as being directly relevant to  the scattering problems in the context of delaminated solid structures discussed in the Introduction. 

We constructed the d'Alembert-type weakly-nonlinear solution of the initial-value problem in terms of solutions of two leading-order Ostrovsky equations. Importantly, we by-passed the zero-mass contradiction for the Ostrovsky equation by considering the deviation from the oscillating mean value, and suggested a novel asymptotic procedure in powers of $\sqrt{\epsilon}$ which is based on the use of fast characteristic variables and two slow time variables.

We then compared the constructed weakly-nonlinear solution to the ``exact'' numerically calculated solution of our initial-value problem, for the initial condition in the form of a localised wave on a constant background, for a number of cases. We carefully investigated the error behaviour for the leading-order solution, the solution including $\O{\sqrt{\epsilon}}$ terms, and the solution including $\O{\epsilon}$ terms. The accuracy improved as more terms from the weakly-nonlinear expansion were included, with the $\O{\sqrt{\epsilon}}$ correction compensating for a phase shift while the $\O{\epsilon}$ correction adjusted the amplitude of the solution and captured higher-order left-propagating waves not captured by the leading-order solution. 

The errors were plotted against $\epsilon$ and we showed that the absolute error generally scales with the order of the next term in the weakly-nonlinear expansion. We also observed that, as $\gamma$ increases, the terms at non-integer powers of $\epsilon$ become small and therefore the absolute error scales with the next integer power. Increasing the mean value of the initial condition reduced this effect as the terms at non-integer powers of $\epsilon$ increase with the mean value, and this behaviour was seen in the tabulated values for the error curves.

We also considered the case of a localised initial condition with non-zero mass on a large (``infinite") domain by using our constructed solution, and reproduced the scenario described in \cite{Grimshaw99} and modelled in \cite{GH}, as well as considering initial conditions in the form of a cnoidal wave of the KdV equation. 

Overall, in all examples in Sections \ref{sec:Num} and \ref{sec:GHRes}  the numerical results showed very good agreement between our constructed weakly-nonlinear solution, direct numerical simulations,  and previously available results.} The developed methodology can be used in many applied contexts, allowing one to by-pass similar contradictions in the studies related, for example, to Kadomtsev-Petviashvili (KP) \cite{KP} and short-pulse \cite{sp} equations, as well as many similar equations and generalisations (see, for example, \cite{KKMS,HNW} and references therein).  

The construction of a semi-analytical weakly-nonlinear solution for the initial-value problem paves the way for the development of efficient semi-analytical numerical procedures  for the complicated scattering problems describing waves in inhomogeneous media, similarly to our previous studies of the scattering of radiating solitary waves in delaminated elastic structures developed for the cases when zero-mass contradiction did not obstruct the development of the solution \cite{KT2017}.
\bigskip

\noindent
{\bf \Large Acknowledgments}
\bigskip

\noindent
We thank D. E. Pelinovsky and Y. A. Stepanyants for useful references and discussions. KRK is grateful to the organisers of  the programme ``Mathematical Aspects of Physical Oceanography" in  the Erwin Schr\"odinger Institute (ESI) in Vienna, Austria in March 2018 for the invitation, and to the ESI  for the financial support of her participation in the programme, where some parts of this paper have been discussed and developed. MRT acknowledges the support of the UK Engineering and Physical Sciences Research Council (EPSRC) during his PhD studentship.

\begin{appendices}

\section{General case}
\label{sec:ExtSol}
In this section we outline the extension of our construction of the d'Alembert-type solution to the general case. Thus, we consider the Cauchy problem (\ref{BousOstOld}) on the domain $\Omega = [-L, L] \times [0, T]$:
\begin{equation}
u_{tt} - c^2 u_{xx} = \epsilon \lsq \frac{\alpha}{2} \lb u^2 \rb_{xx} + \beta u_{ttxx} - \gamma u \rsq,
\label{BousOstOld1}
\end{equation}
\begin{equation}
u |_{t=0} = F(x), \quad u_t |_{t=0} = V(x),
\label{BousOstIC1}
\end{equation}
where $F$ and $V$ are sufficiently smooth $(2L)$-periodic functions, and both functions $F(x)$ and $V(x)$ may have non-zero mean values
\begin{equation}
F_{0} = \frac{1}{2L} \int_{-L}^{L} F(x) \dd{x}  \eqtext{and} V_{0} = \frac{1}{2L} \int_{-L}^{L} V(x) \dd{x}.
\label{MeanValIC1}
\end{equation}
Then, as discussed before,
\begin{equation}
\langle u \rangle (t) := \frac{1}{2L} \int_{-L}^{L} u(x,t) \dd{x} = F_{0} \cos{\lb \sqrt{\epsilon \gamma} t \rb} + V_{0} \frac{\sin{\lb \sqrt{\epsilon \gamma} t \rb}}{\sqrt{\epsilon \gamma}}.
\label{MeanVal1}
\end{equation}
In this general case, the initial-value problem for the deviation from the oscillating mean value $\tilde{u} = u - \langle u \rangle (t)$ takes the form
\begin{equation}
\tilde u_{tt} - c^2 \tilde u_{xx} = \epsilon \lsq \alpha \lb F_0 \cos{\lb \omega t  \rb}  + \frac{1}{\sqrt{\epsilon}} \frac{V_0}{\sqrt{\gamma}} \sin {\lb \omega t \rb} \rb \tilde u_{xx} + \frac{\alpha}{2} \lb \tilde u^2 \rb_{xx} + \beta \tilde u_{ttxx} - \gamma \tilde u \rsq,
\label{BousOstEq1}
\end{equation}
and
\begin{equation}
\tilde u |_{t=0} = F(x) - F_0, \quad \tilde u_t |_{t=0} = V(x) - V_0,
\label{BousOstICnew1}
\end{equation}
where $\omega = \sqrt{\gamma \epsilon}$.

We again look for a weakly-nonlinear solution of the form
\begin{align}
\tilde u \lb x, t \rb &= f^{+} \lb \xi_{+}, \tau, T \rb +  f^{-} \lb \xi_{-}, \tau, T \rb + \sqrt{\epsilon} P \lb \xi_{-}, \xi_{+}, \tau, T \rb + \O{\epsilon},
\label{WNLSol1}
\end{align}
where 
\begin{equation*}
\xi_{\pm} = x \pm c t, \quad \tau = \sqrt{\epsilon} t, \quad T = \epsilon t.
\end{equation*}
Here, we aim to construct the solution up to and including $\O{\sqrt{\epsilon}}$ terms, but the procedure can be continued to any order.

As before, the first non-trivial equation appears at $\O{\sqrt{\epsilon}}$, taking the form
\begin{equation}
- 4 c^2 P_{\xi_- \xi_+} = 2 c f^-_{\xi_- \tau} - 2 c f^+_{\xi_+ \tau} + \frac{\alpha V_0}{\sqrt{\gamma}} \sin \lb \sqrt{\gamma} \tau \rb \lb f^-_{\xi_- \xi_-} + f^+_{\xi_+ \xi_+} \rb.
\label{Peq}
\end{equation}
Averaging with respect to $x$ at constant $\xi_-$ or $\xi_+$ yields the equations
\begin{equation}
2 c f^-_{\xi_- \tau} + \frac{\alpha V_0}{\sqrt{\gamma}} \sin \lb \sqrt{\gamma} \tau \rb f^-_{\xi_- \xi_-} = 0,
\label{f-}
\end{equation}
and
\begin{equation}
2 c f^+_{\xi_+ \tau} - \frac{\alpha V_0}{\sqrt{\gamma}} \sin \lb \sqrt{\gamma} \tau \rb f^+_{\xi_+ \xi_+} = 0,
\label{f+}
\end{equation}
which can be integrated using the method of characteristics, giving
\begin{equation}
f^- = f^- \lb \xi_- + \frac{\alpha V_0}{2 c \gamma} \cos \lb \sqrt{\gamma} \tau \rb, T\rb, \quad f^+ = f^+ \lb \xi_+ - \frac{\alpha V_0}{2 c \gamma} \cos \lb \sqrt{\gamma} \tau \rb, T\rb.
\label{ff}
\end{equation}
The formulae (\ref{ff}) suggest the use of new variables
\begin{equation}
\tilde \xi_- = \xi_- + \frac{\alpha V_0}{2 c \gamma} \cos \lb \sqrt{\gamma} \tau \rb, \quad \tilde \xi_+ = \xi_+ - \frac{\alpha V_0}{2 c \gamma} \cos \lb \sqrt{\gamma} \tau \rb,
\label{change}
\end{equation}
instead of $\xi_-$ and $\xi_+$, and we can now rewrite the equation for $P$ as $P_{\tilde \xi_- \tilde \xi_+} = 0$, yielding
\begin{equation}
P = g^-\lb \tilde \xi_-, \tau, T \rb + g^+ \lb \tilde \xi_+, \tau, T \rb.
\label{P}
\end{equation}

At $\O{\sqrt{\epsilon}}$, using the same averaging procedure,
we obtain
\begin{align}
g^{\pm} _{\tilde \xi_{\pm}} &= - \frac{\alpha V_0}{4 c^2 \sqrt{\gamma}} \sin \lb \sqrt{\gamma} \tau \rb f^{\pm}_{\tilde \xi_{\pm} } + \left [ \mp \frac{\alpha^2 V_0^2}{16 c^3 \gamma} \lb \tau - \frac{\sin \lb 2 \sqrt{\gamma} \tau \rb}{2 \sqrt{\gamma}} \rb - \frac{\alpha F_0}{2 c \sqrt{\gamma}} \sin \lb \sqrt{\gamma} \tau \rb \right ] f^{\pm}_{\tilde \xi_{\pm} \tilde \xi_{\pm}} \nonumber \\
&\pm \frac{1}{2 c} A^{\pm}\lb \tilde \xi_{\pm}, T \rb \tau, 
\label{g-}
\end{align}
where 
\begin{equation}
A^{\pm}  \lb  \tilde \xi_{\pm}, T \rb = \lb \mp 2 c f_{T}^{\pm} + \alpha f^{\pm} f_{\tilde \xi_{\pm}}^{\pm} + \beta c^2 f_{\tilde \xi_{\pm} \tilde \xi_{\pm} \tilde \xi_{\pm}}^{\pm} \rb_{\tilde \xi_{\pm}} - \gamma f^{\pm}.
\end{equation}
Here, we omitted the homogeneous parts of the solutions for $g^{\pm} _{\tilde \xi_{\pm}}$. As before, they can be assumed to be equal to zero. Now, to avoid secular terms we require
\begin{equation}
\lb \mp 2 c f_{T}^{\pm} - \frac{\alpha^2 V_0^2}{8 c^2 \gamma}  f^{\pm}_{\tilde \xi_{\pm}}  + \alpha f^{\pm} f_{\tilde \xi_{\pm}}^{\pm} + \beta c^2 f_{\tilde \xi_{\pm} \tilde \xi_{\pm} \tilde \xi_{\pm}}^{\pm} \rb_{\tilde \xi_{\pm}} - \gamma f^{\pm} = 0.
\label{Ost1}
\end{equation}
Thus, in the general case there is an additional term in each of the Ostrovsky equations for the left- and right-propagating waves, and the equations (\ref{Ost1}) reduce to (\ref{feq}) when $V_0 = 0$. The additional terms provide corrections to the wave speeds. Indeed, the equations (\ref{Ost1}) can be reduced to the standard form of the Ostrovsky equations by the change of variables
$
\displaystyle 
\hat \xi_{\pm} = \tilde \xi_{\pm} \mp \frac{\alpha^2 V_0^2}{16 c^3 \gamma} T.
$

The expansion (\ref{WNLSol1}) is also substituted into the initial conditions (\ref{BousOstICnew1}), which we satisfy at the respective orders of the small parameter, as before.

To summarise, in the general case, the solution of the Cauchy problem (\ref{BousOstOld1}), (\ref{BousOstIC1}) for the original variable $u(x, t)$ up to and including $\O{\sqrt{\epsilon}}$ terms has the form
\begin{align}
u(x, t) &= V_{0} \frac{\sin{\lb \sqrt{\gamma} \tau \rb}}{\sqrt{\epsilon \gamma}} + F_{0} \cos{\lb \sqrt{\gamma} \tau \rb}  \nonumber \\ & + f^-  + f^+  + \sqrt{\epsilon} \left [ -\frac{\alpha V_0}{4 c^2 \sqrt{\gamma}} \sin \sqrt{\gamma} \tau \lb f^- + f^+ \rb  \right . \nonumber \\
&\left . - \frac{\alpha }{2 c \sqrt{\gamma}} \lb F_0 \sin \lb \sqrt{\gamma} \tau \rb + \frac{\alpha V_0^2}{16 c^2 \gamma} \sin \lb 2 \sqrt{\gamma} \tau \rb \rb \lb f^-_{\tilde \xi_-} - f^+_{\tilde \xi_+} \rb  \right ] + \O{\epsilon},
\label{WNLFV}
\end{align}
where the functions $f^{\pm} \lb \tilde \xi_{\pm}, T \rb$ are solutions of the Ostrovsky equations (\ref{Ost1}), which should be solved subject to the initial conditions
\begin{equation}
f^{\pm}|_{T=0} = \frac{1}{2c} \lb c [F \lb \tilde \xi_{\pm} \rb - F_0] \pm \int_{-L}^{\tilde \xi_{\pm}} (V(\sigma) - V_0) \dd{\sigma} \rb.
\end{equation}
Here $\displaystyle \tilde \xi_{\pm} = \xi_{\pm} \mp \frac{\alpha V_0}{2 c \gamma} \cos \lb \sqrt{\gamma} \tau \rb$ i.e the characteristic variables have variable speed when $V_0 \ne 0$.

To illustrate the validity of the constructed solution we use the numerical schemes in Appendix \ref{sec:NumMethod} to solve the BKG equation (\ref{BousOstOld1}) and the modified Ostrovsky equation (\ref{Ost1}). 

\begin{figure}[!htbp]
	\centering
	\includegraphics[width=0.6\textwidth]{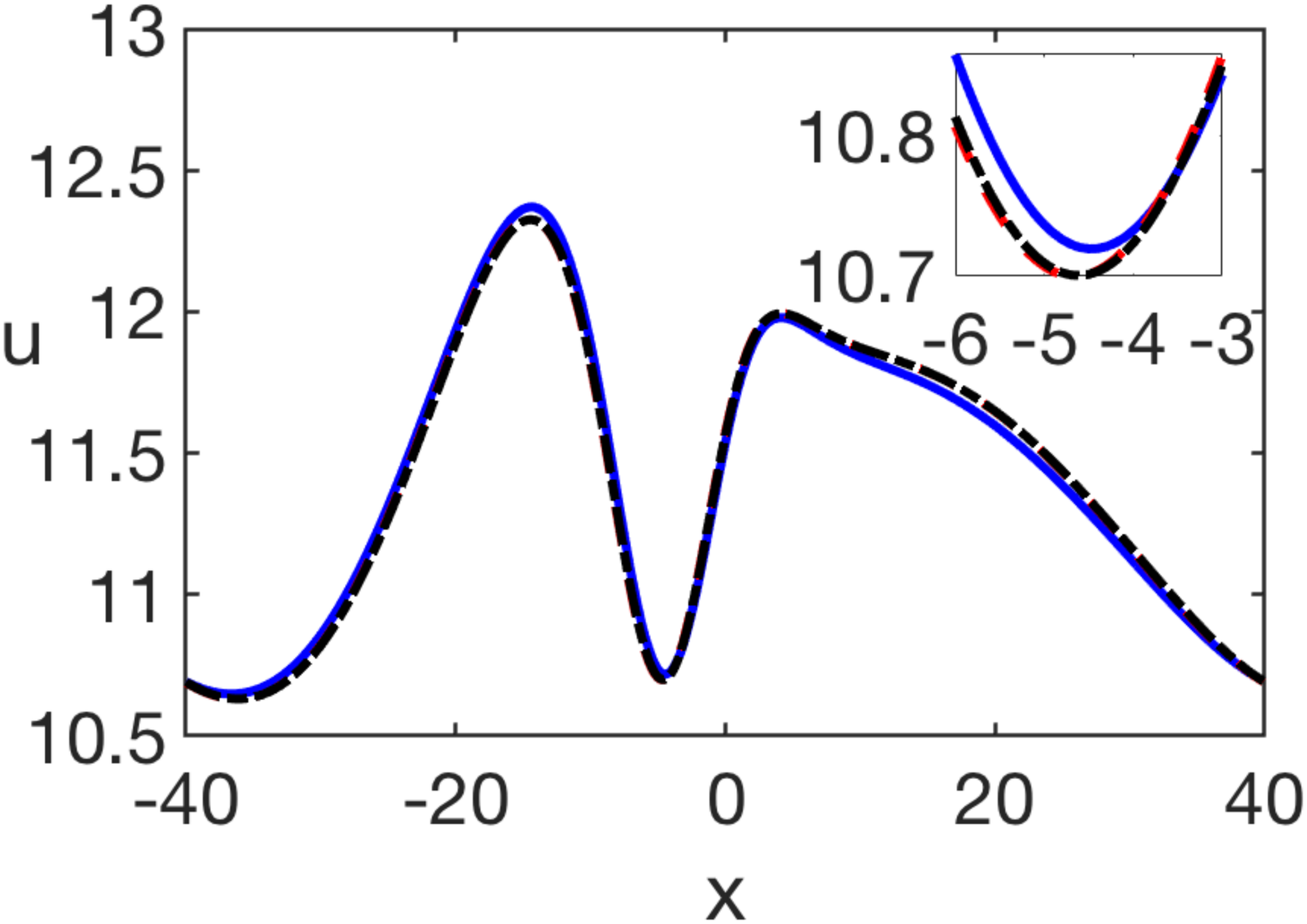}
	\caption{\small A comparison of the numerical solution of the BKG equation (solid, blue) and the constructed weakly-nonlinear solution  including leading-order (dashed, red) and $\O{\sqrt{\epsilon}}$ (dash-dot, black) terms, at $t=1/\epsilon$. Parameters are $L=40$, $N=800$, $k = 1/\sqrt{3}$, $c = \alpha = \beta = \gamma = 2$, $\epsilon = 0.001$, $\Delta t = 0.01$ and $\Delta T = \epsilon \Delta t$, $d_{1} = 5$ and $d_{2} = 0.5$. There is a good agreement between the numerical solution and the constructed weakly-nonlinear solution.}
	\label{fig:FVFigure}
\end{figure}

We take $c = \alpha = \beta = \gamma = 2$ and $\epsilon = 0.001$, with the initial condition defined as
\begin{align}
u(x,0) &= A \sechn{2}{\frac{x}{\Lambda}} + d_{1}, \\
u_t (x,0) &= \frac{2cA}{\Lambda} \sechn{2}{\frac{x}{\Lambda}} \tanhn{ }{\frac{x}{\Lambda}} + d_{2},
\label{ICFV}
\end{align}
where $d_{1}$, $d_{2}$ are constants and we have
\begin{equation}
A = \frac{6ck^2}{\alpha}, \quad \Lambda = \frac{\sqrt{2c \beta}}{k},
\label{BOstCoefsFV}
\end{equation}
with $k = \sqrt{\alpha/3c}$. We take $d_{1} = 5$ and $d_{2} = 0.5$ and present the results at $t = 1/\epsilon$ in Figure \ref{fig:FVFigure}. We can see that there is a reasonable agreement between the results, with a small phase shift between the constructed solution and the exact numerical solution. We note that the inclusion of $\O{\sqrt{\epsilon}}$ terms has not provided a significant improvement on the results, suggesting that the cases when $V_{0} \neq 0$ require either more terms in the expansion, or a smaller value of $\epsilon$, in order to see an improvement in the accuracy of the solution. Thus, the constructed weakly-nonlinear solution invites further studies concerning the range of its validity for $V_0 \ne 0$, when the dynamics is dominated by the large oscillations of the mean value $\langle u \rangle (t)$.

\section{Numerical methods}
\label{sec:NumMethod}
To solve the BKG equation (\ref{BousOst}) we use a pseudospectral method with a 4$^{\text{th}}$-order Runge-Kutta method for time-stepping, as was used in \cite{Engelbrecht, AGK2013,AGK2014}. Let us introduce
\begin{equation}
w = u - \epsilon \beta u_{xx},
\label{BousOstTrans}
\end{equation}
so that we have
\begin{equation}
w_{tt} = c^2 u_{xx} + \epsilon \lsq \frac{\alpha}{2} \lb u^2 \rb_{xx} - \gamma u \rsq.
\label{BousOstuw}
\end{equation}
Taking the Fourier transform of (\ref{BousOstTrans}) we obtain
\begin{equation}
\hat{w} = \lb 1 + \epsilon \beta k^2 \rb \hat{u} \RA \hat{u} = \frac{\hat{w}}{1 + \epsilon \beta k^2}.
\label{uwTransform}
\end{equation}
We take the Fourier transform of (\ref{BousOstuw}) and substitute (\ref{uwTransform}) into this expression to obtain an ODE in $\hat{w}$
\begin{equation}
\hat{w}_{tt} = -\frac{\epsilon \gamma + c^2 k^2}{1 + \epsilon \beta k^2} \hat{w} - \frac{\epsilon \alpha k^2}{2} \mathscr{F} \lset \mathscr{F}^{-1} \lsq \frac{\hat{w}}{1 + \epsilon \beta k^2} \rsq^2 \rset.
\label{wODE}
\end{equation}
We solve this ODE using a 4$^{\text{th}}$-order Runge-Kutta method for time-stepping e.g. \cite{AGK2013, KT2017}). Let us define the following:
\begin{equation}
\hat{w}_{t} = \hat{G}, \quad \hat{G}_{t} = \hat{S} \lb \hat{w} \rb,
\label{PSBOstSplit}
\end{equation}
where we defined $\hat{S}$ as the right-hand side of (\ref{wODE}). We discretise the time domain and functions as $t = t_n$, $\hat{w}(k, t_{n}) = \hat{w}_n$, $\hat{G}(k, t_{n}) = \hat{G}_n$ for $n=0,1,2,\dots$, where $t_{n} = n \Delta t$. Here $k$ discretises the Fourier space. Taking the Fourier transform of the initial conditions as defined in (\ref{IC}) and making use of (\ref{uwTransform}) we obtain initial conditions $\hat{w}_0$ and $\hat{G}_0$ of the form
\begin{align}
\hat{w}_0 &= \lb 1 + \epsilon \beta k^2 \rb \mathscr{F} \lset F(x) \rset, \notag \\
\hat{G}_0 &= \lb 1 + \epsilon \beta k^2 \rb \mathscr{F} \lset V(x) \rset.
\label{BOstWGIC}
\end{align}
Now we have initial conditions, we implement the following 4$^{\mathrm{th}}$-order Runge-Kutta method:
\begin{equation*}
\hat{w}_{n+1} = \hat{w}_n + \frac{1}{6} \lsq k_1 + 2k_2 + 2k_3 + k_4 \rsq, \quad \hat{G}_{n+1} = \hat{G}_n + \frac{1}{6} \lsq l_1 + 2l_2 + 2l_3 + l_4 \rsq,
\end{equation*}
where
\begin{align}
&k_{1} = \Delta t \hat{G}_n, &&l_{1} = \Delta t \hat{S}(\hat{W}_n), \notag \\
&k_{2} = \Delta t \lb \hat{G}_n + \frac{l_1}{2} \rb, &&l_{2} = \Delta t \hat{S} \lb \hat{W}_n + \frac{k_1}{2} \rb, \notag \\
&k_{3} = \Delta t \lb \hat{G}_n + \frac{l_2}{2} \rb, &&l_{3} = \Delta t \hat{S} \lb \hat{W}_n + \frac{k_2}{2} \rb, \notag \\
&k_{4} = \Delta t \lb \hat{G}_n + l_3 \rb, &&l_{4} = \Delta t \hat{S} \lb \hat{W}_n + k_3 \rb.
\label{BOstRK4}
\end{align}
This ystsem has to be solved in pairs i.e. we calculate $k_1$, then $l_1$, followed by $k_2$ and $l_2$, and so on. To obtain the solution in the real domain, we transform $\hat{w}$ back to $u$ through relation (\ref{uwTransform}). Explicitly we have
\begin{equation}
u(x,t) = \mathscr{F}^{-1} \lset \frac{\hat{w}}{1 + \epsilon \beta k^2} \rset.
\label{BOstufromw}
\end{equation}
To remove aliasing effects, we use the truncation 2/3-rule by Orszag in Boyd \cite{Boyd01}. This effect is due to the pollution of the numerically calculated Fourier transform by higher frequencies due to the series being truncated. 

For the Ostrovsky equations we again use a pseudospectral method similar to \cite{AGK2013, KT2017}. We will present the method for the non-homogeneous linearised Ostrovsky equation on non-zero background, specifically (\ref{phieq}). We take the equation for $\phi^{-}$ (as the equation for $\phi^{+}$ takes the same form) so we have
\begin{equation}
\lb 2 c \phi_{T}^{-} + \alpha \lb f^{-} \phi^{-} \rb_{\xi_{-}} + \beta \phi_{\xi_{-} \xi_{-} \xi_{-}}^{-} \rb_{\xi_{-}} = \gamma \phi^{-} + H \lb f^{-}, \xi_{-}, T \rb,
\label{fmOst}
\end{equation}
where $H$ has the form
\begin{equation}
H = f_{TT}^{-} + 2 c \beta f_{\xi_{-} \xi_{-} \xi_{-} T}^{-} + \frac{\gamma \tilde{\theta}^2}{2} f_{\xi_{-} \xi_{-}}^{-} - \frac{\alpha \tilde{\theta}^2}{2} \lb f_{\xi_{-}}^{-^2} \rb_{\xi_{-} \xi_{-}}.
\label{Heq}
\end{equation}
Taking the Fourier transform of (\ref{fmOst}) yields
\begin{equation}
\phi_{T}^{-} = -\frac{i}{2c} \lb \beta k^3 - \frac{\gamma}{k} \rb \phi^{-} - \frac{i \alpha k}{2c} \widehat{f^{-} \phi^{-}} - \frac{i}{2ck} \hat{H}.
\label{fmOstFour}
\end{equation}
We use the approach presented in \cite{Trefethen00} to remove the stiff term from this equation, so we multiply through by the multiplicative factor $M$ and introduce a new function $\Phi$, where $M$ and $\Phi$ take the form
\begin{equation}
M = e^{-\frac{i}{2c} \lb \beta k^3 - \frac{\gamma}{k} \rb T}, \qquad \hat{\Phi}^{-} = e^{-\frac{i}{2c} \lb \beta k^3 - \frac{\gamma}{k} \rb T} \hat{\phi}^{-},
\label{TPhieq}
\end{equation}
which gives an ODE for $\Phi$ of the form
\begin{equation}
\hat{\Phi}_{T}^{-} = - \frac{i \alpha k}{2c} M \mathscr{F} \lset f^{-} \mathscr{F}^{-1} \lsq \frac{\hat{\Phi}^{-}}{M} \rsq \rset - \frac{i}{2ck} M \hat{S}.
\label{PhiODE}
\end{equation}
This yields an optimised 4$^{\text{th}}$ order Runge-Kutta algorithm. Discretising the time domain as $T_i = i \Delta T$ and discretising the functions $\hat{\Phi}_{i}^{-} = \hat{\Phi}^{-} \lb k, T_i \rb$, $\hat{\phi}_{i}^{-} = \hat{\phi}^{-} \lb k, T_i \rb$ and $\hat{f}_{i}^{-} = \hat{f}^{-} \lb k, T_i \rb$, we introduce the function
\begin{equation}
E = e^{\frac{i}{4c} \lb \beta k^3 - \frac{\gamma}{k} \rb \Delta T}
\label{Eeq}
\end{equation}
and therefore we can use the optimised Runge-Kutta algorithm (written in the original variable $\phi^{-}$)
\begin{align}
\hat{\phi}_{i+1} &= E^2 \hat{\phi}_{i} + \frac{1}{6} \lsq E^2 k_1 + 2 E \lb k_2 + k_3 \rb + k_4 \rsq,  \quad \text{where}\notag \\
\quad k_1 &= -i \alpha s \Delta t \mathscr{F} \lset \hat{f}_{i} \mathscr{F}^{-1} \lsq \hat{\phi}_{i} \rsq \rset - \frac{i \Delta t}{sk} \hat{H}, \notag \\
k_2 &= - i \alpha k \Delta t \mathscr{F} \lset \hat{f}_{i} \mathscr{F}^{-1} \lsq E \lb \hat{\phi}_{i} + \frac{k_1}{2} \rb \rsq \rset - \frac{i \Delta t}{sk} \hat{H}, \notag \\
k_3 &= - i \alpha k \Delta t \mathscr{F} \lset \hat{f}_{i} \mathscr{F}^{-1} \lsq E \hat{\phi}_{i} + \frac{k_2}{2} \rsq \rset - \frac{i \Delta t}{sk} \hat{H}, \notag \\
k_4 &= - i \alpha k \Delta t \mathscr{F} \lset \hat{f}_{i} \mathscr{F}^{-1} \lsq E^2 \hat{\phi}_{i} + E k_3 \rsq \rset - \frac{i \Delta t}{sk} \hat{H}.
\label{RK4Opt}
\end{align}
We can apply this algorithm to the case of a homogeneous Ostrovsky equation by setting $\hat{S} = 0$ and replacing the term $f^{-} \phi^{-}$ with $f^{-^2}/2$. Similarly we can apply it to the case of $\phi^{+}$ and $f^{+}$ by changing the appropriate signs as shown in (\ref{feq}) and (\ref{phieq}).

\end{appendices}


\end{document}